\documentclass[doublespace]{article}
\usepackage{arxiv}

\usepackage[utf8]{inputenc} % allow utf-8 input
\usepackage[T1]{fontenc}    % use 8-bit T1 fonts
\usepackage{hyperref}       % hyperlinks
\usepackage{url}            % simple URL typesetting
\usepackage{booktabs}       % professional-quality tables
\usepackage{amsfonts}       % blackboard math symbols
\usepackage{nicefrac}       % compact symbols for 1/2, etc.
\usepackage{microtype}      % microtypography
\usepackage{lipsum}		% Can be removed after putting your text content
\usepackage{graphicx}

 \usepackage{moreverb}
\usepackage{pdflscape}
\usepackage{graphicx}
\usepackage{enumitem}

\usepackage{array}
\usepackage{mathtools} 
\usepackage{chngpage}

\DeclareMathAlphabet{\mathbfsf}{\encodingdefault}{\sfdefault}{bx}{sl}

\DeclareMathOperator{\Corr}{Corr}
\DeclareMathOperator{\Bin}{Bin}
\DeclareMathOperator{\logit}{logit}
\newcommand{\indep}{\raisebox{-0.05em}{$\perp\!\!\!\perp$}}

\usepackage{tikz}
\usetikzlibrary{mindmap}			
\usetikzlibrary{arrows,arrows.meta, calc,positioning, fit}
\usetikzlibrary{shapes,decorations, decorations.pathreplacing}
\definecolor{mylightskyblue}{rgb}{0.53,0.81,0.98}
\definecolor{mydodgerblue3}{rgb}{.09,0.45,0.80}
\definecolor{mynavy}{rgb}{0, 0, 0.5}

\usepackage{amsmath, stackrel, bm}
\usepackage[font=normalsize]{subcaption}
\usepackage{multirow}
\usepackage{acro}
\acsetup{first-style=long-short}
\DeclareAcronym{ATE}{
  short = ATE,
  long  = average treatment effect
}
\DeclareAcronym{ATT}{
  short = ATT,
  long  = average treatment effect of the treated 
}
\DeclareAcronym{CART}{
  short = CART,
  long  = Classification And Regression Tree
}
\DeclareAcronym{DAG}{
  short = DAG,
  long  = directed acyclic graph
}
\DeclareAcronym{IPTW}{
  short = IPTW,
  long  = Inverse Probability of Treatment Weighting
}
\DeclareAcronym{GBSG}{
  short = GBSG,
  long  = German Breast Cancer Study Group
}
\DeclareAcronym{MFP}{
  short = MFP,
  long  = Multivariable Fractional Polynomial
}\DeclareAcronym{MOB}{
  short = MOB,
  long  = Model-based Recursive Partitioning for Subgroup Analysis
}
\DeclareAcronym{MSE}{
  short = MSE,
  long  = Mean Squared Error
}
\DeclareAcronym{palmtree}{
  short = palmtree,
  long  = partially additive (generalized) linear model tree
}
\DeclareAcronym{RFS}{
  short = RFS,
  long  = Relapse-free survival
}\DeclareAcronym{PI}{
  short = PI,
  long  = permutation importance
}

\usepackage[numbers, super]{natbib}
\newcommand\BibTeX{{\rmfamily B\kern-.05em \textsc{i\kern-.025em b}\kern-.08em
T\kern-.1667em\lower.7ex\hbox{E}\kern-.125emX}}

\title{Tree-based exploratory identification of predictive biomarkers in observational data}

\author{{Julia Krzykalla} \\
	Division of Biostatistics\\
	German Cancer Research Center\\
	Im Neuenheimer Feld 280 \\
    D-69120 Heidelberg, GERMANY \\
	\texttt{j.krzykalla@dkfz-heidelberg.de}
	\And
	{Axel Benner\thanks{Both authors contributed equally to this work.}} \\
	Division of Biostatistics\\
	German Cancer Research Center\\
	Im Neuenheimer Feld 280 \\
    D-69120 Heidelberg, GERMANY \\
	\And
	{Annette Kopp-Schneider\footnotemark[1]} \\
	Division of Biostatistics\\
	German Cancer Research Center\\
	Im Neuenheimer Feld 280 \\
    D-69120 Heidelberg, GERMANY \\
}
\newcommand{\headeright}{J. Krzykalla}

\renewcommand{\shorttitle}{Tree-based identification of predictive factors in observational data}

\begin{document}
\maketitle

%\title{Tree-based exploratory identification of predictive biomarkers in observational data}

% \author[1]{Julia Krzykalla*}
% \author[1]{Axel Benner}
% \author[1]{Annette Kopp-Schneider}

% \address[1]{\orgdiv{Division of Biostatistics}, \orgname{German Cancer Research Center (DKFZ)}, \orgaddress{Im Neuenheimer Feld 280, D-69120 Heidelberg, \country{Germany}}}

% %\address[2]{\orgdiv{Org Division}, \orgname{Org name}, \orgaddress{\state{State name}, \country{Country name}}}
% %
% %\address[3]{\orgdiv{Org Division}, \orgname{Org name}, \orgaddress{\state{State name}, \country{Country name}}}

% \corres{*Julia Krzykalla, Division of Biostatistics, German Cancer Research Center, Im Neuenheimer Feld 280, D-69120 Heidelberg, Germany. \email{j.krzykalla@dkfz-heidelberg.de}}

%\presentaddress{Present address}

\begin{abstract}
The idea of ``stratified medicine'' is an important driver of methodological research on the identification of predictive biomarkers. Most methods proposed so far for this purpose have been developed for the use on randomized data only. However, especially for rare cancers, data from clinical registries or observational studies might be the only available data source. For such data, methods for an unbiased estimation of the average treatment effect are well established. Research on confounder adjustment when investigating the heterogeneity of treatment effects and the variables responsible for this is usually restricted to regression modelling. \\
In this paper, we demonstrate how the predMOB, a tree-based method that specifically searches for predictive factors, can be combined with common strategies for confounder adjustment (covariate adjustment, matching, \acf{IPTW}). In an extensive simulation study, we show that covariate adjustment allows the correct identification of predictive factors in the presence of confounding whereas \ac{IPTW} fails in situations in which the true predictive factor is not completely independent of the confounding mechanism. A combination of both, covariate adjustment and \ac{IPTW} performs as well as covariate adjustment alone, but might be more robust in complex settings. An application to the \ac{GBSG} Trial $2$ illustrates these conclusions.
\end{abstract}

%\keywords{Predictive factors, effect modification, predMOB, observational data, confounder adjustment}

%\jnlcitation{\cname{%
%\author{Williams K.}, 
%\author{B. Hoskins}, 
%\author{R. Lee}, 
%\author{G. Masato}, and 
%\author{T. Woollings}} (\cyear{2016}), 
%\ctitle{A regime analysis of Atlantic winter jet variability applied to evaluate HadGEM3-GC2}, \cjournal{Q.J.R. Meteorol. Soc.}, \cvol{2017;00:1--6}.}

\section{Introduction}
Especially in cancer research, the idea of a therapy that is specifically tailored to the individual patient has been increasingly sought in recent decades. The decision as to which is the best treatment option is usually founded on baseline measures (for example demographics, disease characteristics, laboratory parameters or genetic information) that determine whether one or the other treatment is effective to the patient. Those factors are called ``predictive'' factors or ``treatment effect modifiers''. Of particular importance are qualitative predictive factors that define strata with opposing treatment effects, i.e., a major benefit in one group and a harmful effect in the complementary group. On the contrary, quantitative predictive factors only indicate a difference in the size of the effect. Treatment effect modification is not to be confused with a true causal interaction which denotes the situation of two treatments affecting the outcome together.\cite{vanderweele2009} Assessing whether a factor has a causal interaction with the treatment of primary interest would be of relevance if it can be manipulated to achieve a better treatment effect, whereas this work is restricted to the search for factors that define strata with differential treatment effects. \\
Importantly, treatment effect modification only applies to a very specific context. \cite{vanderweele2012} A predictive factor for the relationship of a certain treatment on a certain outcome measure does not necessarily have to be a predictive factor for another drug as well, even if that drug targets the same disease; nor does it imply effect modification with respect to an alternative effect measure. Moreover, effect modification is relative to the population as it might be affected by the distribution of a third factor which is neither the treatment nor the outcome. A very common example for such a factor is the ethnicity of a patient. \cite{ICHE5} \\
In a statistical regression model, predictive factors are represented by an interaction term of the respective variable with treatment. If only a few variables are to be tested, simple interaction tests, possibly with a subsequent adjustment for multiple testing, are sufficient although they may be compromised by low power. \cite{Marshall2007} Yet, in situations in which the pathways of the disease itself or the treatment's mode of action are not fully understood, the number of potential candidates might be large. In these cases, tree-based methods are advantageous for performing an exploratory search across all candidates. \cite{DGHO2016} Besides being able to handle a large number of factors at the same time (especially in combination with ensemble methods like random forests), these methods are also very flexible as they do not require prespecification with regard to the functional form of the interaction or the order of the interaction. Prominent examples of such methods are Interaction Trees, \cite{Su2009} causal forests \cite{Athey2016} or \ac{MOB}.\cite{Seibold2016} Most of these methods provide a precise prediction of the individual treatment effect, but ignore interpretability of the resulting tree or forest. That is, factors with exclusively prognostic effects (i.e. they refine the outcome irrespective of any treatment) may be selected as splitting variables as well. Furthermore, simultaneous selection of splitting variable and cut-off, as is done by most methods, favors variables with multiple cut-offs, particularly continuous variables (``variable selection bias''). \citeauthor{Krzykalla2020} proposed the predMOB, which, as an extension of the \ac{MOB}, is also free of variable selection bias and is specifically tailored to identify predictive factors. \cite{Krzykalla2020} \\
Like the majority of these methods (except for the causal forests), the predMOB has been developed for randomized settings. In practice, data from randomized trials might not always be available so one has to rely on observational data. It is well known that estimating the treatment effect on such kind of data involves the risk of confounding bias and the use of adjustment methods is well established. The purpose of this paper is to investigate how predMOB can be used to correctly identify predictive factors based on observational data. We examine whether adjustment for confounding is necessary and if so, which of the common adjustment strategies (covariate adjustment, \ac{IPTW} or matching) can achieve an appropriate correction. We first describe the causal notation of effect modification in a simple setting and explain how unbiased results can be obtained from a simple interaction model. In the same section, the predMOB approach is introduced and its combination with the mentioned adjustment methods is outlined. In a simulation study, we investigate which of the adjustment strategies allows a correct identification of predictive factors in a non-randomized setting and how individual treatment effects and modifying effects of the predictive factors can be estimated without bias. In section \ref{sec:GBSG}, we apply predMOB in combination with the same adjustment methods to the \ac{GBSG} Trial $2$ data set to identify predictive factors for hormonal therapy with tamoxifen in node-positive breast cancer patients. 

%%%%%%%%%%%%%%%%%%%%%%%%%%%%%%%%%%%%%%%%%%%%%%%%%%%%%%%%%%%%%%%%%%%%%%%%%%%%%%%%%%%%%%%%%%%%%%%%%%%%%%%%%%%%%%%%%%%%%%%

\section{Methods}\label{sec:methods}
This section briefly introduces the basic principles of causal inference and explains how to quantify effect modification in the presence of confounding within a simple regression model. For more details on this topic, we refer the reader to the book of \citet{Hernan2020} and the papers by \citet{Robins2000} and \citeauthor{vanderweele2009}.\cite{vanderweele2009}
We then describe how predictive factors can be identified using the predMOB \cite{Krzykalla2020} and how the adjustment methods mentioned in the first section can be combined with such an approach. 

\subsection{Notation and basic assumptions}\label{subsec:notation}
Consider the following setting where for each patient $i, \ i=1, \ldots, n$, $Y_i$ is the outcome of interest and $T_i$ the administered treatment ($T_i=1$ if the patient received the experimental treatment and $T_i=0$ for placebo/control). The average causal treatment effect is the expected difference in the potential outcomes $\mathbb{E}(Y^1-Y^0)$ whereas $Y^t$ denotes the (hypothetical) outcome under treatment $T=t$. In practice, only one of the potential outcomes can be observed for a patient, namely the outcome under the treatment actually administered. However, it is still possible to estimate this difference in potential outcomes if three conditions are met: positivity, consistency and exchangeability. The positivity condition requires that all patients have a non-zero probability of receiving each of the treatment options. Consistency is the connection of the potential outcome to the observed data. It says that if patient $i$ received treatment $T_i=t$, then $Y_i=Y_i^t$ which ensures that $\mathbb{E}(Y^t|T=t)=\mathbb{E}(Y|T=t)$. This condition seems to be self-evident, but might be violated, for example, if two patients receive different doses of the same treatment but only the substance itself is recorded. The potential outcomes would then also depend on the dosage and thus be different, although $T=1$ for both patients. The third condition, exchangeability (or unconfoundedness), is the independence of $T$ and the potential outcomes $(Y^0, Y^1)$. In other words, potential outcomes are baseline characteristics that are known prior to treatment allocation. In randomized trials, exchangeability (as well as the other two conditions) is ensured by the random assignment of the patients to a treatment arm. Under these conditions, the average causal effect is identifiable
\[
\mathbb{E}(Y|T=1)-\mathbb{E}(Y|T=0) \overset{\text{consistency}}{=}\mathbb{E}(Y^1|T=1)-\mathbb{E}(Y^0|T=0)\stackrel[\text{positivity}]{\text{exchangeability}}{=}\mathbb{E}(Y^1-Y^0).
\]
In the absence of randomization, exchangeability can be moderated to a conditional exchangeability (independence of $T$ and $(Y^0, Y^1)$ given the confounders $\bm{X}$), assuming that there is no unmeasured confounding. Basically all methods of causal inference rely on this assumption. In the non-randomized case, the causal treatment effect can be estimated either conditional on the confounding variables $\bm{X}$
\[
\mathbb{E}(Y|T=1, \bm{X})-\mathbb{E}(Y|T=0, \bm{X}) =\mathbb{E}(Y^1|T=1, \bm{X})-\mathbb{E}(Y^0|T=0, \bm{X})=\mathbb{E}(Y^1-Y^0|\bm{X})
\]
or marginally over the whole population by using appropriate adjustment methods (see section \ref{subsec:confounding}). Both strategies, marginal and conditional estimation, produce valid results, but with a different meaning. Conditional estimates measure the difference in the outcome for an individual if he or she would switch from one treatment to the other, given all other covariates remain fixed. In contrast to this, the marginal estimate gives the difference that is observed on average for the entire population. Even using an appropriate confounder adjustment, these two estimates may not necessarily coincide. In this case, the estimates are called non-collapsible which is true for odds ratios or hazard ratios, for example.

\subsection{Confounder adjustment for the marginal treatment effect}\label{subsec:confounding}
There are three common strategies for confounder adjustment if the marginal treatment effect shall be estimated: standardization, \ac{IPTW} or matching. Standardization essentially means to estimate the average causal effects within the strata defined by the confounder(s) and to use the law of total probability to compute the standardized mean as the weighted mean of the stratum-specific causal effects $\sum_x \mathbb{E}(Y|X=x, T=t) \mathbb{P}(X=x)$. Evidently, the calculation becomes burdensome with increasing numbers of confounders or if the confounding variable is continuous. In contrast to this, \ac{IPTW} involves weighted regression with weights based on the propensity score $e(X)=\mathbb{P}(T=1|X=x)$, the probability of receiving the active treatment given all confounding covariates. The propensity score captures all information such that $T \indep (Y^1,Y^0)|e(X)$. In practice, the relationship between treatment assignment (as an outcome) and confounders is modelled using, for example, logistic regression (``propensity model''). The inverse of the predicted probabilities of receiving the actual treatment (that is, in the simplest case, $1/e(x)$ for a patient in the active treatment arm and $1/(1-e(x))$ for a patient in the control arm) is used to weight the observations in the outcome regression model. The weighting of observations creates a pseudo-population in which all considered covariates are balanced across the treatment arms as in a randomized trial. In order to avoid weights that are too extreme and to limit the increase in variability for the treatment effect estimate, weights can be stabilized and/or trimmed as recommended by \citet{Cole2008} and \citet{Lee2011}\\
Another strategy to achieve balance in the covariate distributions is to match each patient in the experimental arm with one or more patients of the control arm so that they are as similar as possible with respect to the confounding variables. In the ideal case, observations are matched in pairs or sets with identical information in all confounders (``exact matching''). Evidently, this is only feasible when there are relatively few confounders that are strictly categorical. Alternatively, similarity is quantified by means of distance measures, such as the propensity score or the Mahalanobis distance. A matching scheme that minimizes the global distance between the matched patients with respect to the confounders (``optimal matching'') is generally preferred over matching the patients in the order they appear in the data set (``greedy matching''). The matching ratio of treated and controls can either be fixed in advance (``k:1 matching'') or it can be kept variable in order to avoid discarding a large amount of information due to non-matching observations. The latter approach, called ``full matching'' was shown to be particularly effective at reducing bias due to confounding.\cite{Stuart2008} Regardless of the matching procedure used, unless the matching ratio is $1:1$, observations need to be weighted according to the size of the subclass to which they belong. For example, if within one set, $2$ controls have been matched to one treated patient, the treated patient receives a weight of $1$, while the controls both are weighted with $\frac{1}{2}$.\\
As matching selects the control group such that the patients correspond as closely as possible to the treated patients, this only allows the estimation of the \ac{ATT} whereas weighting can be used for estimating the \ac{ATT} as well as the \ac{ATE}. The two only coincide in the absence of selection bias. The major drawback of all matching methods is the risk of losing observations for the analysis.

\subsection{Effect modification in the presence of confounding}\label{subsec:effectmod_confounding}
In the previous sections, methods have been introduced that allow valid estimation of the treatment effect despite possible confounding. As the focus of this paper is on the identification of predictive factors, the interest lies rather on the unbiased quantification of effect modification. Effect modification means that the average causal effect differs across strata of an effect modifier $M$. For the sake of simplicity, let $M$ be binary such that
\[
\mathbb{E}(Y^1-Y^0|M=m_1) \neq \mathbb{E}(Y^1-Y^0|M=m_0).
\]
The common way of assessing whether $M$ is a treatment effect modifier is to test whether $\gamma$ is non-zero in the following conditional regression model
\begin{equation}\label{eq:Intmodel_cond}
\mathbb{E}(Y|T, M)=\alpha + \beta_T T + \beta_M M + \gamma MT.
\end{equation}
In the presence of confounding, all relevant confounder variables $X$ have to be included as covariates in this model.\\
Alternatively, treatment effect modification can be assessed by means of a marginal structural model including an effect for $M$, as described in \citet{vanderweele2009}
\begin{equation}\label{eq:Intmodel_marg}
\mathbb{E}(Y^t|M)=\alpha + \beta_T t + \beta_M M + \gamma Mt.
\end{equation}
Although model \eqref{eq:Intmodel_marg} looks very similar to model \eqref{eq:Intmodel_cond} above, there is one essential difference: it is not a model for the observed outcome $Y$ but for the potential outcome $Y^t$. In order to be able to estimate the model parameters in this model based on observed data, confounding must be ruled out. This is usually achieved by creating a pseudo-randomized population using inverse probability of treatment weights (cf. section \ref{subsec:confounding}). In doing so, it is crucial to include the treatment effect modifier as a covariate in the propensity model, that is, the denominator is given by $\mathbb{P}(T_i=1|X_i=x_i, M_i=m_i)$. By including $\mathbb{P}(T_i=1|M_i=m_i)$ in the numerator of the weights, the estimates are likely to be more efficient, but it is no prerequisite to obtaining valid results.\cite{Robins2000} \\
Another option is to use matching instead of \ac{IPTW}. Again, the treatment effect modifier has to be taken into account as a matching variable. In case of a simple setting that allows for exact matching, let $I$ be the number of matched pairs and $T_{ij}$ the treatment indicator of observation $j \ (j \in \{1,2\})$ belonging to pair $i \ (i = 1, \ldots, I)$. Then the observed outcome difference of treated minus control $D_i=(T_{i1}-T_{i2})(Y_{i1}-Y_{i2})$ can be interpreted as the causal effects given the respective covariate information and so treatment effect modification can be directly tested by means of a simple two-group comparison of the differences in potential outcomes between the strata defined by $M$. However, as mentioned above, exact matching is often not feasible and so treatment effect modification can usually only be investigated on the outcome level rather than exploiting the differences. 

%%%%%%%%%%%%%

\subsection{Identification of predictive factors using the predMOB}\label{subsec:predMOB}
As mentioned in the introduction, the \ac{MOB} \cite{Seibold2016} is one of the popular examples of tree-based methods for subgroup identification. The method starts with a base model that regresses the outcome on treatment 
\begin{equation}\label{eq:MOBbase}
   \mathbb{E}(Y|T)=a+b \cdot T 
\end{equation}
and then recursively splits the population in two subgroups whenever the coefficients of this base model, either the intercept $a$ or the coefficient for the treatment effect $b$, are ``instable'' with respect to one of the potential splitting variables and thus differential across the resulting subgroups. This instability of the model parameters is assessed using the M-fluctuation test.\cite{Zeileis2007} Used in connection with ensemble methods, more precisely with random forests, this approach produces reliable estimates of the individual treatment effects. However, it has a major drawback with regard to the identification of predictive factors as the splitting rule allows splits for prognostic factors as well, as these are factors responsible for ``instabilities'' in the intercept. Consequently, it is not possible to distinguish between prognostic and predictive factors, and variable importance measures rank variables according to both types of effects. To overcome this problem and focus on predictive factors in the construction of the tree, \citet{Krzykalla2020} proposed to reparameterize the base model using effect coding instead of dummy coding for the treatment variable ($T^\ast = 1 \text{ for active treatment and } T^\ast = -1 \text{ for control}$). In doing so, the resulting working model consists only of the treatment effect without intercept
\begin{equation}\label{eq:MOBpred}
\mathbb{E}(Y|T^\ast)= b \cdot T^\ast /2    
\end{equation}
such that each split can be attributed to a predictive factor, while the estimate $\hat{b}$ for the treatment effect in model \eqref{eq:MOBpred} is the same as the estimate for $b$ in model \eqref{eq:MOBbase}. In order to decide whether a variable is predictive, multiple predMOB trees are grown - each on a subsample of the original data - and the importance of all potential predictive factors on the construction of this random forest can be assessed using variable importance measures such as the permutation importance \cite{Breiman2001} or the mean minimal depth. \cite{Ishwaran2010} Whereas permutation importance describes the loss in prediction accuracy caused by breaking the relationship between the variable of interest and the outcome via permutation, mean minimal depth only takes into account the structure of the tree and gives the shortest distance of the first node in which the variable of interest is used for splitting and the root node. The use of subsampling (without replacement) rather than bootstrapping (with replacement) is especially preferable when using random forests in combination with permutation importance to avoid the introduction of bias. \cite{Strobl2007}\\
This modification of the \ac{MOB} achieves a concentration on predictive factors, reducing the number of variables that are falsely identified due to a purely prognostic effect.
An additional advantage of the modification \eqref{eq:MOBpred} is that predictions for the individual treatment effect can be deduced directly from the parameter estimate $\hat{b}$ for the base model in the terminal node into which an observation is classified. If multiple trees are grown, the final prediction is obtained by aggregating all single-tree predictions, e.g. by calculating the mean.\\
The modification is inspired by the modified covariates approach for (generalized) linear regression models of \citet{Tian2014} In addition to using an effect-coded treatment $T^\ast$, the biomarkers that are assumed to be predictive are multiplied by $T^\ast /2$ in order to directly estimate the corresponding interaction terms with treatment. 

\subsection{Confounder adjustment when using predMOB}\label{subsec:adjpredMOB}
Combining the adjustment methods for simple regression models introduced in sections \ref{subsec:confounding} and \ref{subsec:effectmod_confounding} with tree-based methods or methods based on random forests is not straightforward. The analogue of covariate adjustment in an ordinary regression model for a \ac{MOB} tree would be a so-called \ac{palmtree}. \cite{Seibold2019} The fit of a MOB tree can be expressed via a regression model with linear predictor 
\[
\boldsymbol{\eta} = I(s=1)\boldsymbol{x}^T\boldsymbol{\beta}_1 + I(s=2)\boldsymbol{x}^T\boldsymbol{\beta}_2 + \ldots = \boldsymbol{x}^T\boldsymbol{\beta}(s)
\]
where $\boldsymbol{x}^T\boldsymbol{\beta}$ is equal to the linear predictor of the base model as in \eqref{eq:MOBbase} and $s=1,2,\ldots$ represents the assignment to the subgroup that corresponds to the leaf node in the tree. This means that all model parameters $(\boldsymbol{\beta}^T_1, \boldsymbol{\beta}^T_2, \ldots)$ are depending on the structure of the tree. In contrast to this, \ac{palmtree}s allow the inclusion of factors with a global effect ($\boldsymbol{x_F}$) on the outcome, that is, their effects $\boldsymbol{\gamma}$ are fixed and do not depend on the tree structure: 
\[
\boldsymbol{\eta} =  \boldsymbol{x}^T\boldsymbol{\beta}(s) + \boldsymbol{x_F}^T\boldsymbol{\gamma}.
\]
As the only difference between \ac{MOB} and predMOB is the parameterization of the base model, the combination of covariate adjustment and a predMOB tree is analogous.\\
Combining the predMOB with \ac{IPTW} means to use the weights obtained from an appropriate propensity model both in fitting the base model as well as in computing the M-fluctuation test statistic for the splitting decision throughout the entire construction of the tree. Just as weighting, also matching is done separately from the construction of the tree itself. The matching procedure is applied to the full population and subsampling is performed on the matched pairs or sets. The matching weights (for optimal matching, and for exact matching as well) are used again as case weights in the construction of the trees.

%%%%%%%%%%%%%%%%%%%%%%%%%%%%%%%%%%%%%%%%%%%%%%%%%%%%%%%%%%%%%%%%%%%%%%%%%%%%%%%%%%%%%%%%%%%%%%%%%%%%%%%%%%%%%%%%%%%%%%%

\section{Simulation study}\label{sec:sim}
We investigate whether predMOB in combination with common adjustment methods (cf. section \ref{subsec:adjpredMOB}) yields reliable results when applied to non-randomized data. The adjustment methods explored in this simulation study are covariate adjustment, matching, \ac{IPTW}, and a doubly robust approach combining covariate adjustment and \ac{IPTW} (for details, see section \ref{subsec:confounding}). Since for the sake of simplicity, biomarkers are generated as binary, exact matching can be investigated here as ``ideal'', while full optimal matching is examined as an alternative that is more widely applicable. The calculation of the \ac{IPTW} weights as well as the optimal matching procedure are done once for the entire data set and passed to the subsamples for the construction of the single trees. Re-calculation of the weights or repeating the matching procedure within the subsamples has also been tested, but results are similar and thus not presented. All weights are rescaled such that they sum up to the original number of observations in the generated data set and thus, variance estimates can be compared. In order to reflect a real application when the true set of confounders is not known with certainty, all biomarkers are used in the matching procedure and for fitting the propensity model.\\
The evaluation is made with regard to the correct identification of the predictive factor(s) as well as the accuracy of the predicted individual treatment effects and the estimated modifying effect of the predictive factor itself. A factor is considered to be predictive if it shows high variable importance (high permutation importance or low mean minimal depth). \\
Throughout the simulation studies, every forest is an ensemble of $100$ predMOB trees and every tree is fit to a subsample of the original data comprising $63.2 \%$ of all observations (to mimic the amount of information in a bootstrap sample drawn with replacement, cf. \citet{Binder2008}). For the splitting decision, the raw p-values of the M-fluctuation test (without adjustment for multiplicity) are compared against a significance level of $\alpha=0.05$ in order not to be too restrictive. All results are based on $1000$ simulation runs. \\

\subsection{Identification of predictive factors}
For the evaluation of predMOB in identifying the true predictive factor(s), data has been generated according to the following data-generating process and outcome generating models using the simstudy package in R: 
\begin{itemize}
    \item Ten independent and identically distributed (i.i.d.) biomarkers $X_1, \ldots, X_{10}$; binary with $X_i\sim \Bin(1,0.5)$,
    \item Binary treatment variable $T \sim \Bin(1,p)$; depending on biomarkers $X_1, \ldots, X_7$ via the following logistic regression model
    \[
    \logit(\mathbb{P}(T=1))= \beta_0+\log(1.25)X_1 + \log(1.5)X_2 + \log(1.75)X_3 + \log(1.25)X_4 + \log(1.5)X_5 + \log(1.75)X_6 + \log(2)X_7,
    \]
    with $\beta_0$ being chosen so that treatment groups are equally sized  ($p=0.5$), 
    \item Normally distributed outcome variable $Y \sim N(\mu, 0.25)$ with expectation $\mu$ as defined in Table \ref{tab:simsettings_basic},
    \item Sample size $n=1000$ (resulting in $632$ observations per subsample). 
\end{itemize}
The number of potential splitting variables is set to $10$ such that all biomarkers are eligible for each split. 

\begin{table}[ht!]
\caption{Parameter configurations for the simulation study concerning the identification of predictive factors}
\begin{adjustwidth}{-.5in}{-.5in}  
    \label{tab:simsettings_basic}
	\centering
    \fontsize{9}{15}\selectfont
	\begin{tabular}{l l}
    \toprule
    Description & Parameter configuration \\
    \midrule
    0 : Null scenario & $\mu=0$\\
    A : Prognostic effects only & $\mu=0.5T + 0.2X_4 + 0.3X_5 + 0.4X_6 + 0.5X_7 + 0.4X_8 + 0.2X_9 + 0.3 X_{10} + 0 X_{10}\cdot T$\\
       &  B.1 \ $\mu=0.5T + 0.2X_4 + 0.3X_5 + 0.4X_6 + 0.5X_7 + 0.4X_8 + 0.2X_9 + 0.3 X_{10} - 1 X_{10}\cdot T$ \\
    \multirow{-2}{*}{B: $X_{10}$ is prognostic and predictive} &  B.2 \ $\mu=0.5T + 0.2X_4 + 0.3X_5 + 0.4X_6 + 0.5X_7 + 0.4X_8 + 0.2X_9 + 0.3 X_{10} + 1 X_{10}\cdot T$\\
      & C.1 \ $\mu=0.5T + 0.2X_4 + 0.3X_5 + 0.4X_6 + 0.5X_7 + 0.4X_8 + 0.2X_9 + 0 X_{10} - 1 X_{10}\cdot T$\\
      \multirow{-2}{*}{C: $X_{10}$ is predictive only} & C.2 \ $\mu=0.5T + 0.2X_4 + 0.3X_5 + 0.4X_6 + 0.5X_7 + 0.4X_8 + 0.2X_9 + 0 X_{10} + 1 X_{10}\cdot T$ \\
       &  D.1 \ $\mu=0.5T + 0.2X_4 + 0.3X_5 + 0.4X_6 + 0.5X_7 + 0.4X_8 + 0.2X_9 + 0.3 X_{10} - 1 X_{3}\cdot T$ \\
    \multirow{-2}{*}{D: $X_{3}$ is predictive only} &  D.2 \ $\mu=0.5T + 0.2X_4 + 0.3X_5 + 0.4X_6 + 0.5X_7 + 0.4X_8 + 0.2X_9 + 0.3 X_{10} + 1 X_{3}\cdot T$\\
      & E.1 \ $\mu=0.5T + 0.2X_4 + 0.3X_5 + 0.4X_6 + 0.5X_7 + 0.4X_8 + 0.2X_9 + 0.3 X_{10} - 1 X_{7}\cdot T$\\
      \multirow{-2}{*}{E: $X_{7}$ is prognostic and predictive} & E.2 \ $\mu=0.5T + 0.2X_4 + 0.3X_5 + 0.4X_6 + 0.5X_7 + 0.4X_8 + 0.2X_9 + 0.3 X_{10} + 1 X_{7}\cdot T$ \\
      &  F.1 \ $\mu=0.25T + 0.2X_4 + 0.3X_5 + 0.4X_6 + 0.5X_7 + 0.4X_8 + 0.2X_9 + 0.3 X_{10} - 0.5 X_{10}\cdot T$\\
    \multirow{-2}{*}{\shortstack{F: Scenario B, but smaller \\ interaction effects}} &  F.2 \ $\mu=0.25T + 0.2X_4 + 0.3X_5 + 0.4X_6 + 0.5X_7 + 0.4X_8 + 0.2X_9 + 0.3 X_{10} - 0.25 X_{10}\cdot T$ \\
     &  G.1 \ scenario~B.1, but $\Corr(X_7,X_{10})=0.5$ \\
   \multirow{-2}{*}{G: Correlated biomarkers}&  G.2 \ scenario~B.1, but $\Corr(X_7,X_{10})=-0.7$ \\
     & H.1 \ $\mu=0.5T + 0.2X_4 + 0.3X_5 + 0.4X_6 + 0.5X_7 + 0.4X_8 - 1.2X_9T - 1X_{10}T$ \\
     \multirow{-2}{*}{H: Multiple predictive biomarkers}  & H.2 \ $\mu=0.5T + 0.2X_4 + 0.3X_5 + 0.4X_6 + 0.5X_7 + 0.4X_8 + 0.6X_9 + 0.3X_{10} - 1X_9 T - 1X_{10} \cdot T$ \\
    I: Higher order predictive pattern & $\mu=0.5T + 0.2X_4 + 0.3X_5 + 0.4X_6 + 0.5X_7 + 0.4X_8 + 0.2X_9 - 1X_9 \cdot X_{10} \cdot T$, $n=2500$ \\
    J: Higher dimensions &  add $V_1, \ldots, V_{20}, \ V_i\sim \Bin(1,0.5)$ to scenario~C.1\\
    \bottomrule
    \end{tabular}
    \end{adjustwidth}
\end{table}

\begin{figure}
\begin{center}
\begin{subfigure}[b]{.33\textwidth}
\subcaption{Scenarios  B, C, F-H}
\begin{tikzpicture}[node distance=1cm, thick]
\node (i) [draw, fill=mylightskyblue, align=center] {$X_1$-$X_3$};
\node (c) [draw, fill=mydodgerblue3, align=center, below = 1.5cm of i] {$X_4$-$X_7$};
\node (p) [draw, fill=mynavy, text=white,  align=center, below = .75cm of c] {$X_8$-$X_9$};
\node (pred) [draw, align=center, fill=red, below = 1.5cm of c] {$X_{10}$};
\node (t) [draw, align=center, right= .75cm  of c] {$T$};
\node (y) [draw, align=center, right= 1.5cm  of t] {$Y$};

\draw[-{Latex[length=5mm, width=2mm]}]  (i) edge (t)
        (c) edge (t)
        (c) edge[bend right] (y)
        (p) edge[bend right] (y) 
        (pred) edge[bend right] (y) 
        (t) edge (y);
\draw[-{Latex[length=5mm, width=2mm]}, dashed]  (pred) edge[bend right]  ($ (t) !.5! (y) $);
\end{tikzpicture}
\end{subfigure}\hfill
\begin{subfigure}[b]{.33\textwidth}
\subcaption{Scenario D}
\begin{tikzpicture}[node distance=1cm, thick]
\node (i) [draw, fill=mylightskyblue, align=center] {$X_1$-$X_2$};
\node (pred) [draw, align=center, fill=red, below = 0.1cm of i] {$X_3$};
\node (c) [draw, fill=mydodgerblue3, align=center, below = 1.5cm of i] {$X_4$-$X_7$};
\node (p) [draw, fill=mynavy, text=white, align=center, below = 1.5cm of c] {$X_8$-$X_{10}$};
\node (t) [draw, align=center, right= 0.75cm  of c] {$T$};
\node (y) [draw, align=center, right= 1.5cm  of t] {$Y$};

\draw[-{Latex[length=5mm, width=2mm]}]  (i) edge[bend left] (t)
        (pred) edge[bend left] (t) 
        (c) edge (t)
        (c) edge[bend right] (y)
        (p) edge[bend right] (y) 
        (t) edge (y);
\draw[-{Latex[length=5mm, width=2mm]}, dashed]  (pred) edge[bend left] ($ (t) !.5! (y) $);
\end{tikzpicture}
\end{subfigure}\hfill
\begin{subfigure}[b]{.33\textwidth}
\subcaption{Scenario E}
\begin{tikzpicture}[node distance=1cm, thick]
\node (i) [draw, fill=mylightskyblue, align=center] {$X_1$-$X_3$};
\node (c) [draw, fill=mydodgerblue3, align=center, below = 1.5cm of i] {$X_4$-$X_6$};
\node (pred) [draw, align=center, fill=red, below = 2.25cm of i] {$X_7$};
\node (p) [draw, fill=mynavy, text=white, align=center, below = 1.5cm of c] {$X_8$-$X_{10}$};
\node (t) [draw, align=center, right= 0.75cm  of c] {$T$};
\node (y) [draw, align=center, right= 1.5cm  of t] {$Y$};

\draw[-{Latex[length=5mm, width=2mm]}]  (i) edge (t)
        (c) edge (t)
        (c) edge[bend left] (y)
        (pred) edge (t) 
        (pred) edge[bend right] (y) 
        (p) edge[bend right] (y) 
        (t) edge (y);
\draw[-{Latex[length=5mm, width=2mm]}, dashed]  (pred) edge[bend right] ($ (t) !.5! (y) $);
\end{tikzpicture}
\end{subfigure}
\end{center}
\caption[Graphical illustration of the simulation scenarios in the style of causal \aclp{DAG}]{Graphical illustration of the simulation scenarios in the style of causal \acp{DAG}. Effect-modification is marked as dashed arrows on the egde representing the treatment effect (cf. \citet{Weinberg2007}). Instrumental variables are shown in light blue, true confounders in medium blue and factors only associated with outcome in dark blue; predictive factors are highlighted in red.}
\label{fig:Simscenarios}
 \vspace{-.3cm}
\end{figure}
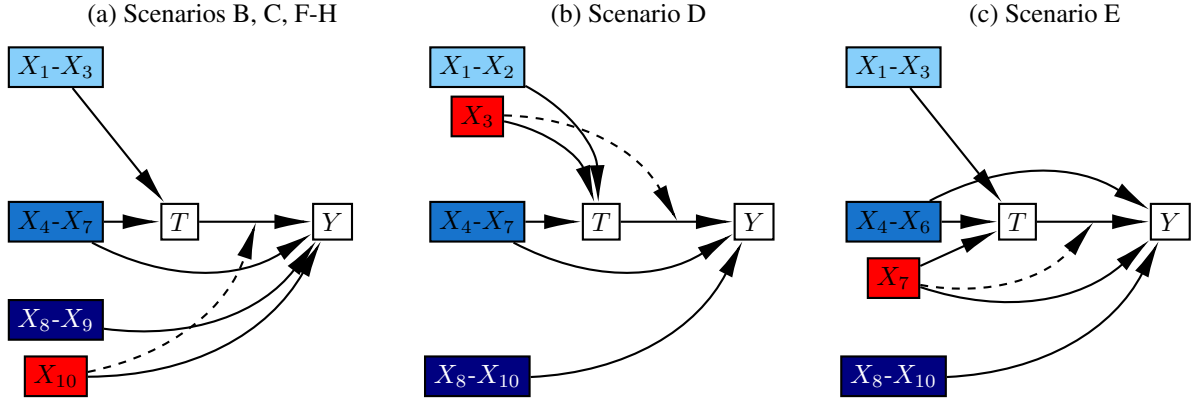

While the first two scenarios serve as reference scenarios with none of the biomarkers being predictive, in scenarios~B and C, biomarker $X_{10}$ is the true predictive factor. Apart from being predictive, it has at most a prognostic effect, but is not associated with the treatment assignment. In contrast to this, in scenario~D, the predictive factor $X_{3}$ is an instrumental variable, that is, it is associated with the treatment assignment, but has no direct relationship to the outcome. Finally, in scenario~E, the confounding variable $X_{7}$ (associated with the treatment assignment as well as the outcome) is chosen as the true predictive factor. Scenarios~F, G and J describe more complex settings in order to investigate how results are affected by design modifications such as non-zero correlations between the true predictive factor and other prognostic factors, small interaction effects (``predictive effects'') or an increasing number of candidate biomarkers. Furthermore, scenarios~H and I add complexity in the effect modification pattern: either two biomarkers are altering the treatment effect independently or act in combination. An illustration of the simulation scenarios is given in Figure \ref{fig:Simscenarios}. Since the results for some of the scenarios are very similar, only parts of the simulation study are discussed in detail in the main part of the paper (see Figure \ref{fig:Simresults_VI}). The results of the remaining scenarios are presented in the supplement (see Figure S.1 in supplement~S~1).\\
The variable importance values over all simulation runs are summarized in boxplots. A colour code is used to distinguish the biomarkers according to their function: instrumental variables are shown in light blue, true confounders in medium blue and factors only associated with outcome (prognostic effect) are coloured in dark blue. The boxplot for a true predictive factor is highlighted in red. Since the conclusions for the two variable importance measures are consistent, results for mean minimal depth are presented only in the supplement (cf. Figure S.2 in supplement~S~2).  \\

\begin{figure}
\begin{center}
% \begin{subfigure}{.49\textwidth}
% \subcaption{\textbf{Scenario 0}: null scenario}
% \centering
% \includegraphics[height=5cm]{figures/predMOBpermimp_obs00_n1000.pdf}
% \label{fig:Simresults_VI_0}
% \end{subfigure}\hfill
\begin{subfigure}[b]{.49\textwidth}
\subcaption{\textbf{Scenario~A}: Prognostic effects only \newline}
\centering
\includegraphics[height=3.8cm]{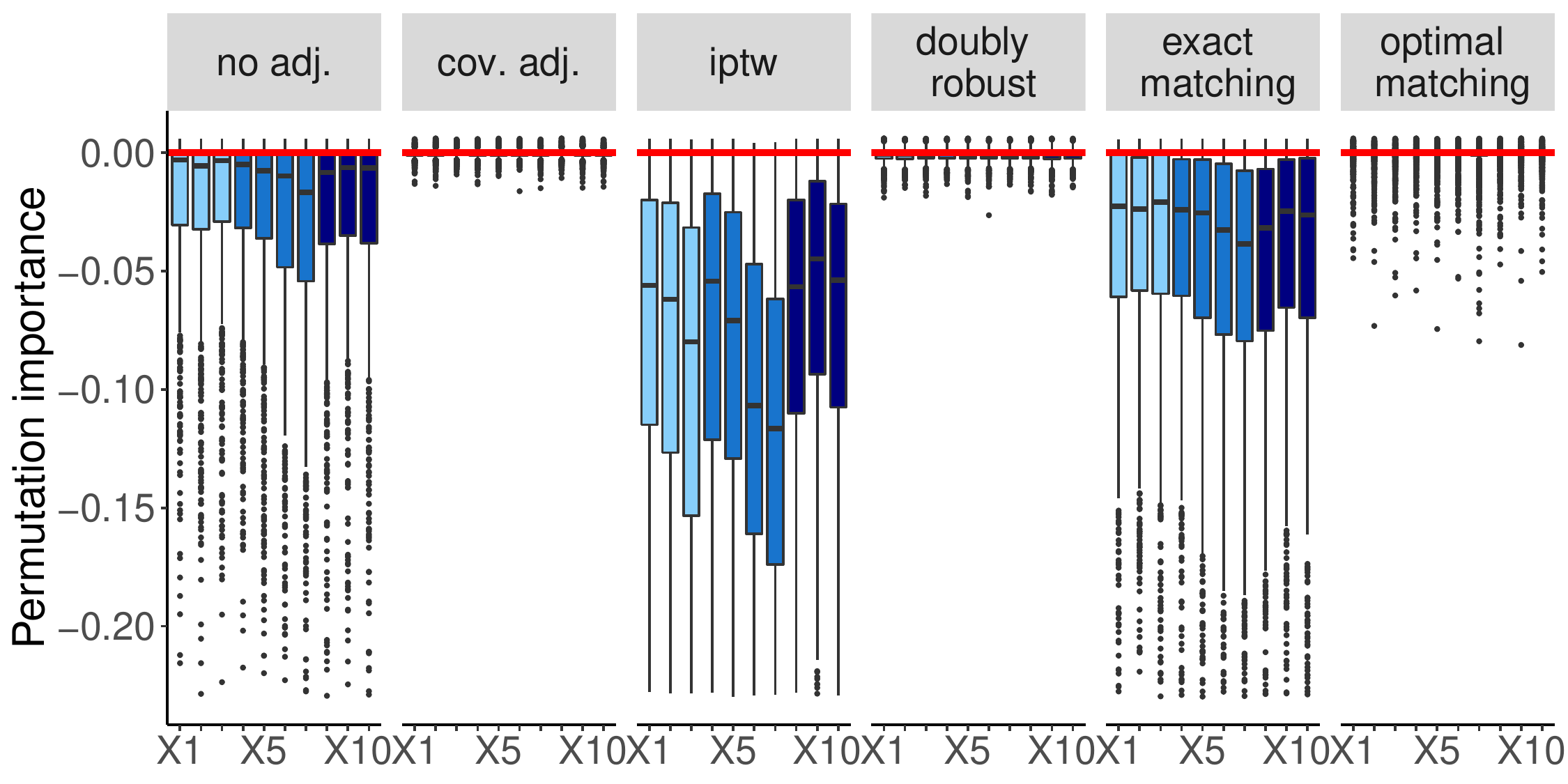}
\label{fig:Simresults_VI_A}
\end{subfigure}\hfill
% \begin{subfigure}{.49\textwidth}
% \subcaption{\textbf{Scenario B.1}: $X_{10}$ has both a prognostic and a qualitative predictive effect}
% \centering
% \includegraphics[height=5cm]{figures/predMOBpermimp_obs01_n1000.pdf}
% \label{fig:Simresults_VI_B1}
% \end{subfigure}\hfill
% \begin{subfigure}{.49\textwidth}
% \subcaption{\textbf{Scenario B.2}: $X_{10}$ has both a prognostic and a quantitative predictive effect}
% \centering
% \includegraphics[height=5cm]{figures/predMOBpermimp_obs02_n1000.pdf}
% \label{fig:Simresults_VI_B2}
% \end{subfigure}\\[3ex]
\begin{subfigure}[b]{.49\textwidth}
\subcaption{\textbf{Scenario~C.1}: $X_{10}$ has a qualitative predictive effect only}
\centering
\includegraphics[height=3.8cm]{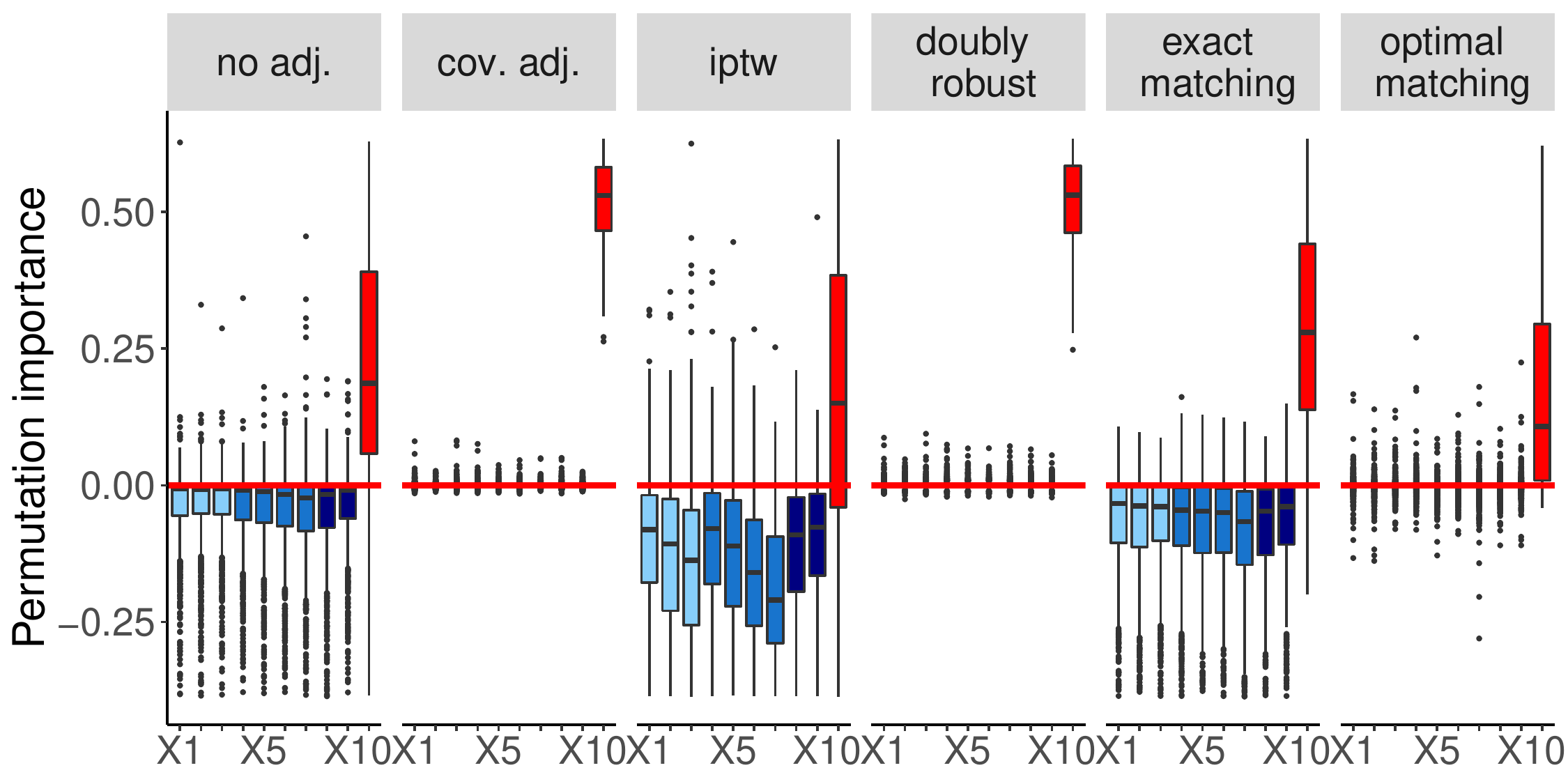}
\label{fig:Simresults_VI_C1}
\end{subfigure}\\[2ex]
% \begin{subfigure}{.49\textwidth}
% \subcaption{\textbf{Scenario C.2}: $X_{10}$ has a quantitative predictive effect only}
% \centering
% \includegraphics[height=5cm]{figures/predMOBpermimp_obs04_n1000.pdf}
% \label{fig:Simresults_VI_C2}
% \end{subfigure}
% \end{center}
% \end{figure}
% \begin{figure}\ContinuedFloat
% \begin{center}
% \begin{subfigure}{.49\textwidth}
% \subcaption{\textbf{Scenario D.1}: $X_{3}$ has a qualitative predictive effect}
% \centering
% \includegraphics[height=5cm]{figures/predMOBpermimp_obs05_n1000.pdf}
% \label{fig:Simresults_VI_D1}
% \end{subfigure}\hfill
\begin{subfigure}[b]{.49\textwidth}
\subcaption{\textbf{Scenario~D.2}: $X_{3}$ has a quantitative predictive effect \newline}
\centering
\includegraphics[height=3.8cm]{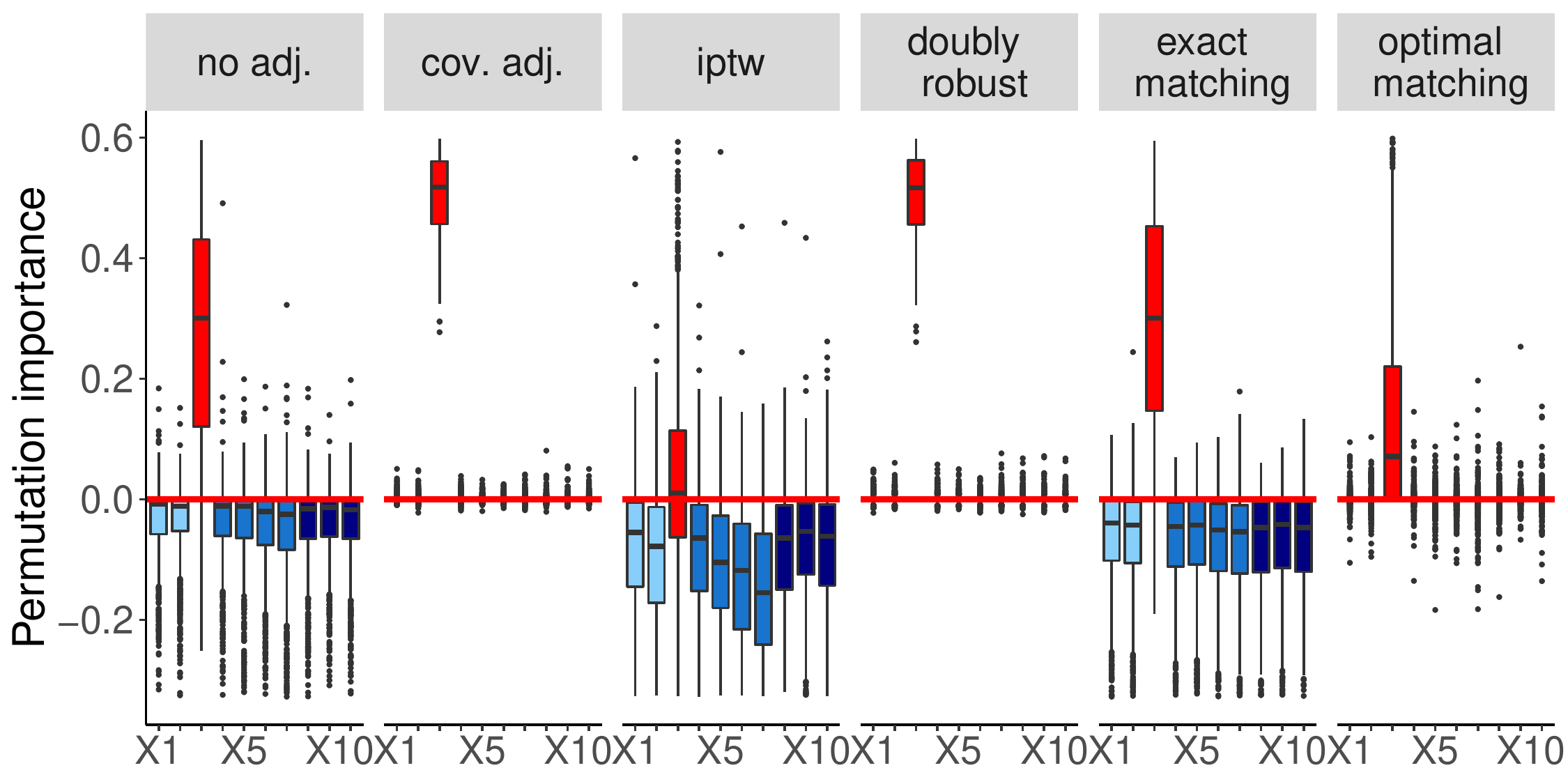}
\label{fig:Simresults_VI_D2}
\end{subfigure}\hfill
% \begin{subfigure}{.49\textwidth}
% \subcaption{\textbf{Scenario E.1}: $X_{7}$ has both a prognostic and a qualitative predictive effect}
% \centering
% \includegraphics[height=5cm]{figures/predMOBpermimp_obs07_n1000.pdf}
% \label{fig:Simresults_VI_E1}
% \end{subfigure}\hfill
\begin{subfigure}[b]{.49\textwidth}
\subcaption{\textbf{Scenario~E.2}: $X_{7}$ has both a prognostic and a quantitative predictive effect}
\centering
\includegraphics[height=3.8cm]{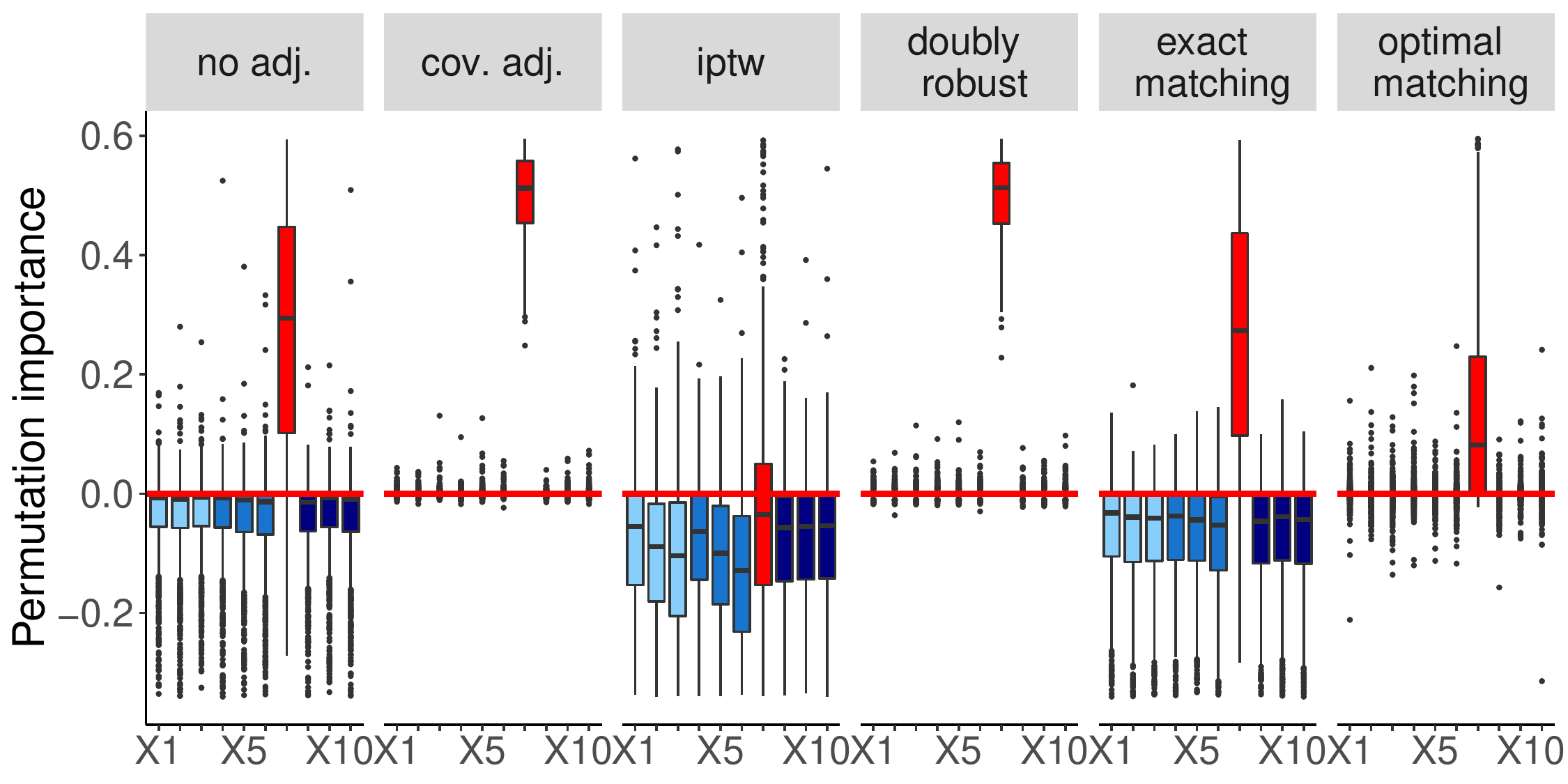}
\label{fig:Simresults_VI_E2}
\end{subfigure}\\[2ex]
\begin{subfigure}[b]{.49\textwidth}
\subcaption{\textbf{Scenario~F.1}: $\beta_T=0.25, \ \beta_{10}=0.3, \ \beta_{\text{int}}=-0.5$ \newline \  \newline}
\centering
\includegraphics[height=3.8cm]{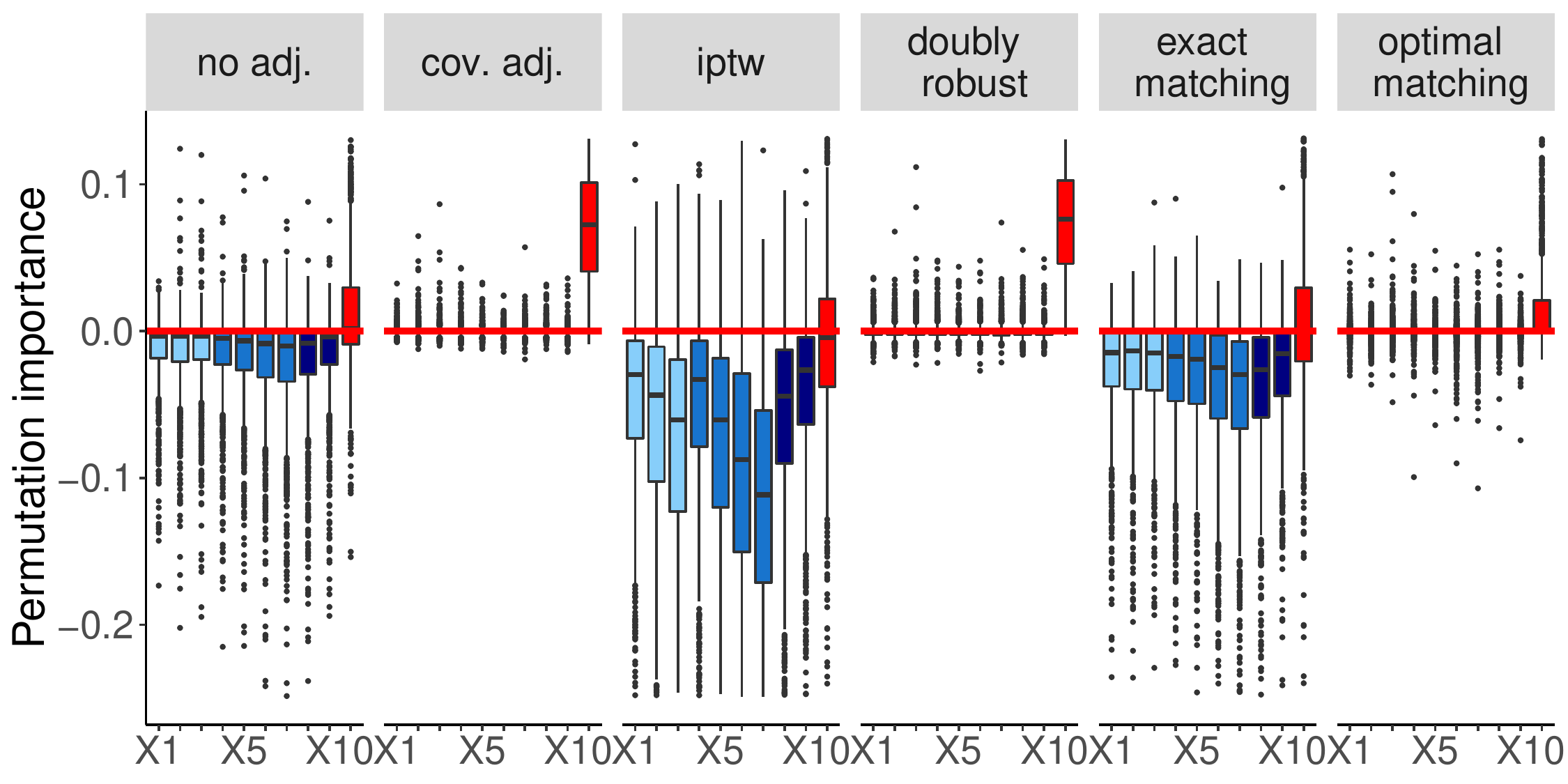}
\label{fig:Simresults_VI_F1}
\end{subfigure}\hfill
\begin{subfigure}[b]{.49\textwidth}
\subcaption{\textbf{Scenario~G.2}: Confounding variable $X_7$ and predictive factor $X_{10}$ are negatively correlated ($\Corr(X_7,X_{10})=-0.7$)}
\centering
\includegraphics[height=3.8cm]{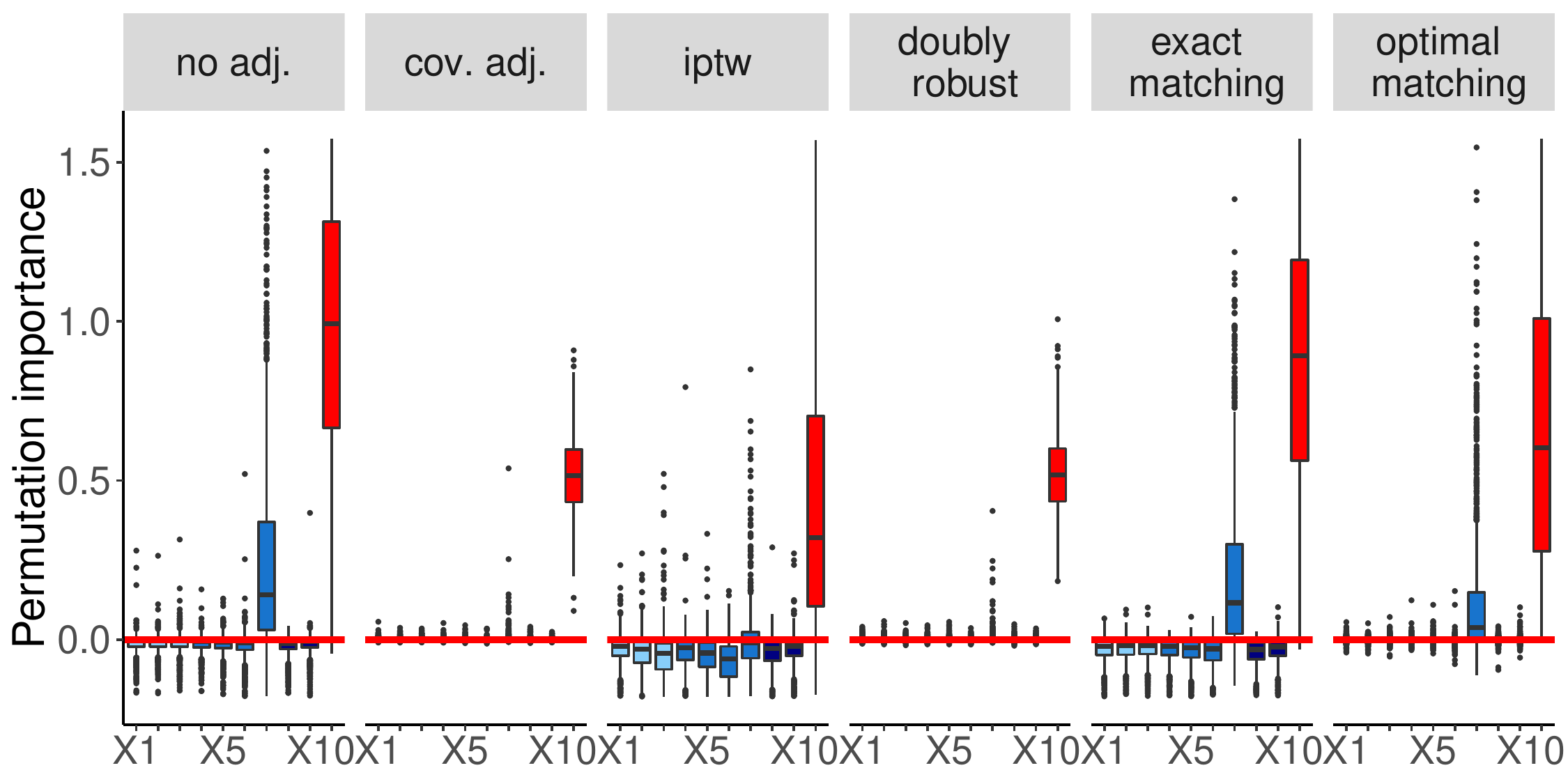}
\label{fig:Simresults_VI_G2}
\end{subfigure}\\[2ex]
\begin{subfigure}[b]{.8\textwidth}
\subcaption{\textbf{Scenario~J}: Additional nuisance biomarkers $V_1, \ldots, V_{20}$}
\centering
\includegraphics[height=3.8cm]{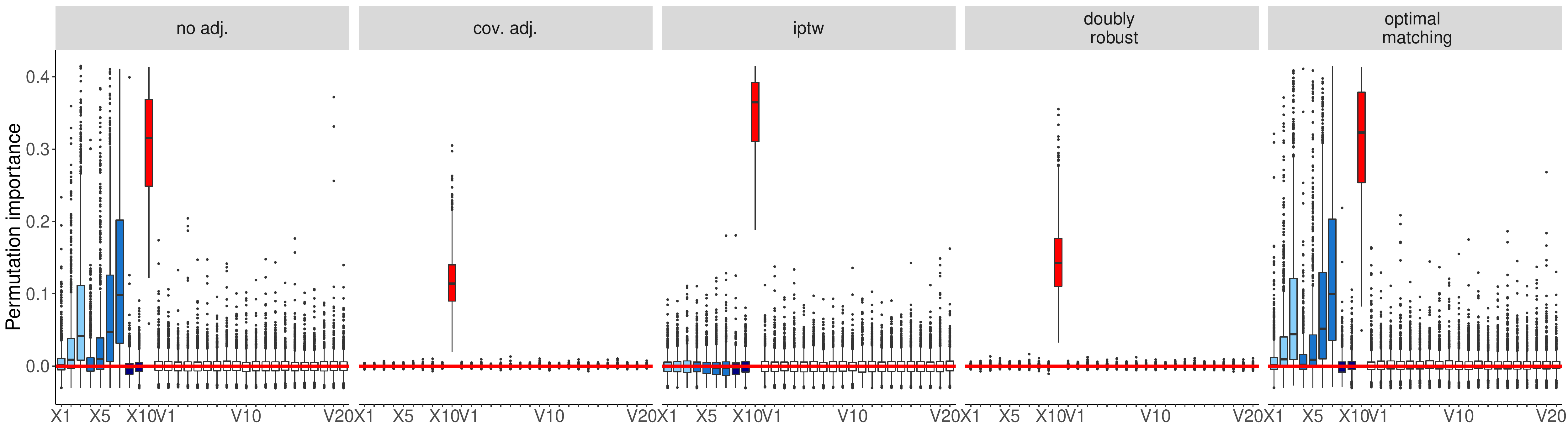}
\label{fig:Simresults_VI_J}
\end{subfigure}
\caption{Permutation importance for the predMOB in combination with various adjustment methods. Instrumental variables are shown in light blue, true confounders in medium blue and factors only associated with outcome in dark blue. The boxplot for a true predictive factor is highlighted in red.\\
Without any adjustment, the superiority of the true predictive biomarker over the other variables is blurred and sometimes not even perceptible (e.g. in scenario~F.1). For some scenarios, \ac{IPTW} and matching are able to slightly improve the results. But especially \ac{IPTW} may also completely obscure the predictive effect (cf. scenario~D.2~\&~E.2). The only adjustment strategy that allows a reliable identification of the true predictive factor(s) in all scenarios is the covariate adjustment.}
\label{fig:Simresults_VI}
\end{center}
\vspace{-18pt}
\end{figure}

\paragraph{Scenarios 0 \& A}
If none of the biomarkers is associated with the outcome (scenario~0, results are depicted in Figure~S.1a), there are only instrumental variables and random variables that are neither associated with the treatment assignment nor with outcome. In this case, no confounding bias is observed. The permutation importance for all variables is close to zero regardless of whether and which adjustment strategy was used. Variability is also comparable across all analysis strategies. In contrast to this, scenario~A includes three types of variables: instrumental variables, true confounders and variables with a prognostic effect on the outcome. As still none of the variables is predictive, all permutation importance measures should be equal to zero, like in scenario~0. In the corresponding Figure~\ref{fig:Simresults_VI_A}, one can observe negative permutation importance values in case of no adjustment as well as for the use of \ac{IPTW} or exact matching. The absolute deviation from zero is proportional to the impact of the respective variables on the treatment assignment. Furthermore, the results of these strategies also show the greatest variability in the permutation importance values. Yet, these results still lead to the right conclusions, since negative permutation importance is interpreted as having no relevant contribution to the construction of the trees. For all analysis strategies involving covariate adjustment (covariate adjustment alone or in a doubly robust approach) as well as optimal matching, the permutation importance for all variables fluctuates randomly around zero. The doubly robust approach shows no advantage over covariate adjustment alone. 

\paragraph{Scenario B \& C}
In these four scenarios, biomarker $X_{10}$ has been generated with a large predictive effect. Figure~\ref{fig:Simresults_VI_C1} representatively shows the results for scenario~C.1. Since the plots for the other scenarios are very similar, they are only included in the supplement (cf. Figures~S.1b-d). Like in scenario~A, negative permutation importance values are observed for all variables that are not predictive if \ac{IPTW} or exact matching is used. However, $X_{10}$ is the only variable with clearly positive permutation importance values and thus the only variable identified as a predictive factor, irrespective of the adjustment. In general, covariate adjustment seems to be the best adjustment strategy, as this is the method that yields the strongest superiority of $X_{10}$ over all non-predictive variables in terms of permutation importance. The use of \ac{IPTW} in addition to covariate adjustment does not provide any further improvement. Optimal matching works comparably well if $X_{10}$ has an additional prognostic effect, but shows lower ``power'' for the detection of the predictive effect if $X_{10}$ is predictive only. Apart from that, an additional prognostic effect of $X_{10}$ does not alter the results.

\paragraph{Scenario D \& E}
In contrast to the scenarios~B \& C, the results differ between settings with a qualitative and a quantitative predictive effect if the true predictive biomarker is also associated with the treatment assignment. In both cases, the permutation importance for the unadjusted predMOB and the predMOB together with exact matching is lower than with covariate adjustment alone or using a doubly robust approach. But, while adjustment using \ac{IPTW} or optimal matching shows good results in case of a qualitative predictive effect (cf. Figures~S.1e-f), $X_{10}$ is no longer identifiable in the scenario with a quantitative predictive effect (see Figures~\ref{fig:Simresults_VI_D2} and \ref{fig:Simresults_VI_E2}). Therefore, covariate adjustment is the only method that appropriately adjusts for confounding in case the quantitative predictive factor might also be associated with the treatment assignment.

\paragraph{Scenario F}
Scenario~F basically describes the same setting as scenario~C.1, but with smaller effect sizes (cf. Figure~\ref{fig:Simresults_VI_F1} and Figure~S.1h) and hence, a lower chance to detect the true predictive factor. As a result, $X_{10}$ is no longer correctly identified by all adjustment methods under investigation. Only when using covariate adjustment, the permutation importance of $X_{10}$ is clearly positive and thus $X_{10}$ is reliably identified as the true predictive factor. If any of the other adjustment strategies is used, none of the biomarkers is identified as a predictive factor. Increasing sample sizes allow the detection of the true predictive factor $X_{10}$ also by means of \ac{IPTW} or matching, but lead to positive permutation importance values for all variables associated with the treatment assignment (cf. Figures~S.1g+i) in case of no adjustment. 

\paragraph{Scenario G}
Scenario~G examines the implications of the true predictive factor being correlated with a confounder variable. To generate a worst case scenario, the strongest prognostic factor $X_7$ was chosen to be correlated with the true predictive factor $X_{10}$. Again, covariate adjustment shows the best performance overall. Without adjustment, the predictive factor $X_{10}$ is correctly detected, but $X_7$ as well shows increased permutation importance. Exact or optimal matching are not able to achieve a correction in this regard. \ac{IPTW} is able to lower the permutation importance of $X_7$ almost back down to zero, but at the same time, also the permutation importance for the true predictive factor $X_{10}$ is considerably diminished. Results are visualized in Figure~\ref{fig:Simresults_VI_G2} for the scenario in which $X_7$ and $X_{10}$ are negatively correlated. The scenario describing a positive correlation between $X_7$ and $X_{10}$ shows similar results (Figure~S.1j).

\paragraph{Scenario H \& I}
Although there is more than one predictive factor in scenarios~H \& I, the results are very similar to the scenarios~B \& C. Both predictive factors are correctly identified no matter which adjustment strategy is used, and covariate adjustment is generally more effective than the other approaches (especially in scenario~H.2). As these conclusions are very much in line with the other scenarios, results are shown in supplement~S~1 (Figures~S.1k-m).

\paragraph{Scenario J}
The results for scenario~J are essentially the same as for scenario~C.1. $X_{10}$ can be correctly identified as the only predictive factor if either covariate adjustment, \ac{IPTW} or both are used. Unadjusted results also suggest other variables as predictive factors and the magnitude of the permutation importance values is proportional to the influence of the corresponding variables on the treatment assignment. The only difference can be observed in the results using predMOB in combination with optimal matching. With a larger number of biomarkers, and thus also a larger number of potential confounders to be taken into account for matching, the correction by optimal matching becomes worse. Exact matching is no longer feasible due to too many subgroups ($p^2=(10+20)^2=900$ subgroups).\\

In summary, \ac{IPTW} fails to correctly identify the true predictive factor in scenarios in which the true predictive factor is not completely independent of the confounding mechanism (either because it is correlated with a confounder or because it is associated with the treatment assignment itself). As expected, exact matching is superior to optimal matching, but both matching methods are not suitable in the presence of correlations and generally require a higher sample size for the detection of predictive effects than covariate adjustment, as does \ac{IPTW}. Combining covariate adjustment with \ac{IPTW} shows no advantage over covariate adjustment alone in the scenarios investigated here.
Hence, covariate adjustment seems to be the only appropriate adjustment strategy. Furthermore, covariate adjustment also maximizes the superiority of the true predictive factor and thus, the ``power'' to identify it as such. The variability in the permutation importance values is greatest without any adjustment. All scenarios that have not been discussed in detail here support the superiority of covariate adjustment, so the results are shown only in supplement~S~1 (Figure~S.1). 

\subsection{Accuracy of predicted individual treatment effects and estimated modifying effects of predictive factors}
The second part of the simulation study is designed to investigate the amount of bias in the predicted individual treatment effects and the estimated modifying effects of predictive factors (``predictive effect''). To be better able to interpret the results and investigate the potential sources of bias, a reduced setting with only two biomarkers is explored. Again, the data is generated using the simstudy package in R using the following data-generating process and outcome generating models: 
\begin{itemize}
    \item 2 i.i.d. biomarkers $X_1, X_2$; binary with $X_i\sim \Bin(1,0.5)$,
    \item Binary treatment variable $T \sim \Bin(1,p)$; depending on biomarkers $X_1$ and $X_2$ via a logistic model: $\logit(\mathbb{P}(T=1))= \beta_0+\eta$, with $\eta$ as defined in Table \ref{tab:simsettings_predacc} and where $\beta_0$ is chosen such that treatment group sizes are equal ($p=\mathbb{P}(T=1)=0.5$), 
    \item Normally distributed outcome variable $Y \sim N(\mu, 0.25)$ with expectation $\mu$ as defined in Table \ref{tab:simsettings_predacc}.
\end{itemize}

\begin{table}[ht!]
    \caption{Parameter configurations for the scenarios investigating prediction accuracy}
 \begin{adjustwidth}{-.5in}{-.5in}  
   \label{tab:simsettings_predacc}
	\centering
	\begin{tabular}{l l l}
    \toprule
    Description &  Outcome model & Propensity model\\
    \midrule
    1: $X_2$ quantitative predictive \& positive prognostic &  $\mu=0.5T+1X_2+2X_2T$ &  \multirow{3}{*}{$\left. \begin{tabular}{@{}l@{}} \\ \\ \end{tabular}\right\}  \eta=log(1.5)*X_2$} \\
    2: $X_2$ quantitative predictive \& not prognostic &  $\mu=0.5T+2X_2T$  & \\
    3: $X_2$ quantitative predictive \& negative prognostic &  $\mu=0.5T-2X_2+2X_2T$  & \\
    4: $X_2$ quantitative predictive \& negative prognostic &  $\mu=0.5T-2X_2+2X_2T$  & $\eta=log(1.5)*X_1+log(1.5)*X_2$ \\
    5: $X_2$ qualitative predictive \&  negative prognostic &  $\mu=0.5T-1X_2-2X_2T$  & \multirow{2}{*}{$\left. \begin{tabular}{@{}l@{}} \\ \end{tabular}\right\}  \eta=log(1.5)*X_2$} \\
    6: $X_2$ qualitative predictive \&  positive prognostic &  $\mu=0.5T+1X_2-2X_2T$  & \\
    7: $X_2$ qualitative predictive \&  $X_1$ positive prognostic &  $\mu=0.5T+3X_1-2X_2T$  & $\eta=log(1.5)*X_1$\\
    8: $X_2$ qualitative predictive \&  $X_1$ positive prognostic &  $\mu=0.5T+3X_1-2X_2T$  & $\eta=log(1.5)*X_1+log(1.5)*X_2$\\
   \bottomrule
    \end{tabular}
    \end{adjustwidth}
\end{table}

\begin{figure}
\begin{center}
\begin{subfigure}[b]{.49\textwidth}
\subcaption{\textbf{Scenario 1}: $X_2$ is quantitative predictive  \& positive \newline prognostic, $X_2$ confounder}
\centering
\includegraphics[height=4cm]{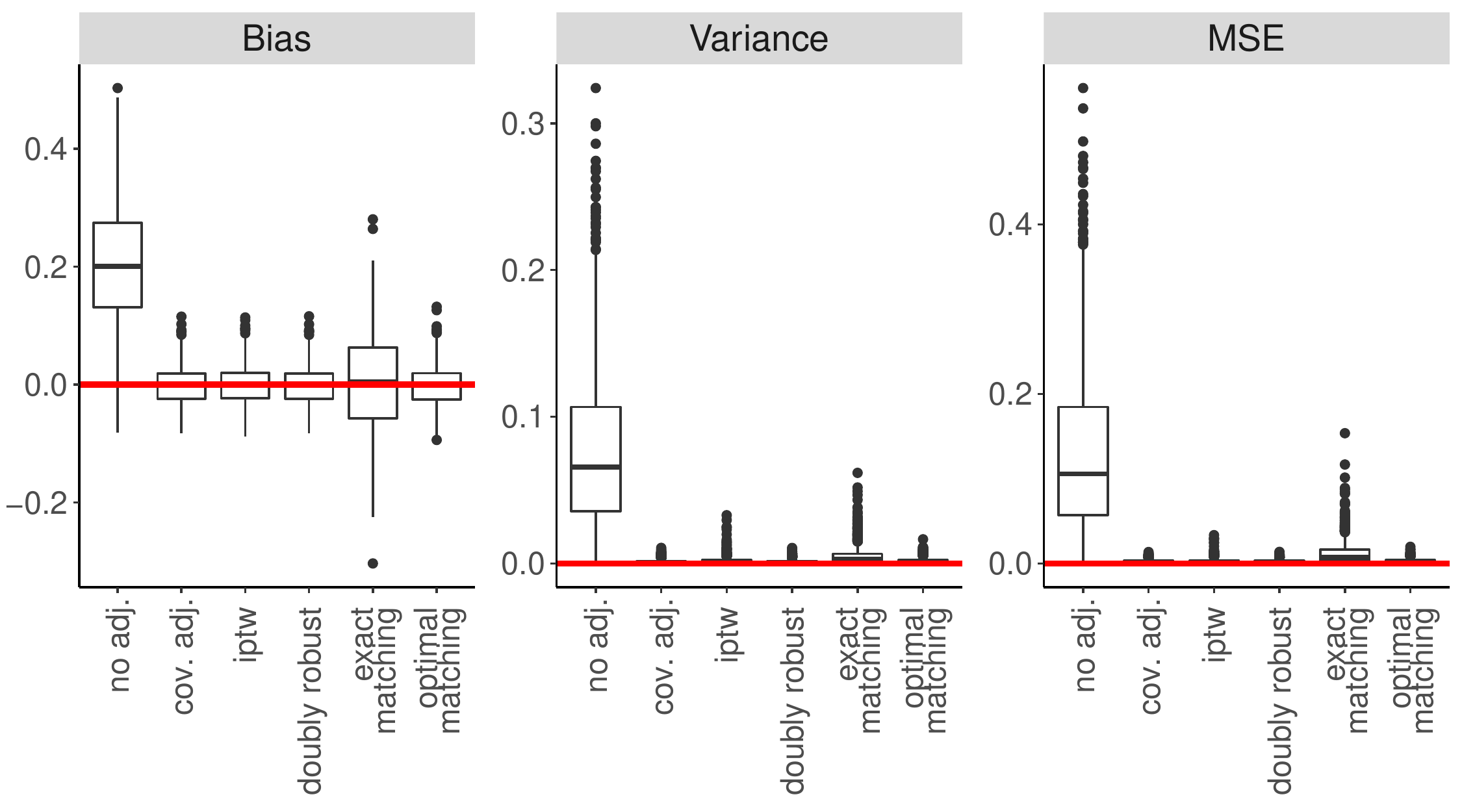}
\label{fig:Simresults_predacc_indtrt1}
\end{subfigure}\hfill
\begin{subfigure}[b]{.49\textwidth}
\subcaption{\textbf{Scenario 2}: $X_2$ is quantitative predictive \& not prognostic, $X_2$ confounder}
\centering
\includegraphics[height=4cm]{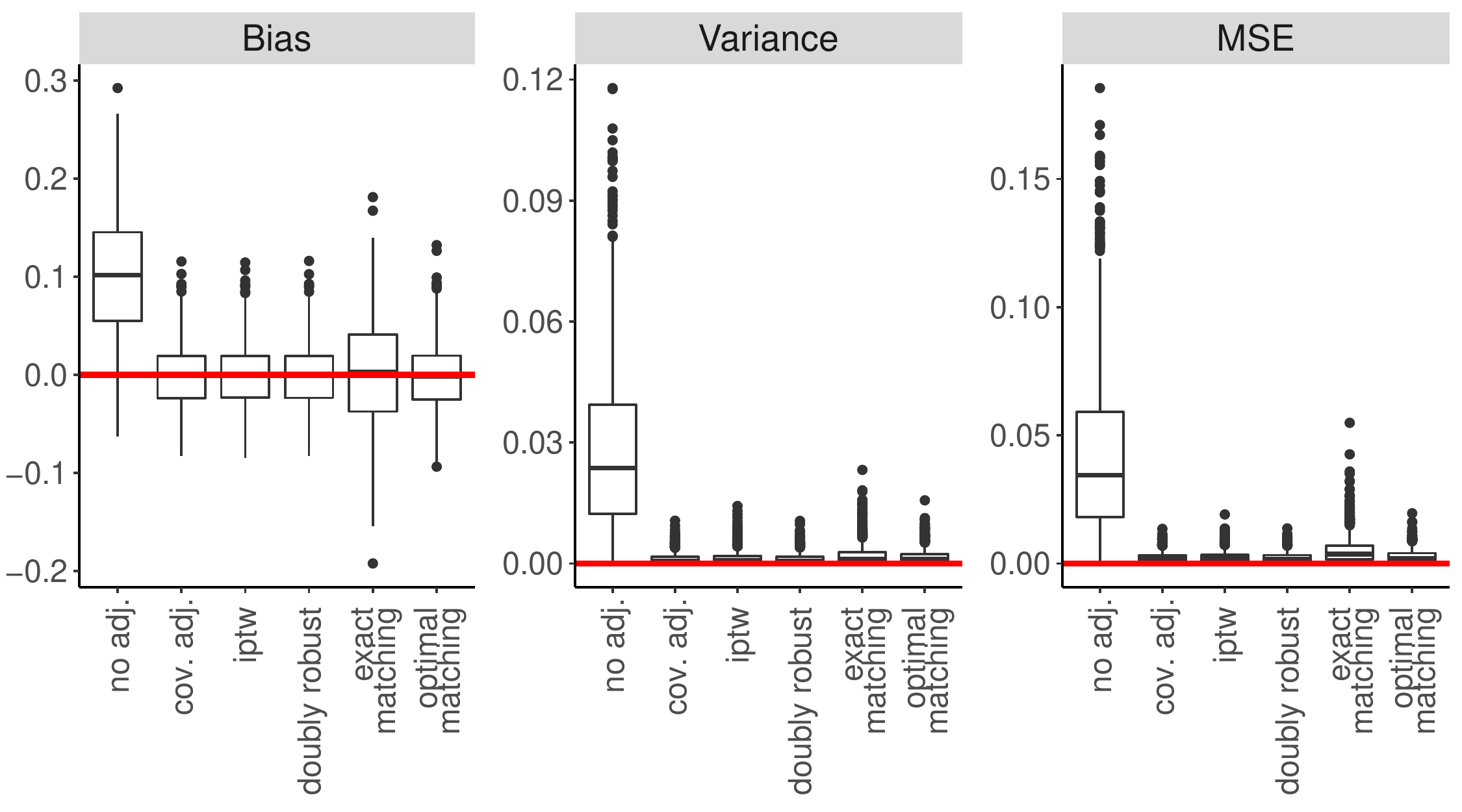}
\label{fig:Simresults_predacc_indtrt2}
\end{subfigure}\\[2ex]
\begin{subfigure}[b]{.49\textwidth}
\subcaption{\textbf{Scenario 3}: $X_2$ is quantitative predictive  \& negative \newline prognostic, $X_2$ confounder}
\centering
\includegraphics[height=4cm]{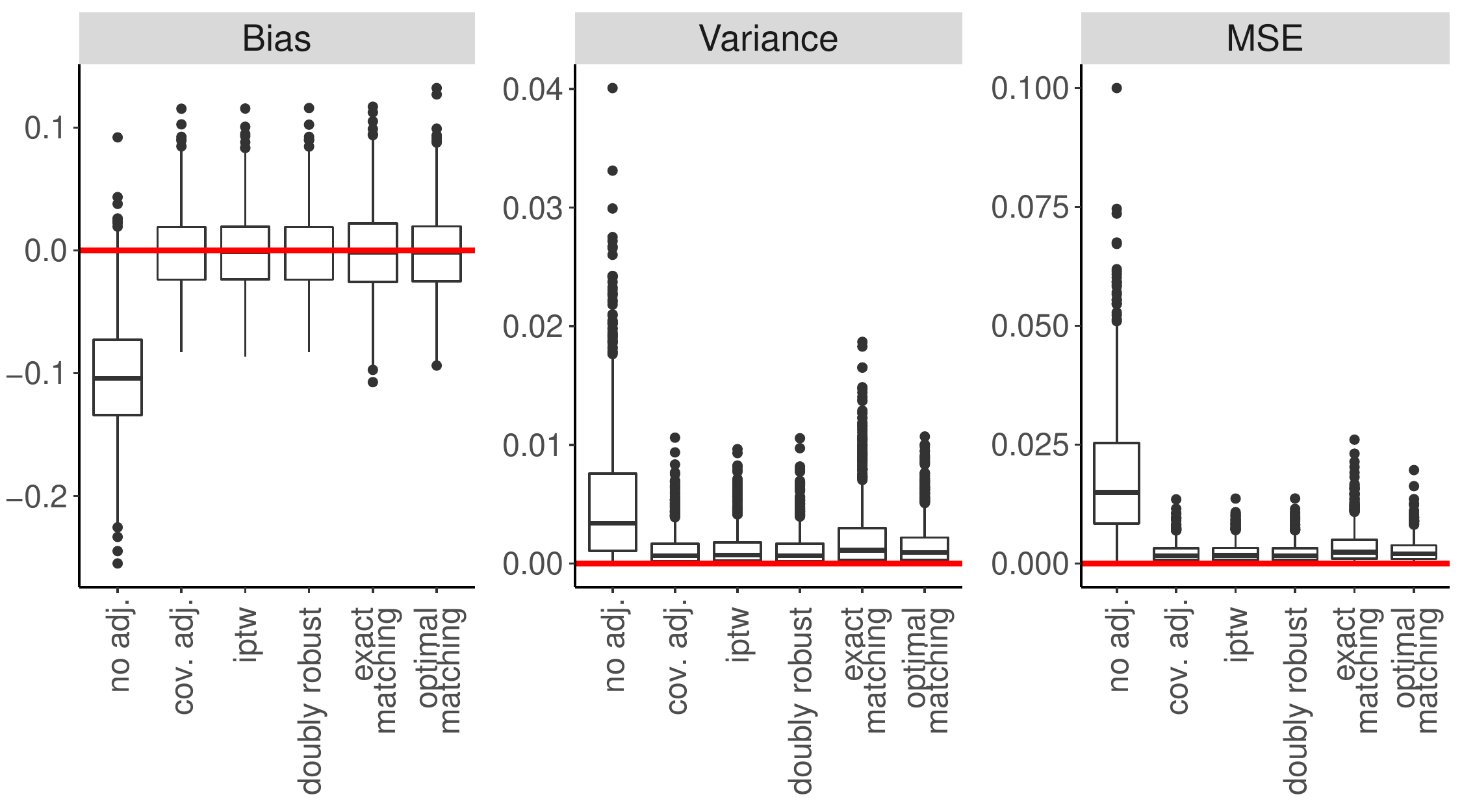}
\label{fig:Simresults_predacc_indtrt3}
\end{subfigure}\hfill
\begin{subfigure}[b]{.49\textwidth}
\subcaption{\textbf{Scenario 4}: $X_2$ is quantitative predictive  \& negative \newline prognostic, $X_1$ \& $X_2$ confounder}
\centering
\includegraphics[height=4cm]{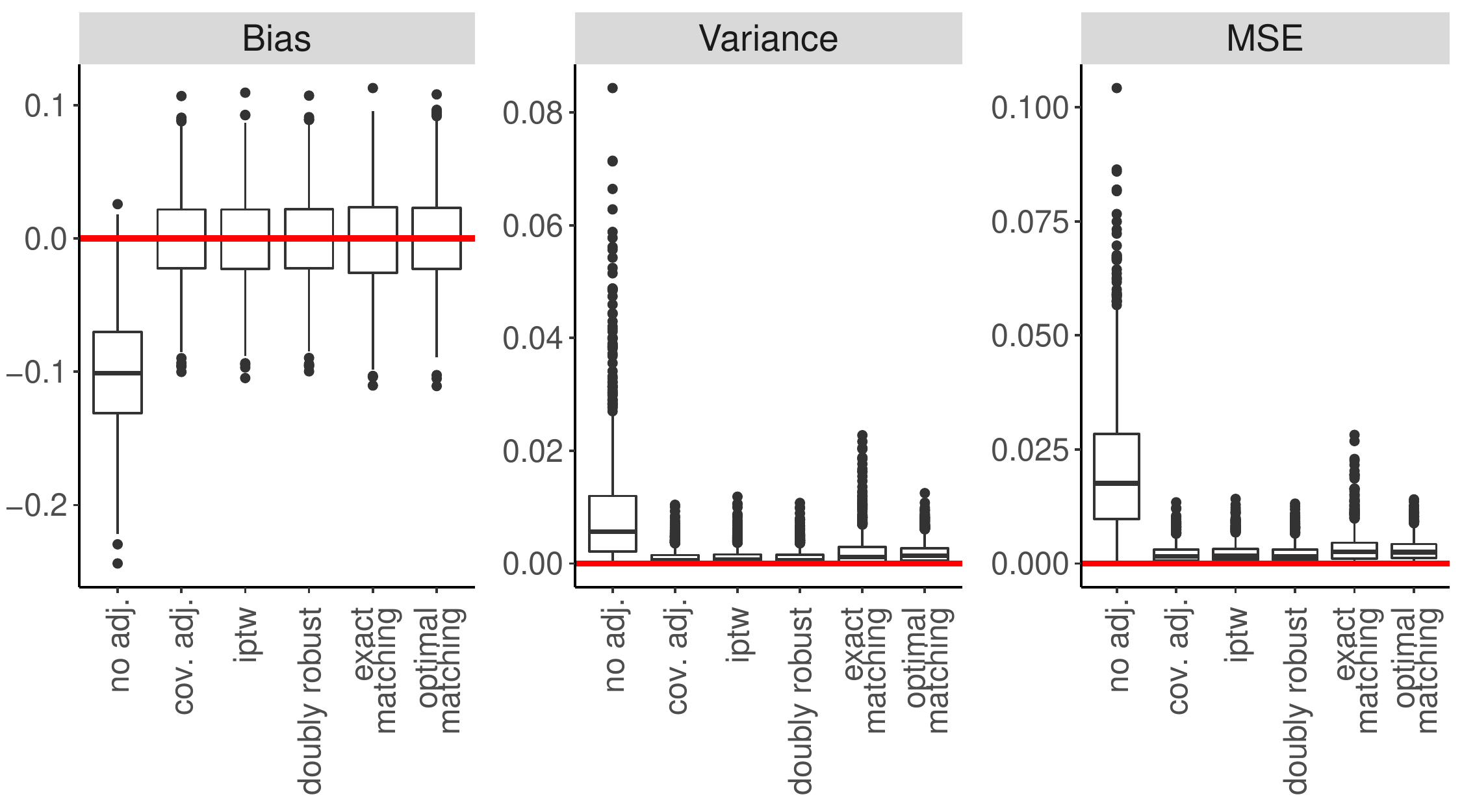}
\label{fig:Simresults_predacc_indtrt4}
\end{subfigure}\\[2ex]
\begin{subfigure}[b]{.49\textwidth}
\subcaption{\textbf{Scenario 5}: $X_2$ is qualitative predictive \&  negative \newline prognostic, $X_2$ confounder}
\centering
\includegraphics[height=4cm]{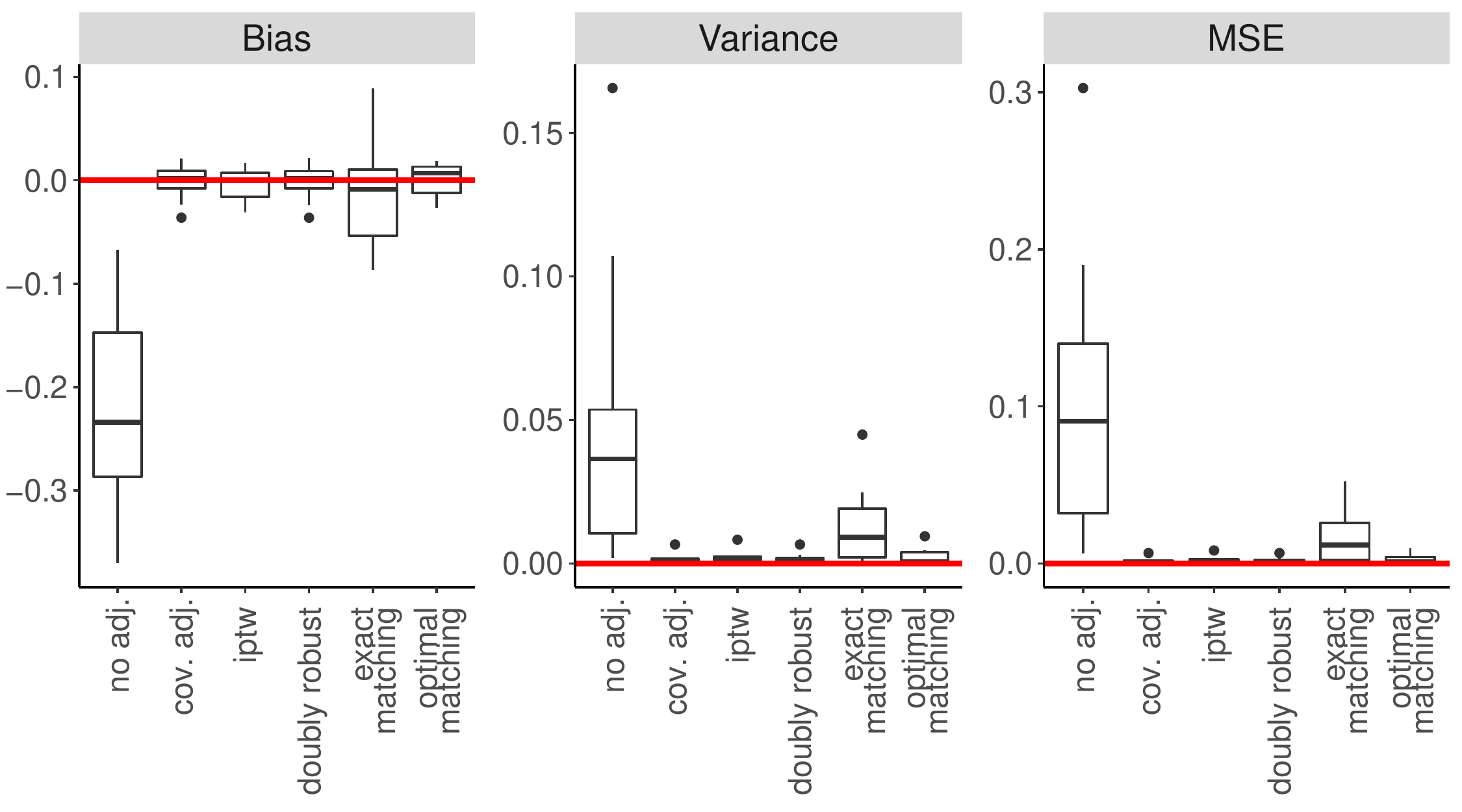}
\label{fig:Simresults_predacc_indtrt5}
\end{subfigure}\hfill
\begin{subfigure}[b]{.49\textwidth}
\subcaption{\textbf{Scenario 6}: $X_2$ is qualitative predictive \&  positive \newline prognostic, $X_2$ confounder}
\centering
\includegraphics[height=4cm]{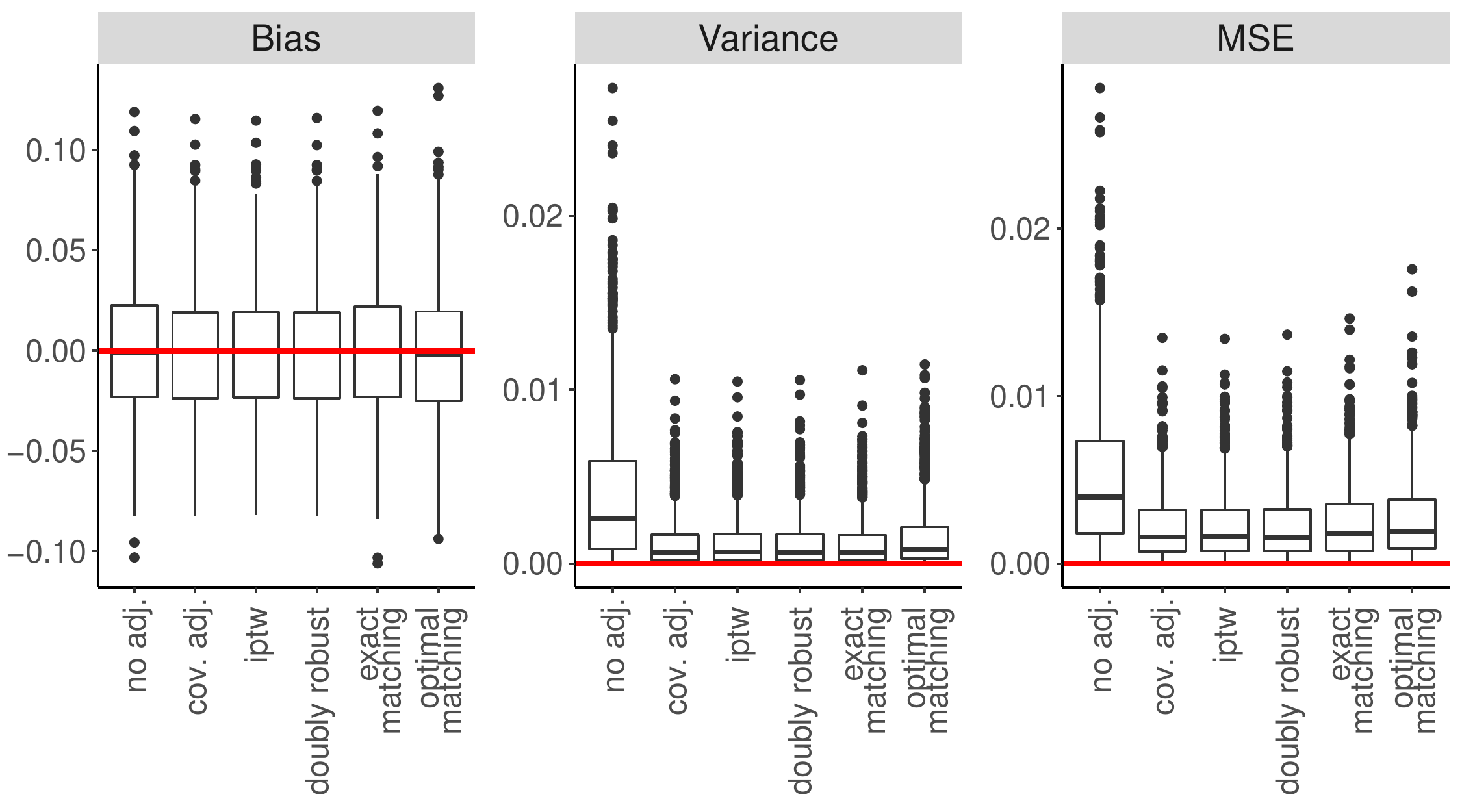}
\label{fig:Simresults_predacc_indtrt6}
\end{subfigure}\\[2ex]
\begin{subfigure}[b]{.49\textwidth}
\subcaption{\textbf{Scenario 7}: $X_2$ is qualitative predictive  \& \newline $X_1$ is positive prognostic, $X_1$ confounder}
\centering
\includegraphics[height=4cm]{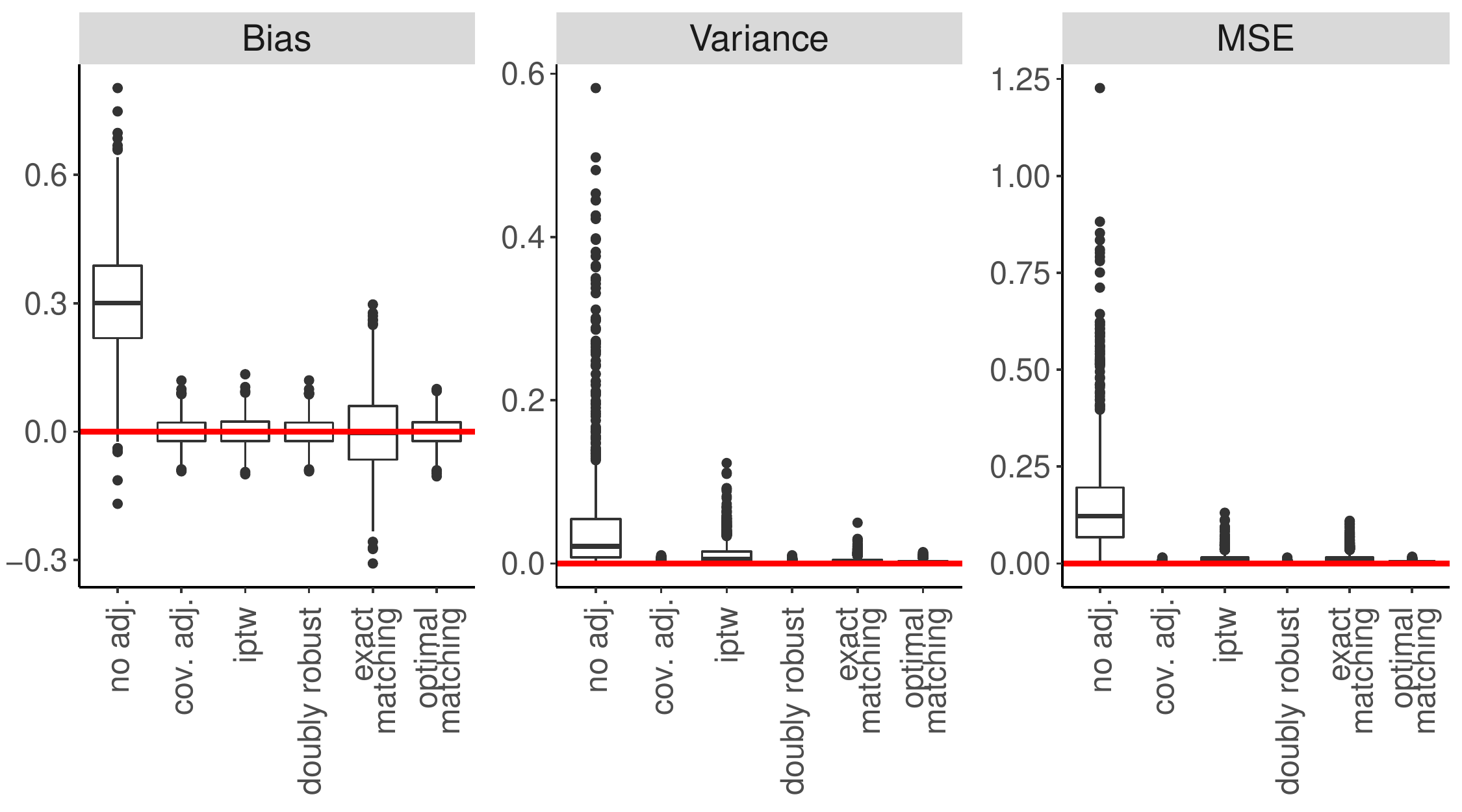}
\label{fig:Simresults_predacc_indtrt7}
\end{subfigure}\hfill
\begin{subfigure}[b]{.49\textwidth}
\subcaption{\textbf{Scenario 8}: $X_2$ is qualitative predictive  \& \newline $X_1$ is positive prognostic, $X_1$ \& $X_2$ confounder}
\centering
\includegraphics[height=4cm]{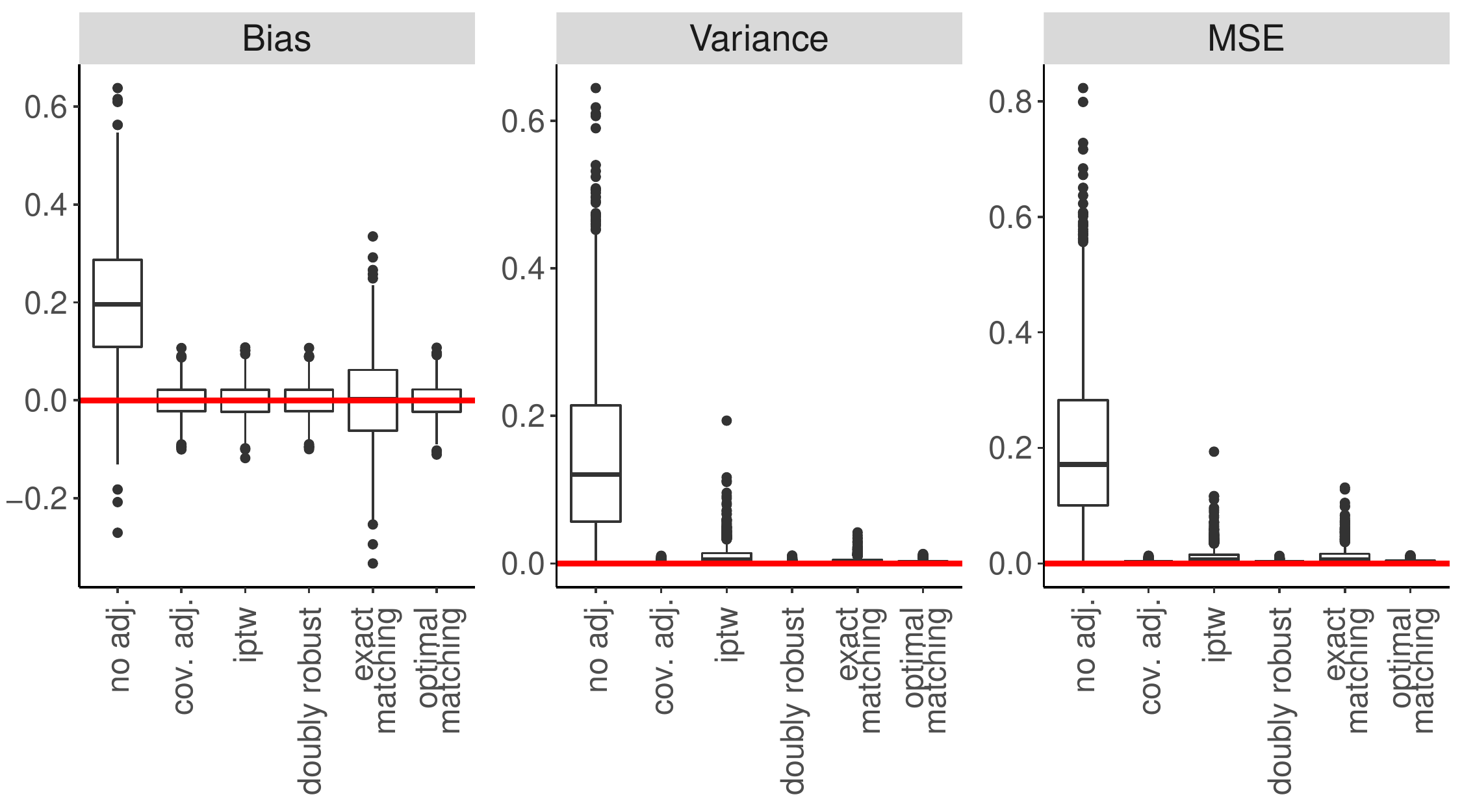}
\label{fig:Simresults_predacc_indtrt8}
\end{subfigure}\\[2ex]
\caption{Bias, Variance and \ac{MSE} of the predicted individual treatment effect. The predictions for the individual treatment effect obtained by the unadjusted predMOB are biased throughout all scenarios. All adjustment methods achieve an appropriate correction and, with the exception of exact matching, also show similar variability.}
\label{fig:Simresults_predacc_indtrt}
\end{center}
\vspace{-11pt}
\end{figure}

\begin{figure}
\begin{center}
\begin{subfigure}[b]{.49\textwidth}
\subcaption{\textbf{Scenario 1}: $X_2$ is quantitative predictive  \& positive \newline prognostic, $X_2$ confounder}
\centering
\includegraphics[height=4cm]{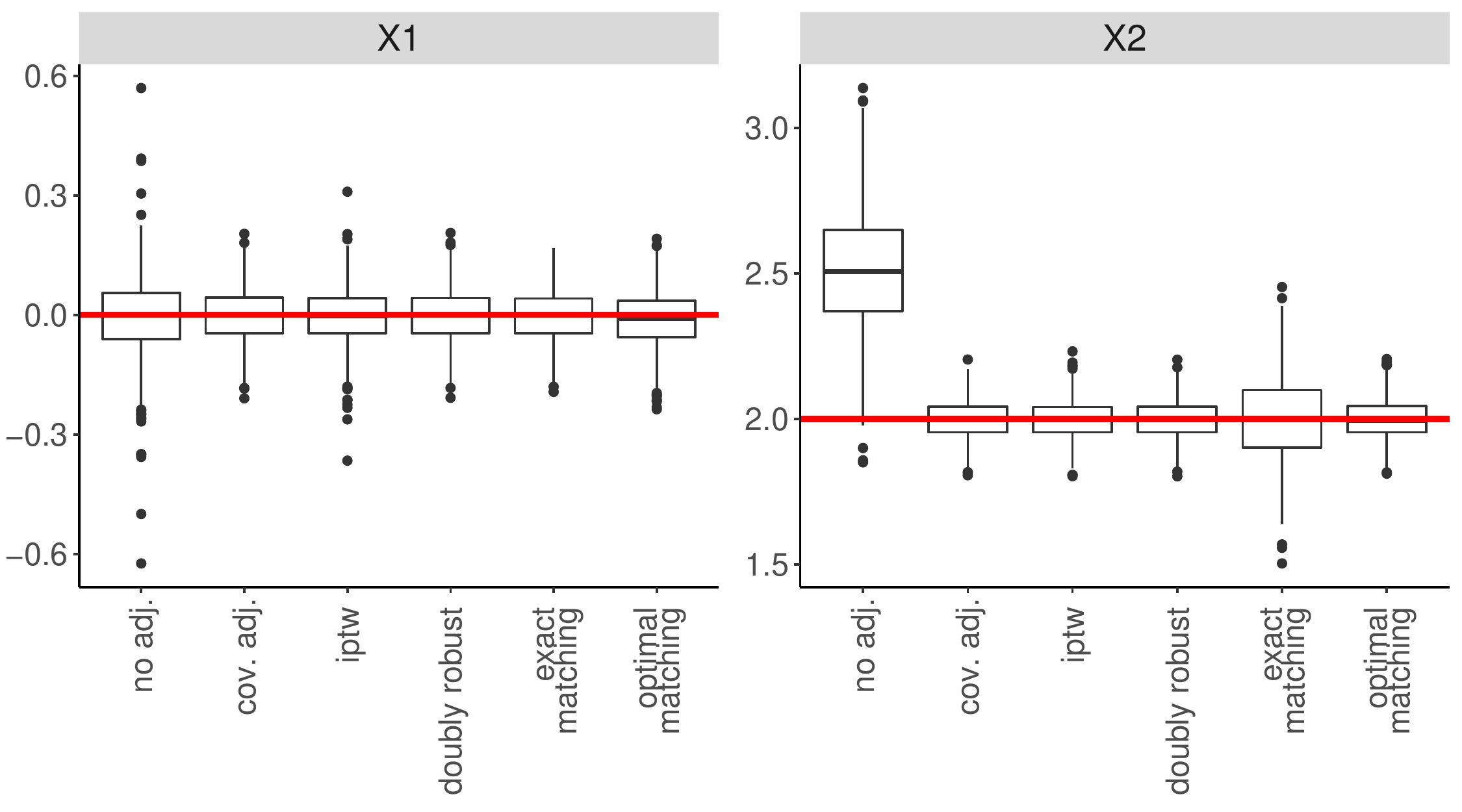}
\label{fig:Simresults_predacc_predeff1}
\end{subfigure}\hfill
\begin{subfigure}[b]{.49\textwidth}
\subcaption{\textbf{Scenario 2}: $X_2$ is quantitative predictive \& not prognostic, $X_2$ confounder}
\centering
\includegraphics[height=4cm]{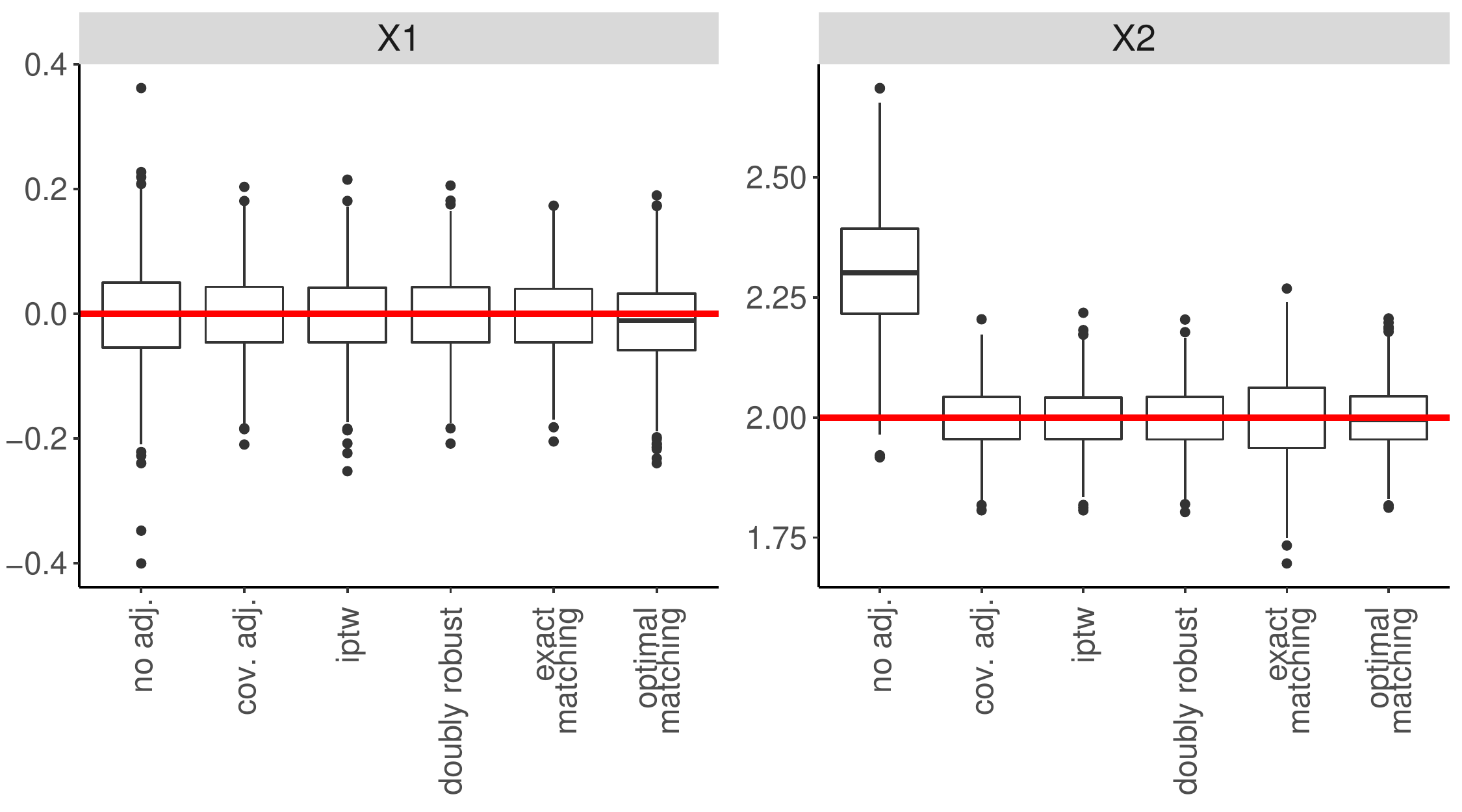}
\label{fig:Simresults_predacc_predeff2}
\end{subfigure}\\[2ex]
\begin{subfigure}[b]{.49\textwidth}
\subcaption{\textbf{Scenario 3}: $X_2$ is quantitative predictive  \& negative \newline prognostic, $X_2$ confounder}
\centering
\includegraphics[height=4cm]{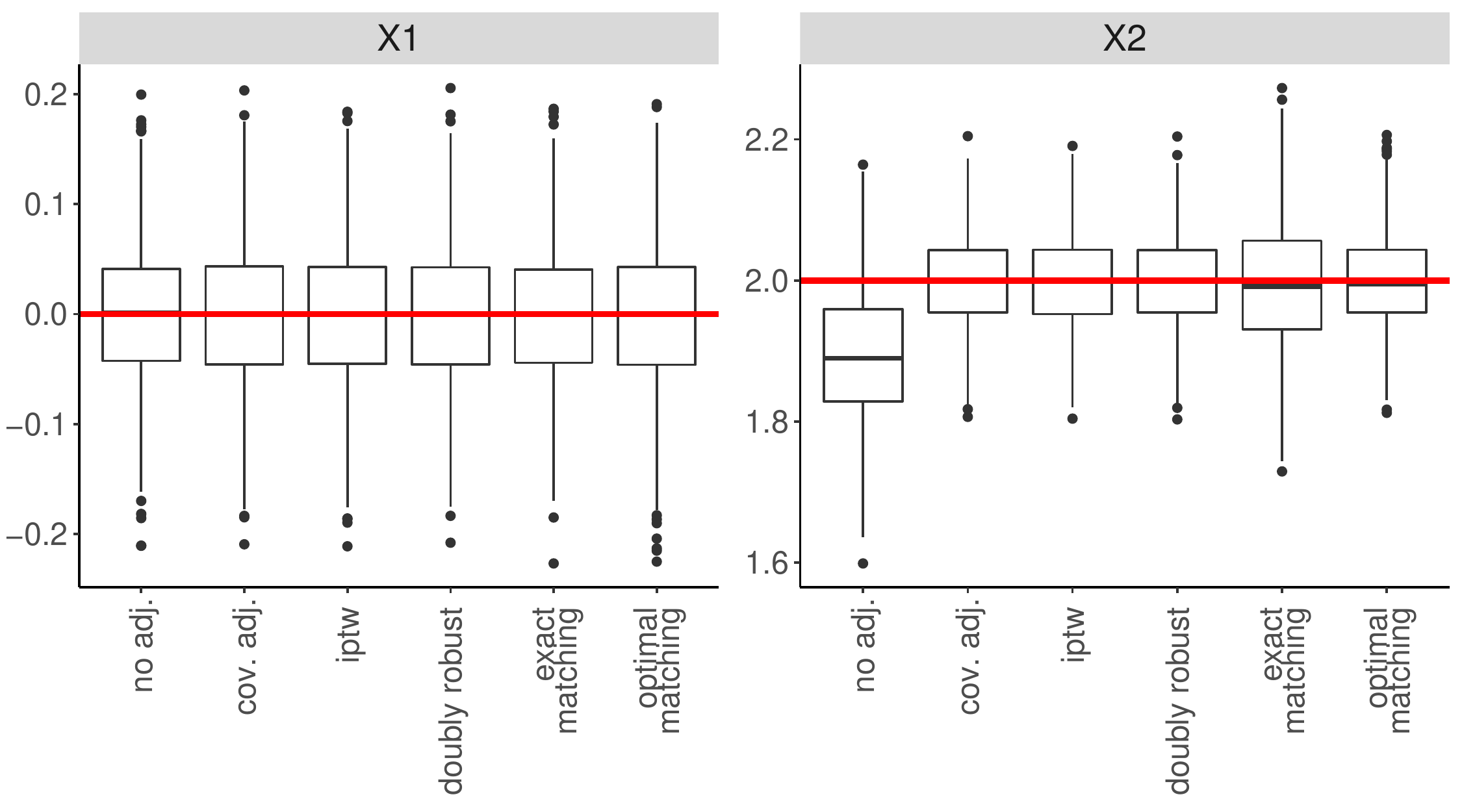}
\label{fig:Simresults_predacc_predeff3}
\end{subfigure}\hfill
\begin{subfigure}[b]{.49\textwidth}
\subcaption{\textbf{Scenario 4}: $X_2$ is quantitative predictive  \& negative \newline prognostic, $X_1$ \& $X_2$ confounder}
\centering
\includegraphics[height=4cm]{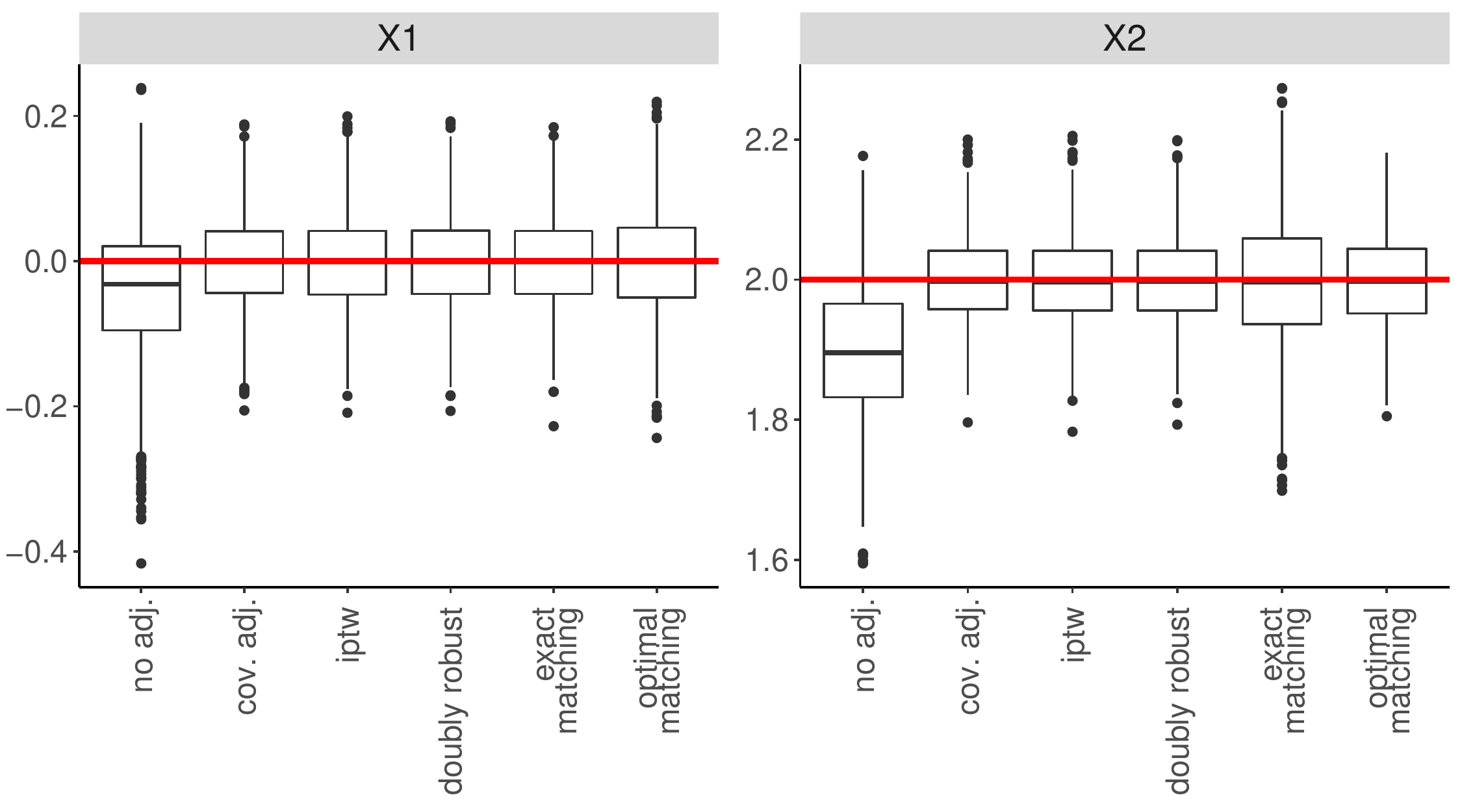}
\label{fig:Simresults_predacc_predeff4}
\end{subfigure}\\[2ex]
\begin{subfigure}[b]{.49\textwidth}
\subcaption{\textbf{Scenario 5}: $X_2$ is qualitative predictive \&  negative \newline prognostic, $X_2$ confounder}
\centering
\includegraphics[height=4cm]{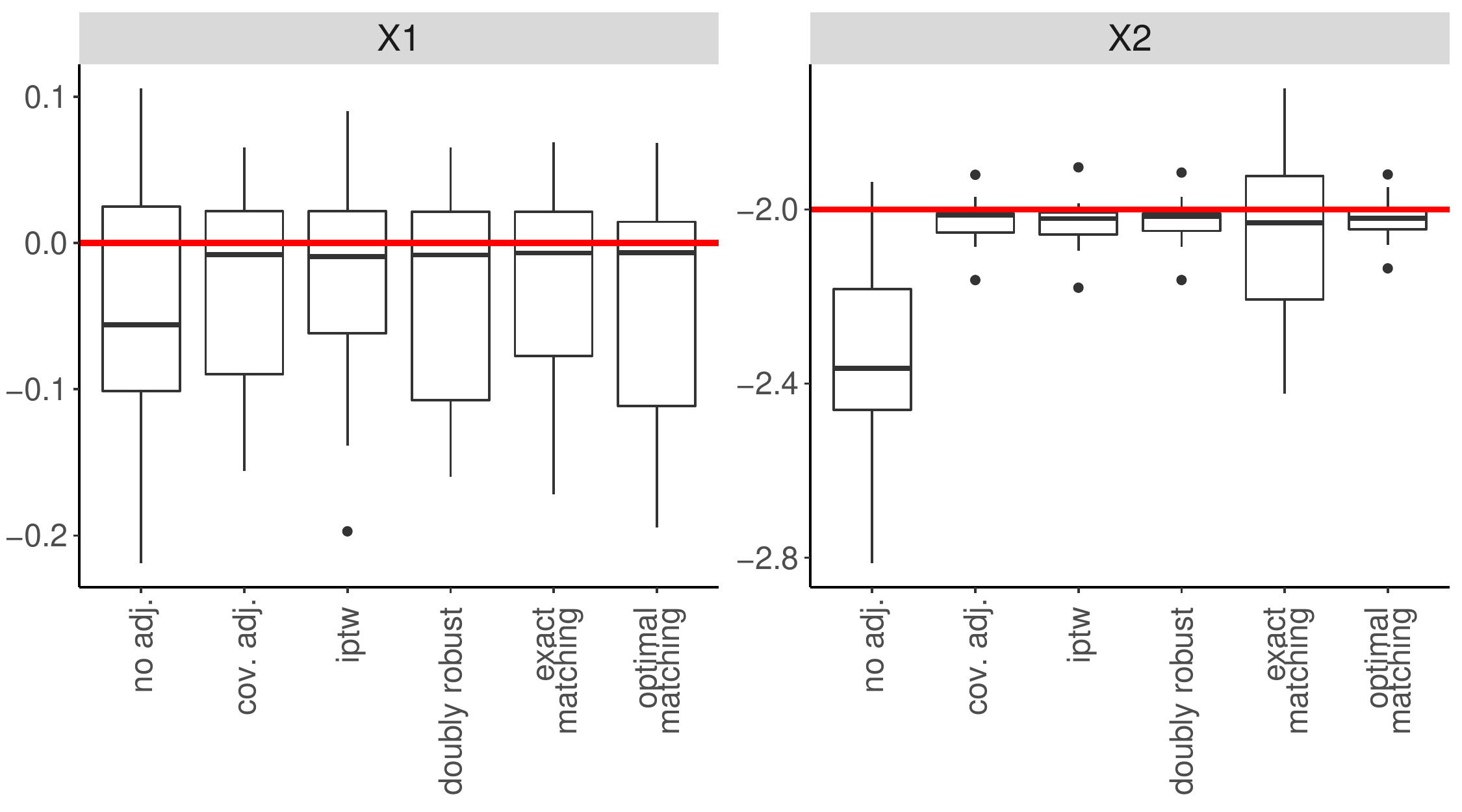}
\label{fig:Simresults_predacc_predeff5}
\end{subfigure}\hfill
\begin{subfigure}[b]{.49\textwidth}
\subcaption{\textbf{Scenario 6}: $X_2$ is qualitative predictive \&  positive \newline prognostic, $X_2$ confounder}
\centering
\includegraphics[height=4cm]{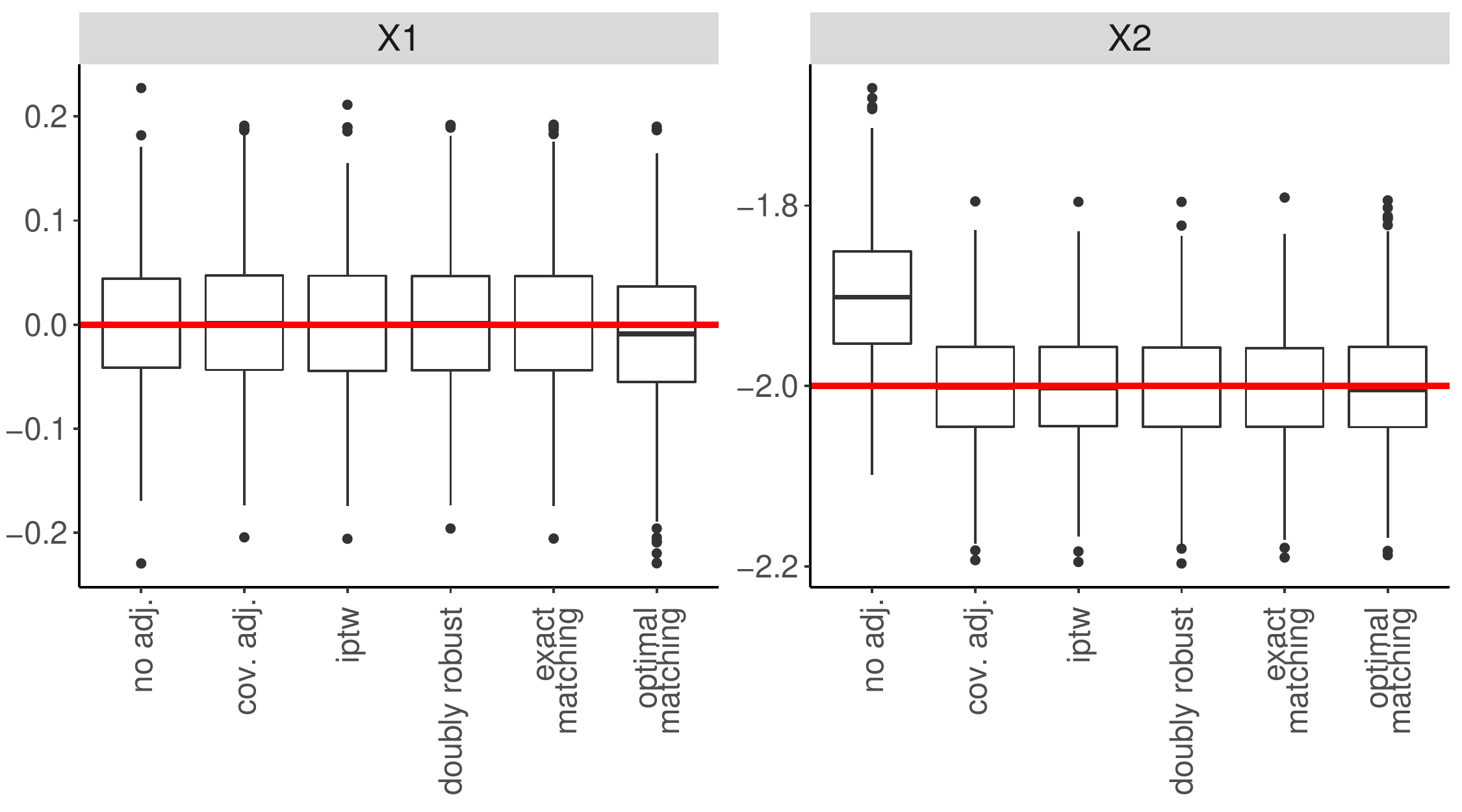}
\label{fig:Simresults_predacc_predeff6}
\end{subfigure}\\[2ex]
\begin{subfigure}[b]{.49\textwidth}
\subcaption{\textbf{Scenario 7}: $X_2$ is qualitative predictive  \& \newline $X_1$ is positive prognostic, $X_1$ confounder}
\centering
\includegraphics[height=4cm]{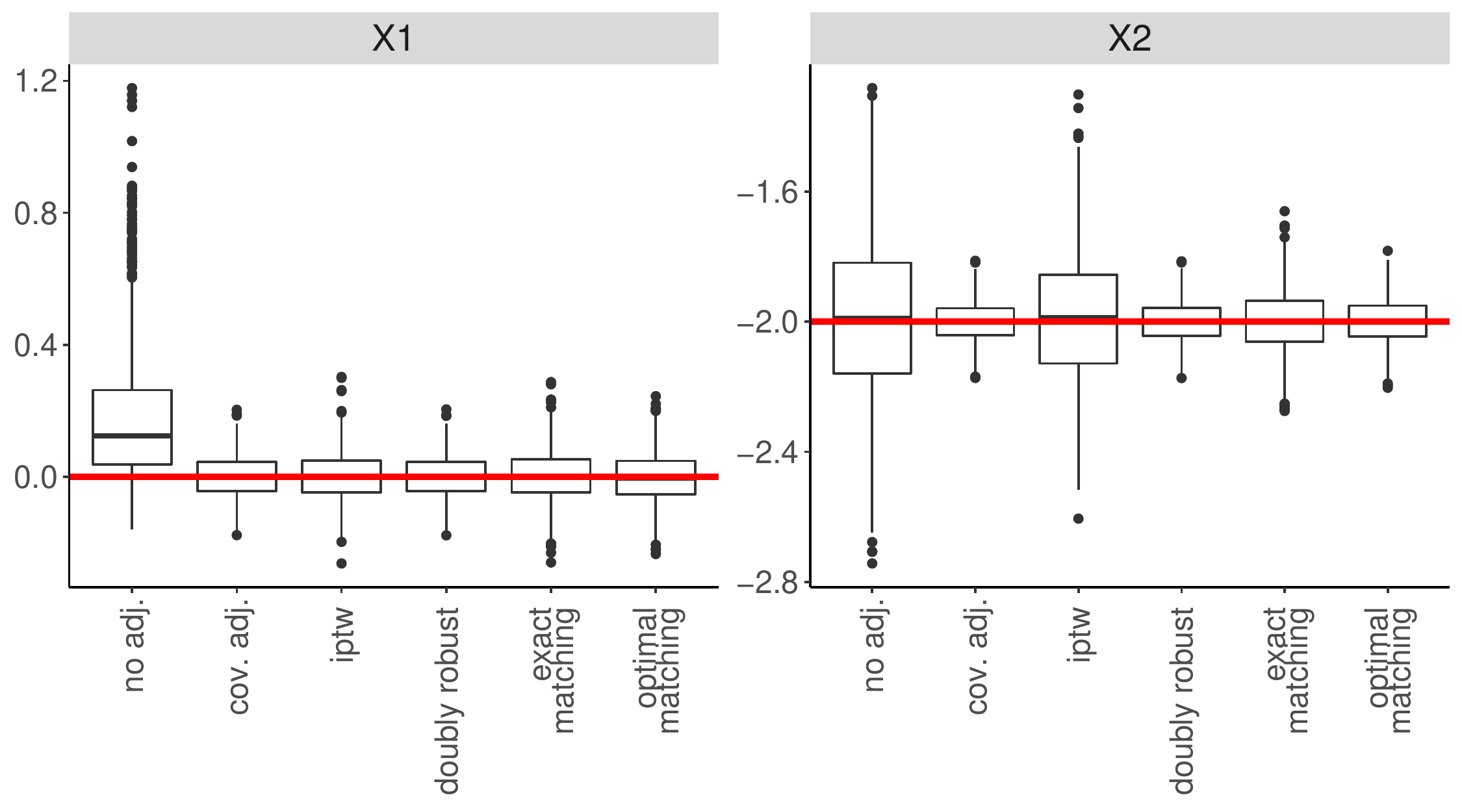}
\label{fig:Simresults_predacc_predeff7}
\end{subfigure}\hfill
\begin{subfigure}[b]{.49\textwidth}
\subcaption{\textbf{Scenario 8}: $X_2$ is qualitative predictive  \& \newline $X_1$ is positive prognostic, $X_1$ \& $X_2$ confounder}
\centering
\includegraphics[height=4cm]{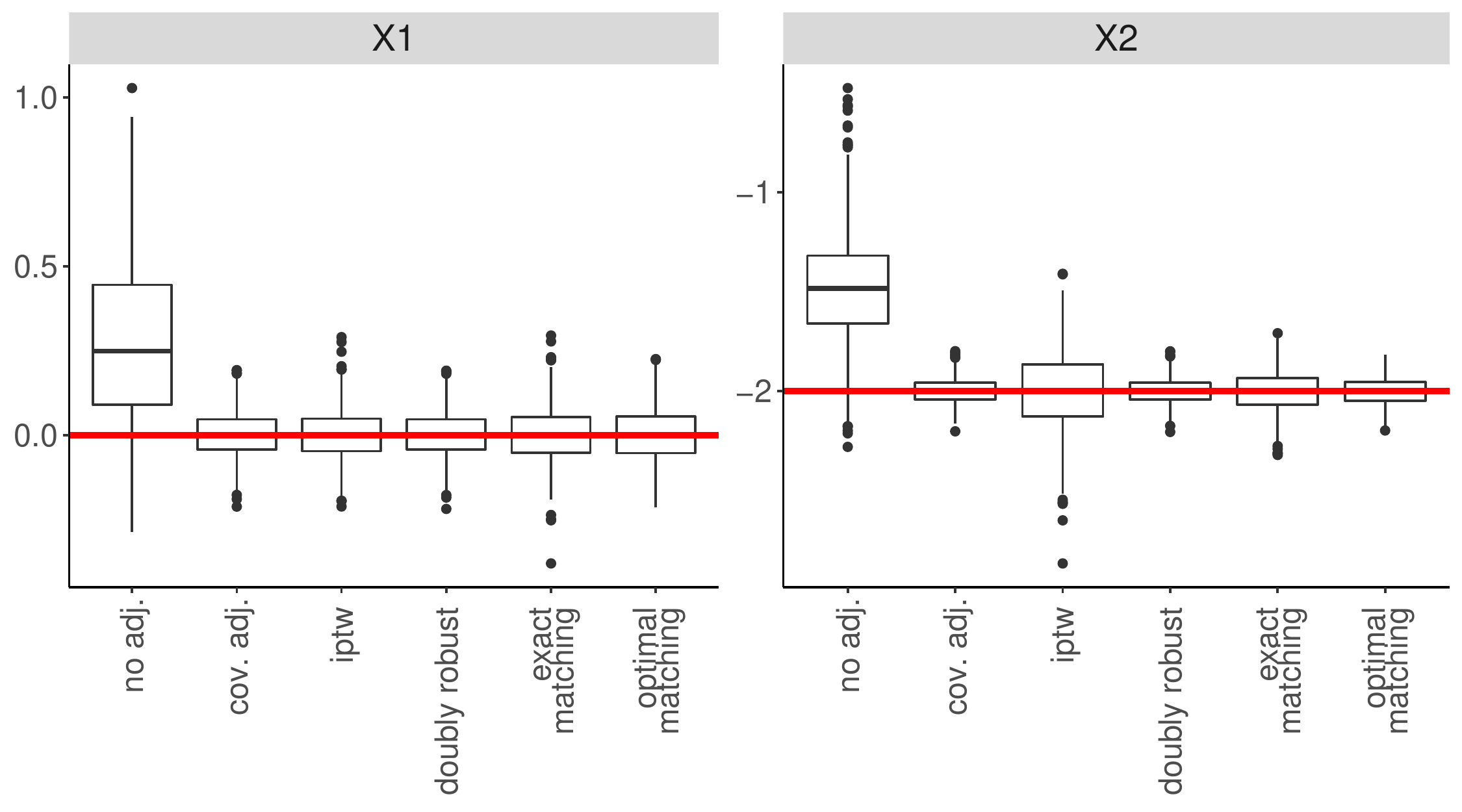}
\label{fig:Simresults_predacc_predeff8}
\end{subfigure}\\[2ex]
\caption{Estimated predictive effects for $X_1$ and $X_2$. The predictive effect is biased whenever the corresponding biomarker is associated with the treatment assignment. All investigated adjustment strategies achieve an appropriate correction.}
\label{fig:Simresults_predacc_predeff}
\end{center}
\end{figure}

The individual treatment effect is obtained as described in section \ref{subsec:predMOB}. These predictions are compared against the true individual treatment effects from the outcome-generating models, and bias, variance and \ac{MSE} are assessed. The results are shown in Figure~\ref{fig:Simresults_predacc_indtrt}.\\

Without any adjustment, the individual treatment effect is biased across almost all scenarios and shows the highest variability in the estimates. The direction of the bias depends on the direction of the prognostic effect. Interestingly, no bias is observed for the unadjusted analysis in scenario~6. Although $X_2$ has a prognostic effect in the sense that it affects the outcome in the absence of treatment, the marginal effect of $X_2$ is zero in this scenario because the effect in the treated and in the untreated neutralize each other. Unbiased estimates are obtained for all scenarios by all of the adjustment methods under investigation. The variability of the predictions is comparable across all methods except exact matching. In summary, covariate adjustment, \ac{IPTW}, the combination of both and optimal matching perform equally well.\\

By averaging the individual treatment effect within the biomarker groups $X_i=0$ or $X_i=1, \ i=1,2$, and calculating the difference of these means, estimates for the modifying effect of the respective predictive biomarker are obtained. The distribution of these estimates across all simulation runs is displayed in Figure \ref{fig:Simresults_predacc_predeff}. The red line marks the true predictive effect in the respective scenario. The results are in line with the observations for the individual treatment effects. The bias that is observed for the estimates from the unadjusted predMOB can be corrected using any of the adjustment methods. The predictive effect is biased whenever the corresponding biomarker is associated with the treatment assignment. Having a prognostic effect on the outcome, that is, being a real confounder is no prerequisite (cf. Figure~\ref{fig:Simresults_predacc_predeff2}). A bias is also observed for $X_1$ in scenario~7,  although $X_1$ has no predictive effect according to the outcome-generating model. The direction of the bias, for both  $X_1$ and $X_2$, depends on the constellation of the direction of the prognostic and predictive effects in the outcome-generating model. More specifically, a bias toward the null is observed in the estimation of the predictive effect of $X_2$ in the scenarios in which the prognostic and the predictive effect of $X_2$ are opposite, and an overestimation of the predictive effect is observed when the effects point in the same direction. Analogously to the predicted individual treatment effects, the variability is the largest for the unbiased predMOB. Among the adjusted analyses, the estimates obtained using exact matching are most variable; for all other methods, the variability is comparable.\\
The direction of the results reported within this section is correct only when the presence of the biomarkers is positively associated with the assignment to the active treatment arm. When the association is reversed, the direction of the bias is also inverted.

%%%%%%%%%%%%%%%%%%%%%%%%%%%%%%%%%%%%%%%%%%%%%%%%%%%%%%%%%%%%%%%%%%%%%%%%%%%%%%%%%%%%%%%%%%%%%%%%%%%%%%%%%%%%%%%%%%%%%%%

\section{Application example: GBSG2 BREAST CANCER STUDY}\label{sec:GBSG}
\begin{figure}[ht!]
\begin{center}
\includegraphics[height=5cm]{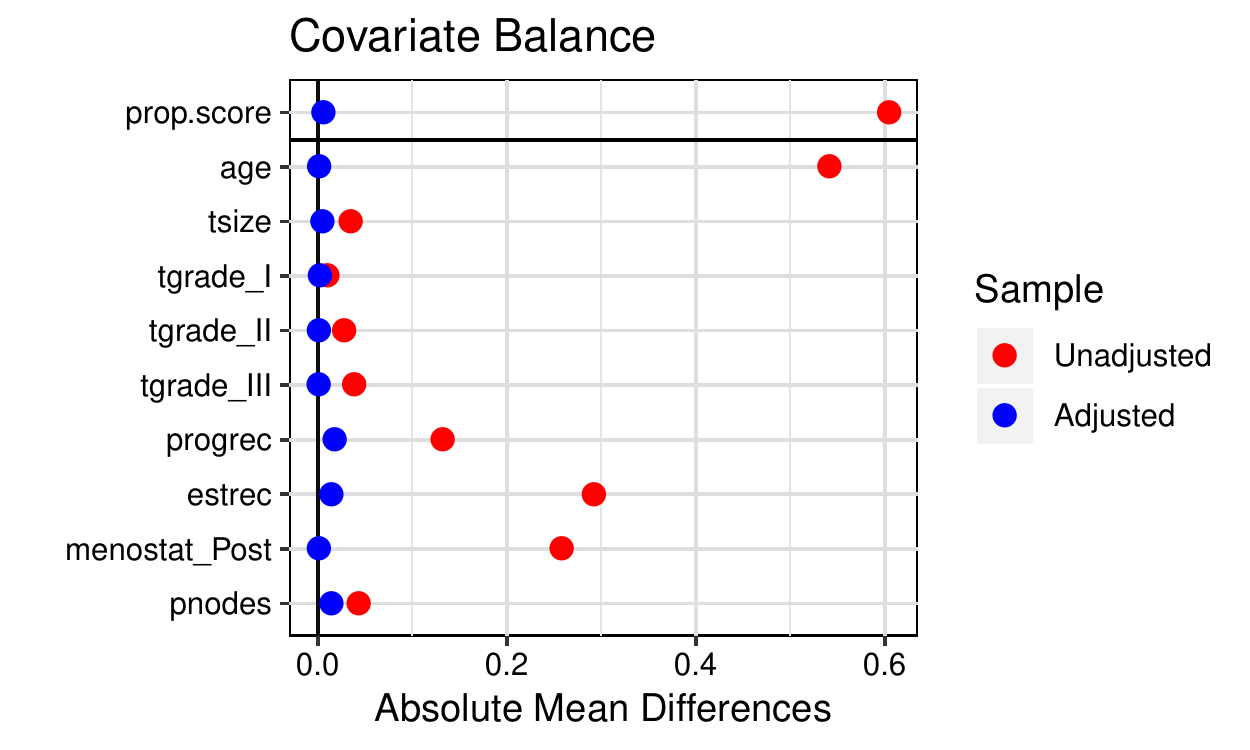}
\caption{Comparison of difference in means between patients treated with or without tamoxifen for potential confounder variables. Red circles represent the unadjusted data while blue circles display the differences after \ac{IPTW}. The differences in age, estrogen receptor and menopausal status can be balanced out by \ac{IPTW}.}
\label{fig:GBCS_balance}
\end{center}
\end{figure}

\begin{landscape}
\begin{figure}[t]
\begin{center}
\begin{subfigure}[t]{.25\textwidth}
\centering
\includegraphics[height=4.5cm, trim=0 20 0 20, clip]{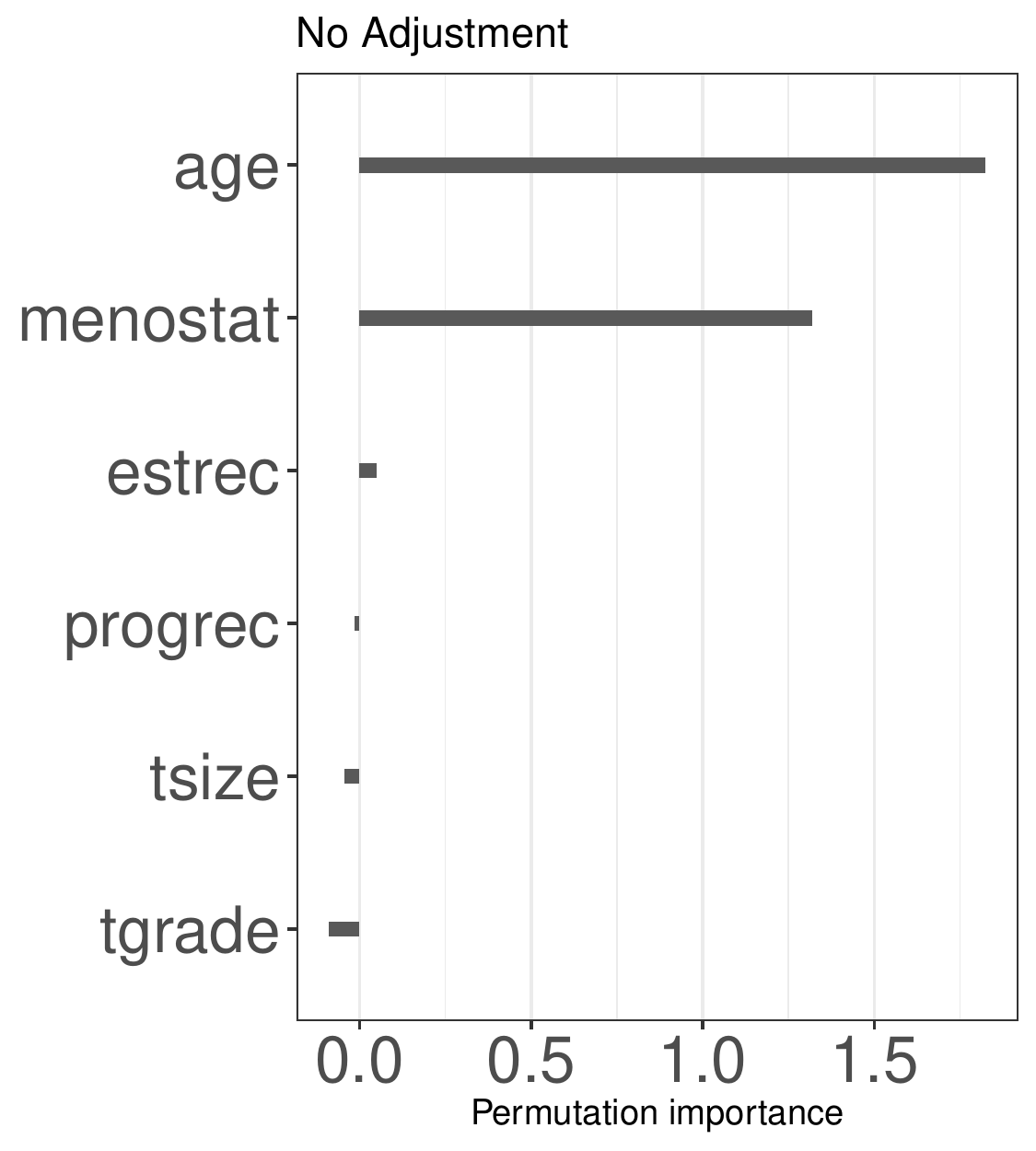}
\subcaption{\ac{PI} for unadjusted predMOB}
\end{subfigure}\hfill
\begin{subfigure}[t]{.25\textwidth}
\includegraphics[height=4.5cm, trim=0 20 0 20, clip]{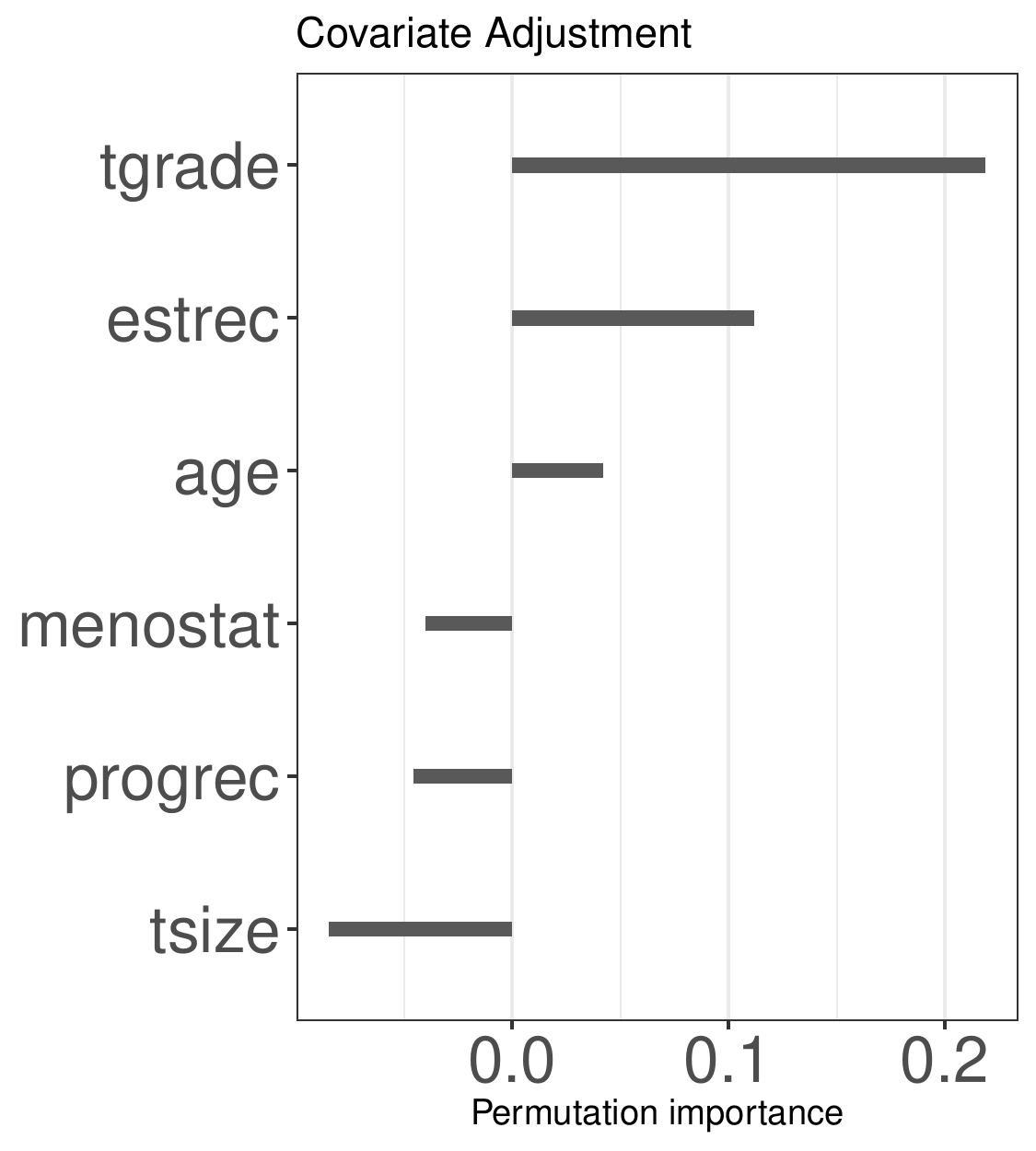}
\subcaption{\ac{PI} for predMOB with covariate adjustment}
\end{subfigure}\hfill
\begin{subfigure}[t]{.25\textwidth}
\includegraphics[height=4.5cm, trim=0 20 0 20, clip]{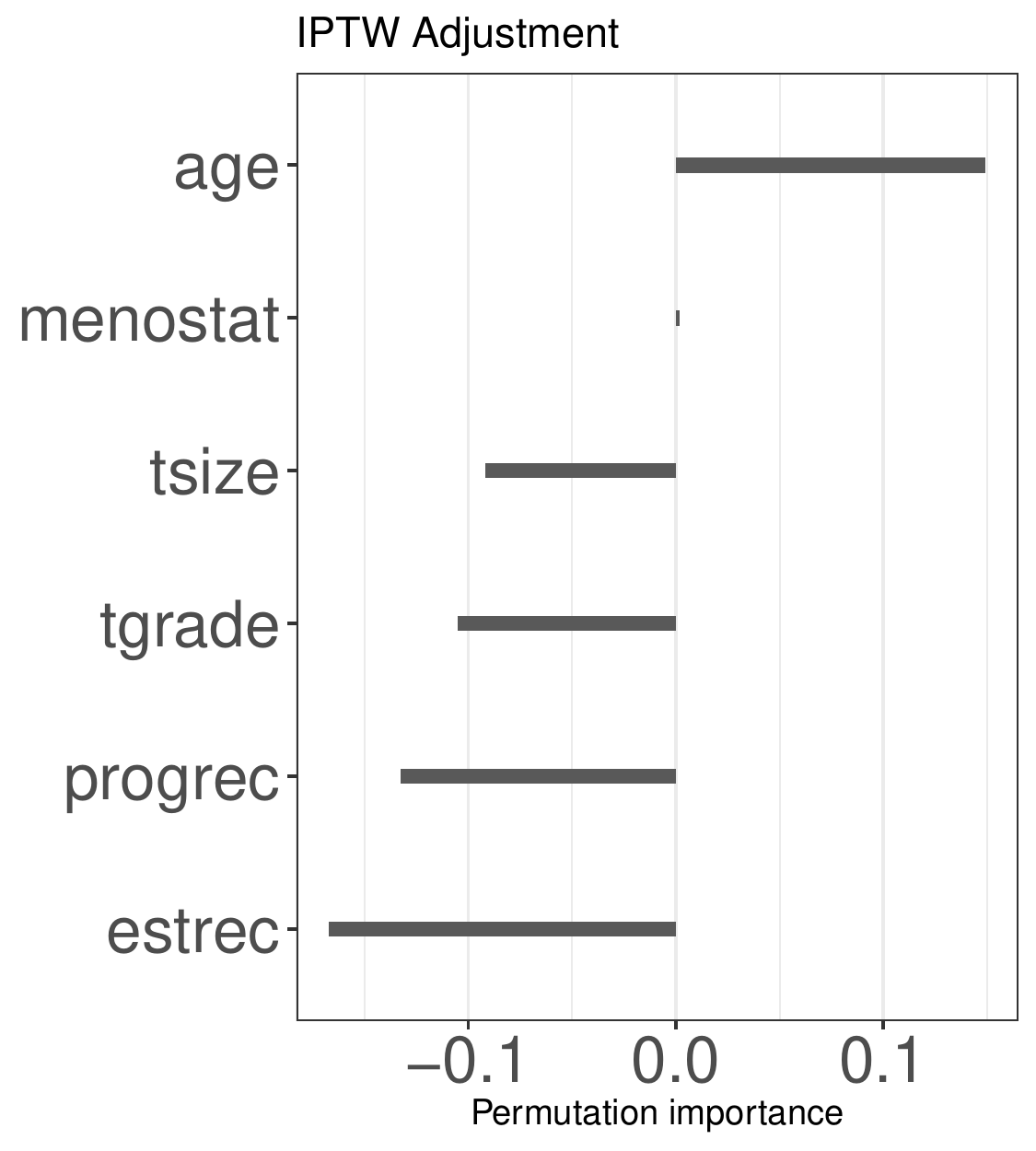}
\subcaption{\ac{PI} for predMOB with \ac{IPTW}}
\end{subfigure}\hfill
\begin{subfigure}[t]{.25\textwidth}
\includegraphics[height=4.5cm, trim=0 20 0 20, clip]{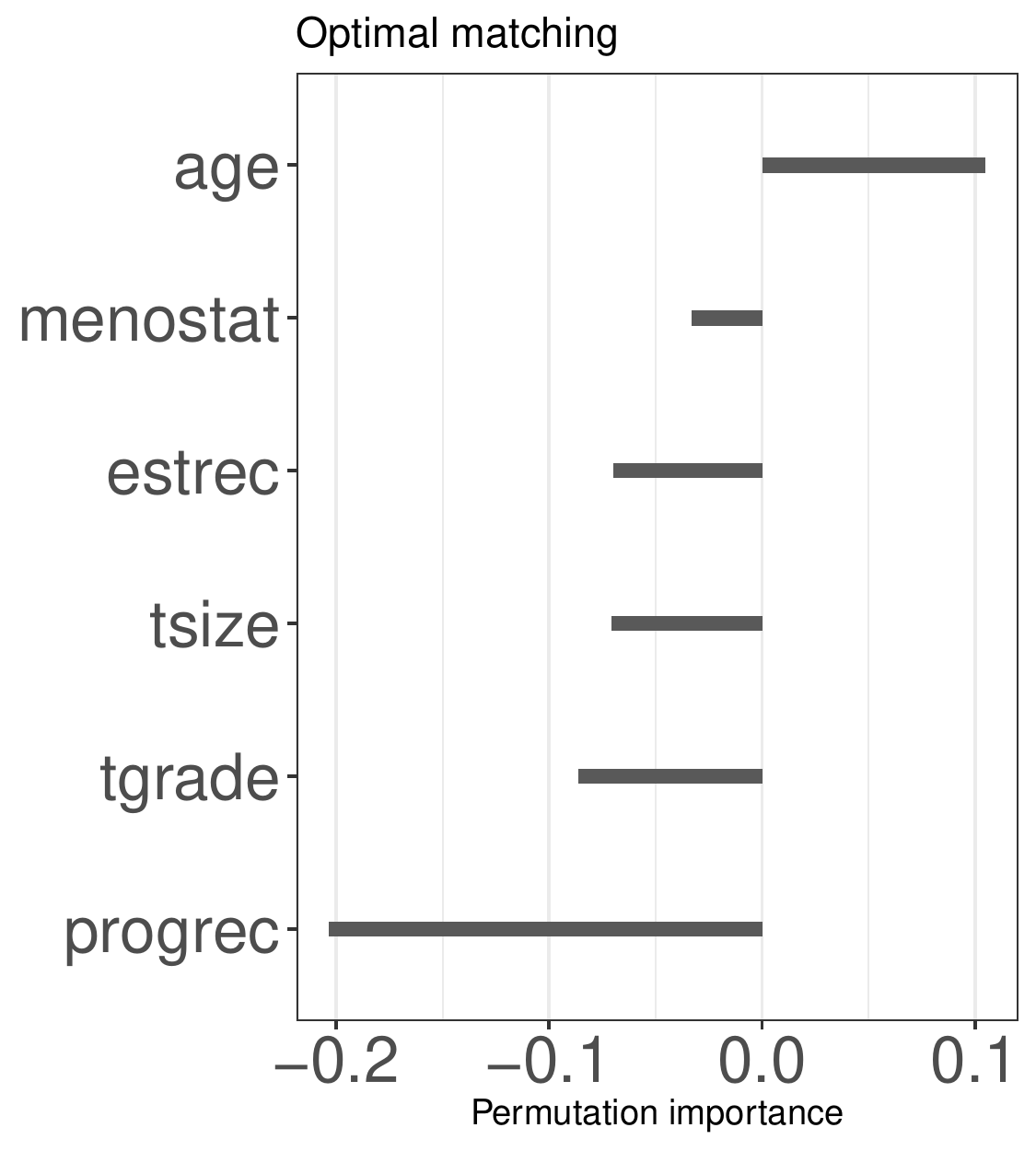}
\subcaption{\ac{PI} for predMOB with optimal matching}
\end{subfigure}\hfill
\begin{subfigure}[t]{.25\textwidth}
\includegraphics[height=4.5cm, trim=0 20 0 20, clip]{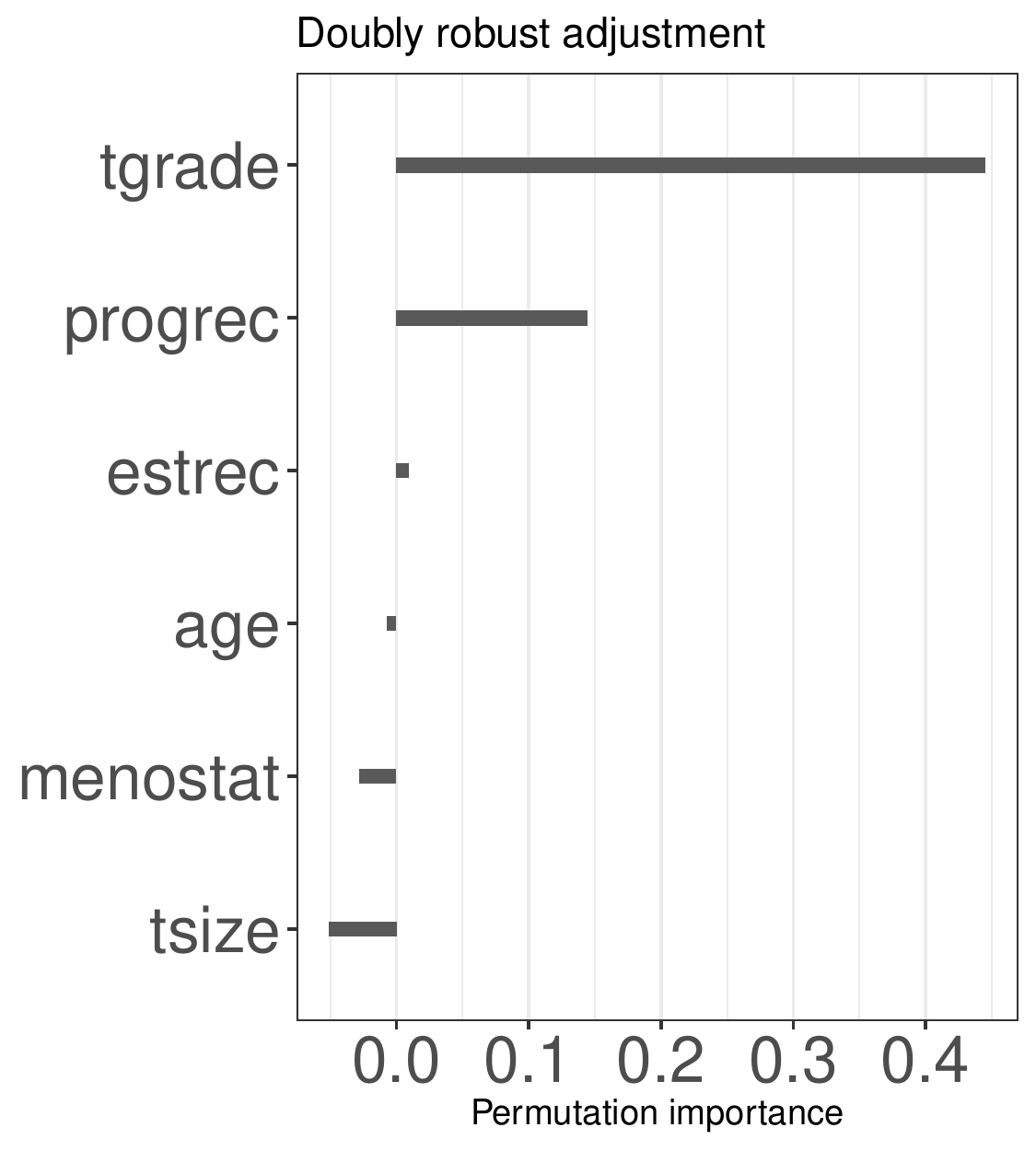}
\subcaption{\ac{PI} for predMOB with doubly robust adjustment}
\end{subfigure}\\[3ex]
\begin{subfigure}[t]{.25\textwidth}
\centering
\includegraphics[height=4.5cm, trim=0 20 0 20, clip]{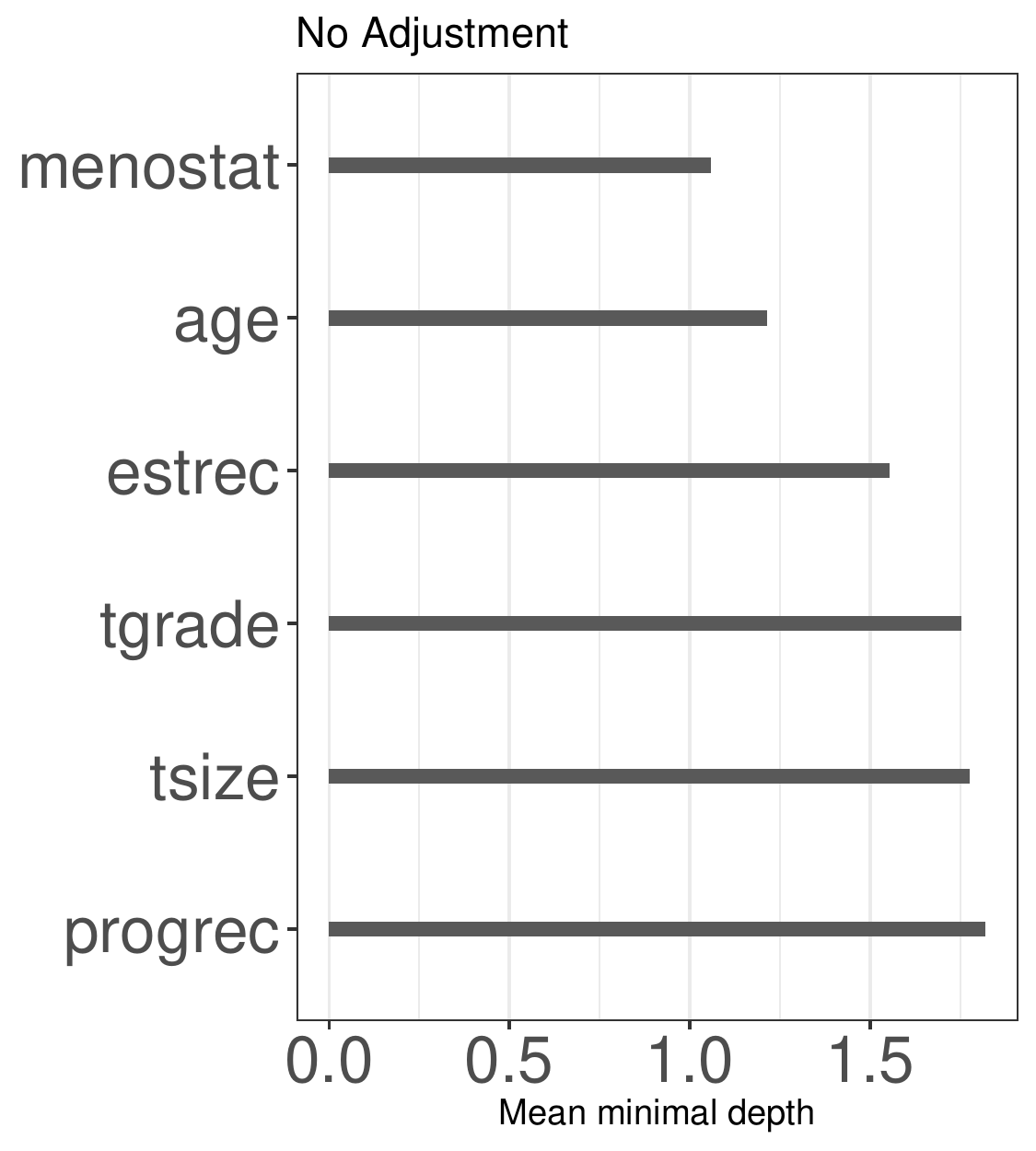}
\subcaption{Mean minimal depth for unadjusted predMOB}
\end{subfigure}\hfill
\begin{subfigure}[t]{.25\textwidth}
\includegraphics[height=4.5cm, trim=0 20 0 20, clip]{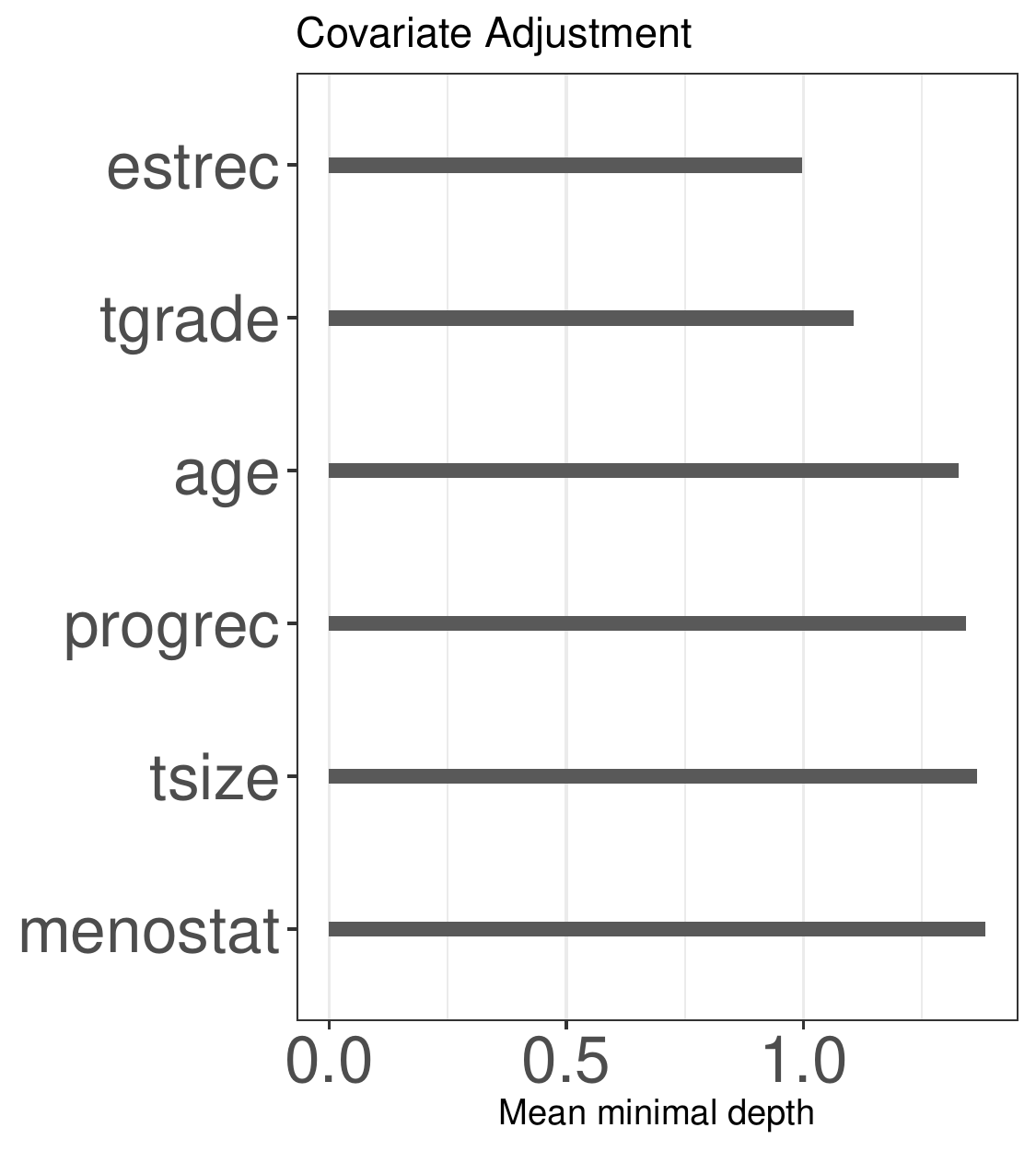}
\subcaption{Mean minimal depth for predMOB with covariate adjustment}
\end{subfigure}\hfill
\begin{subfigure}[t]{.25\textwidth}
\includegraphics[height=4.5cm, trim=0 20 0 20, clip]{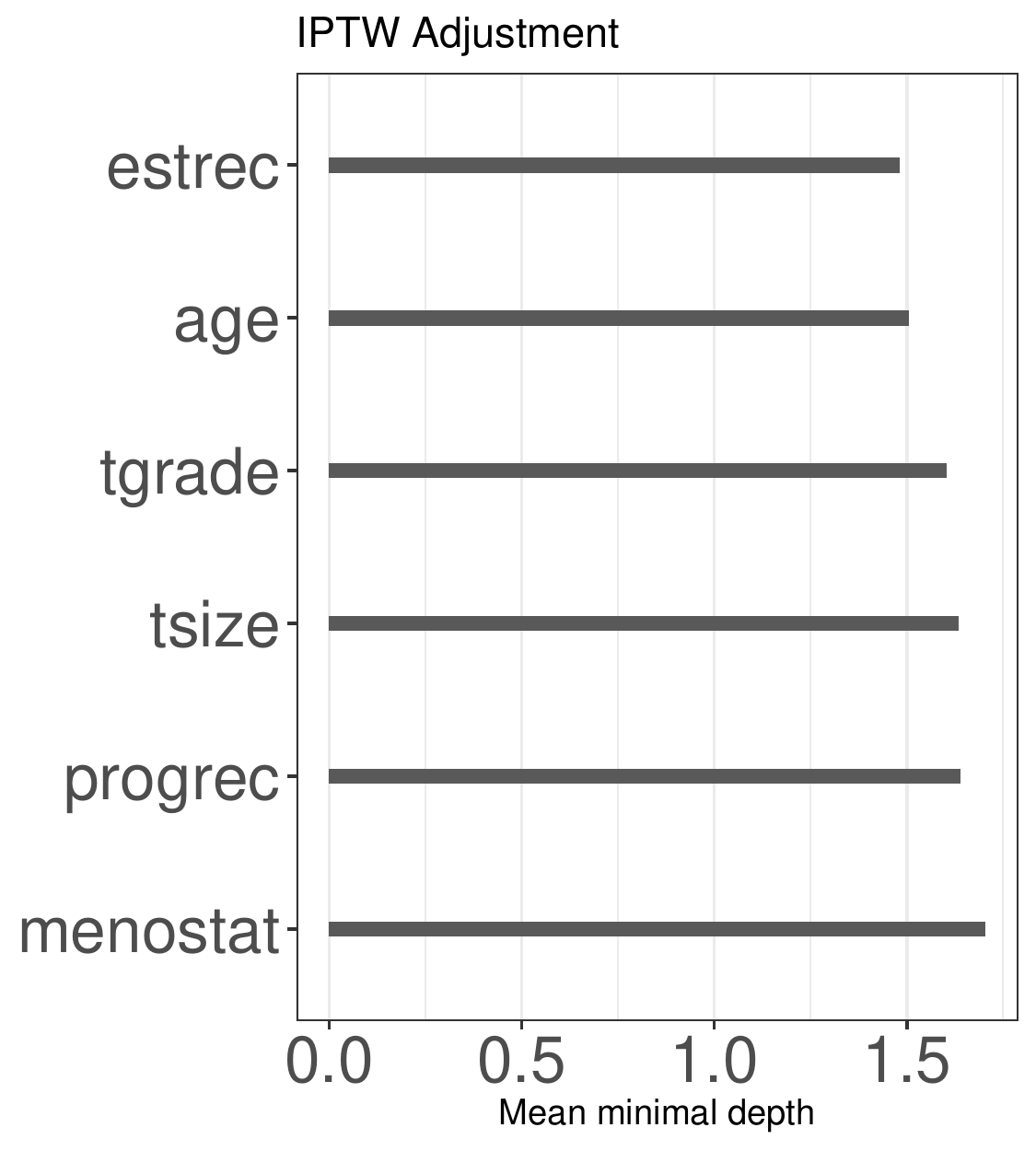}
\subcaption{Mean minimal depth for predMOB with \ac{IPTW}}
\end{subfigure}\hfill
\begin{subfigure}[t]{.25\textwidth}
\includegraphics[height=4.5cm, trim=0 20 0 20, clip]{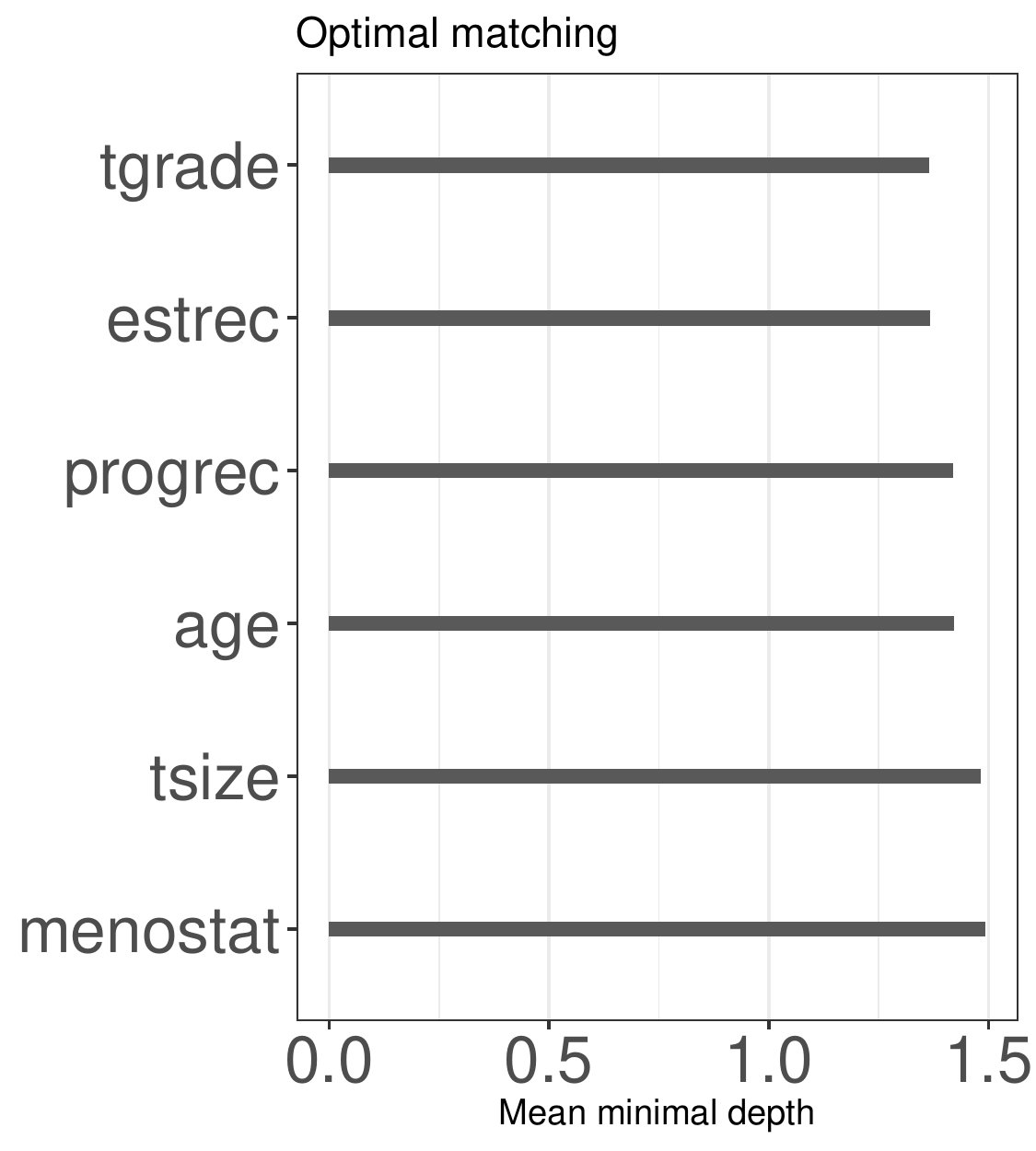}
\subcaption{Mean minimal depth for predMOB with optimal matching}
\end{subfigure}\hfill
\begin{subfigure}[t]{.25\textwidth}
\includegraphics[height=4.5cm, trim=0 20 0 20, clip]{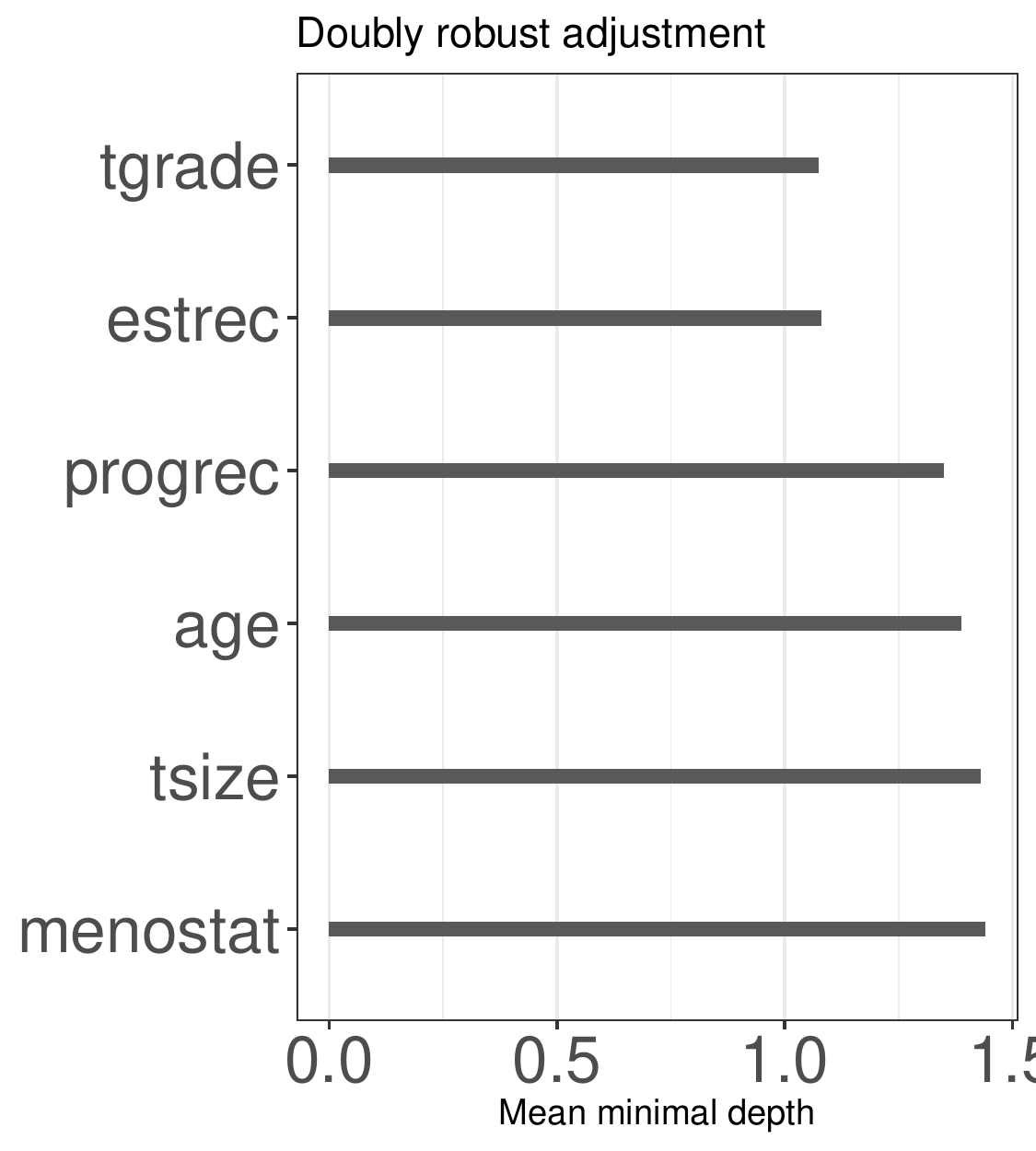}
\subcaption{Mean minimal depth for predMOB with doubly robust adjustment}
\end{subfigure}
\caption{Variable importance for the unadjusted predMOB versus predMOB in combination with common methods for confounder adjustment. Without adjustment, the predMOB identifies two major confounders as predictive factors. Using either \ac{IPTW} or optimal matching produces permutation importance values close to zero and a comparable mean minimal depth across all variables. The two strategies involving covariate adjustment propose tumor grade and the hormone receptor status as variables that are most likely to have a predictive effect.}
\label{fig:GBCS_varimp}
\end{center}
\end{figure}
\end{landscape}

\begin{figure}[h]
\begin{center}
\begin{subfigure}[t]{.5\textwidth}
\centering
\includegraphics[height=5cm]{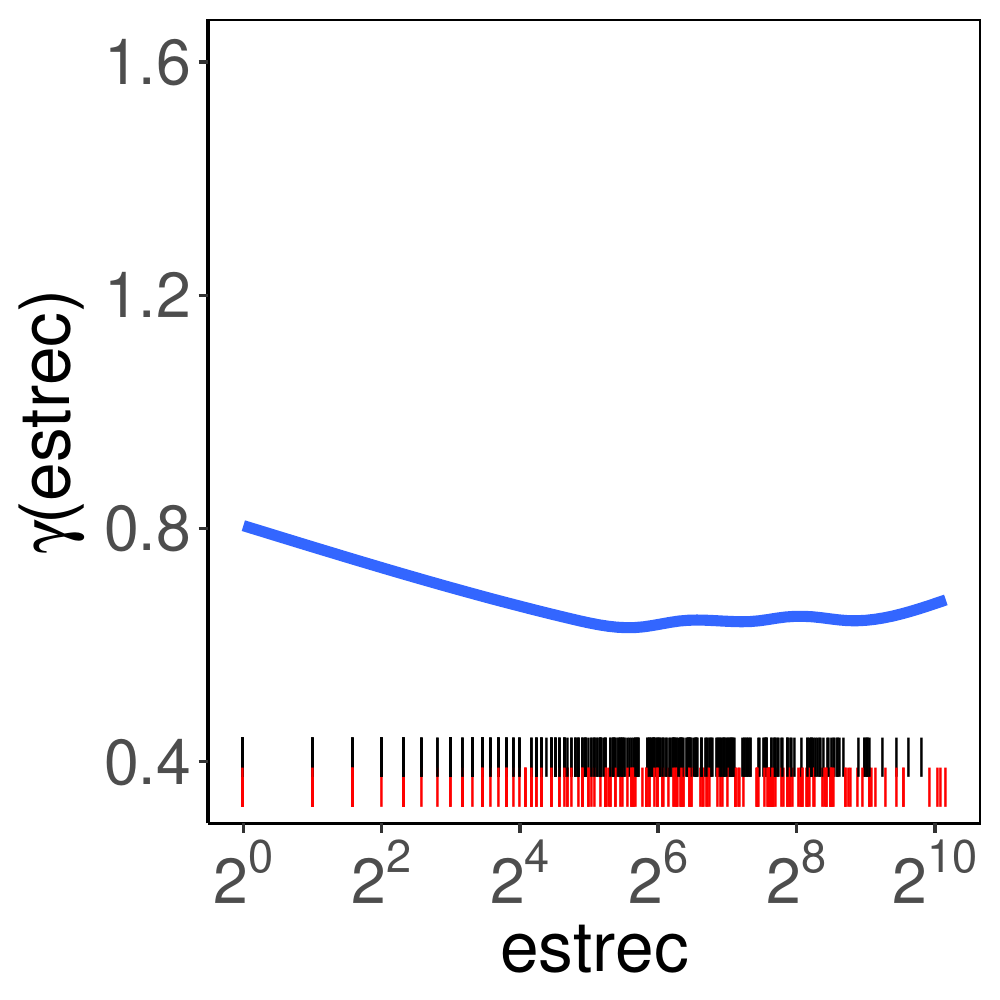}
\end{subfigure}\hfill
\begin{subfigure}[t]{.5\textwidth}
\centering
\includegraphics[height=5cm]{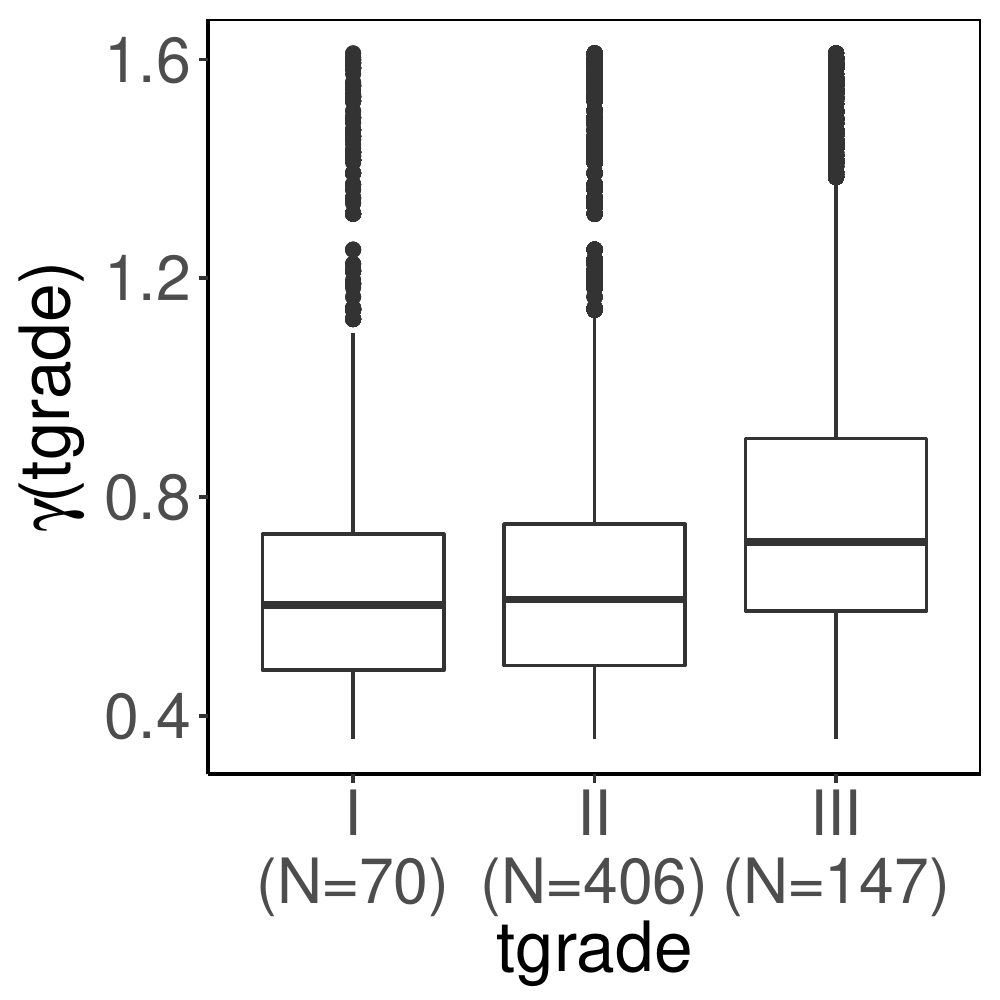}
\end{subfigure}\\
\begin{subfigure}[t]{.5\textwidth}
\centering
\includegraphics[height=5cm]{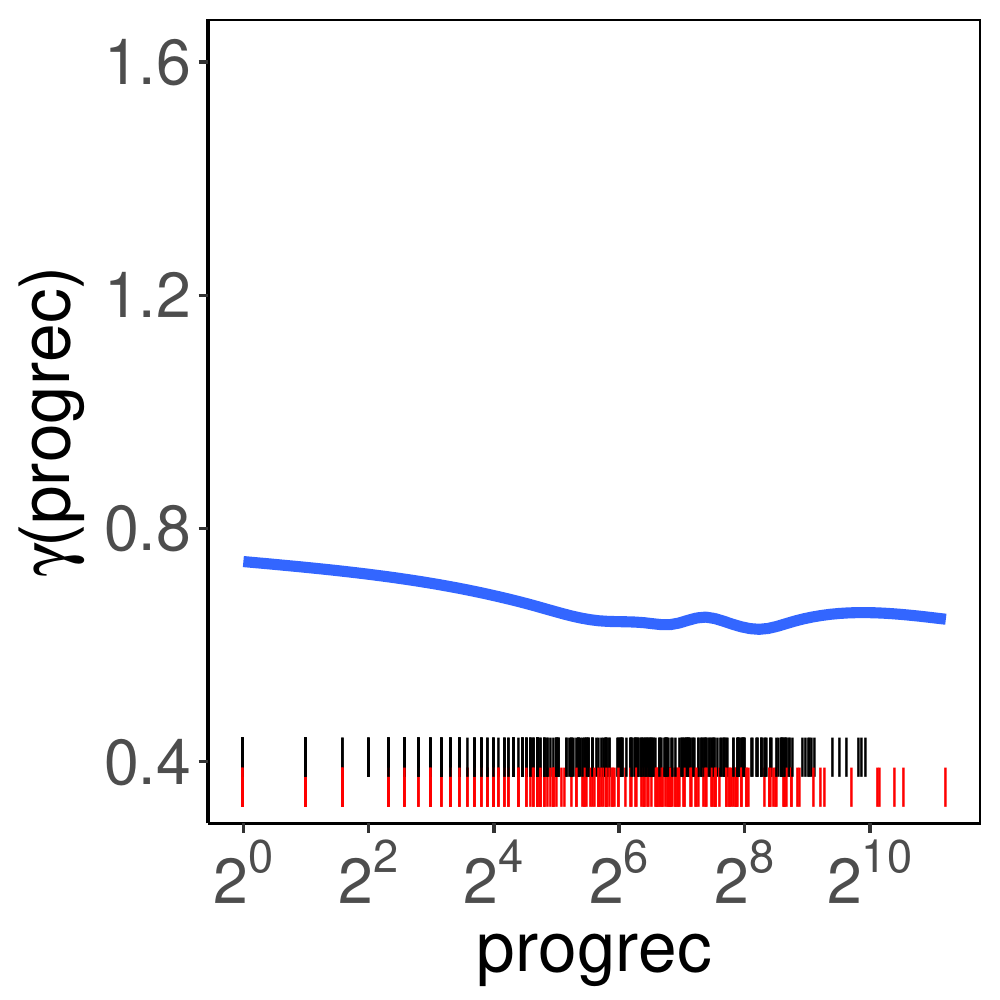}
\end{subfigure}\hfill
\begin{subfigure}[t]{.5\textwidth}
\centering
\includegraphics[height=5cm]{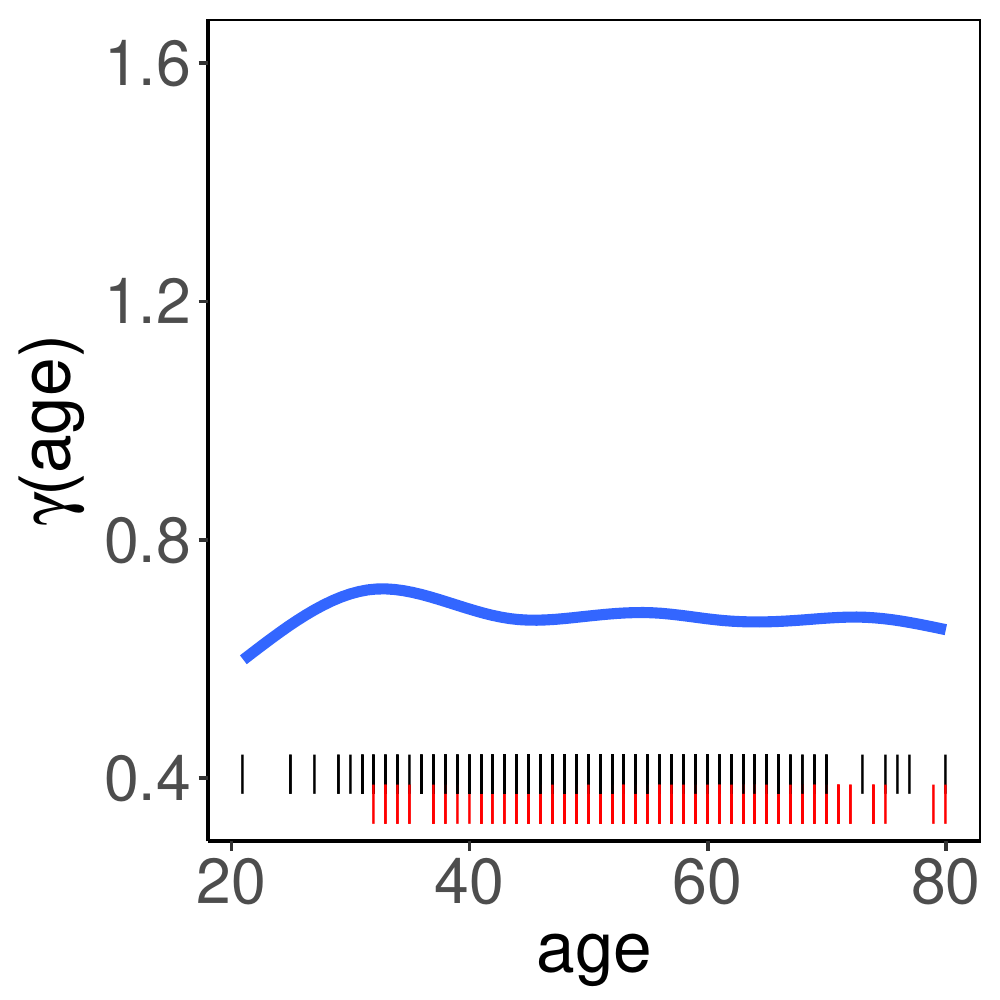}
\end{subfigure}\\
\begin{subfigure}[t]{.5\textwidth}
\centering
\includegraphics[height=5cm]{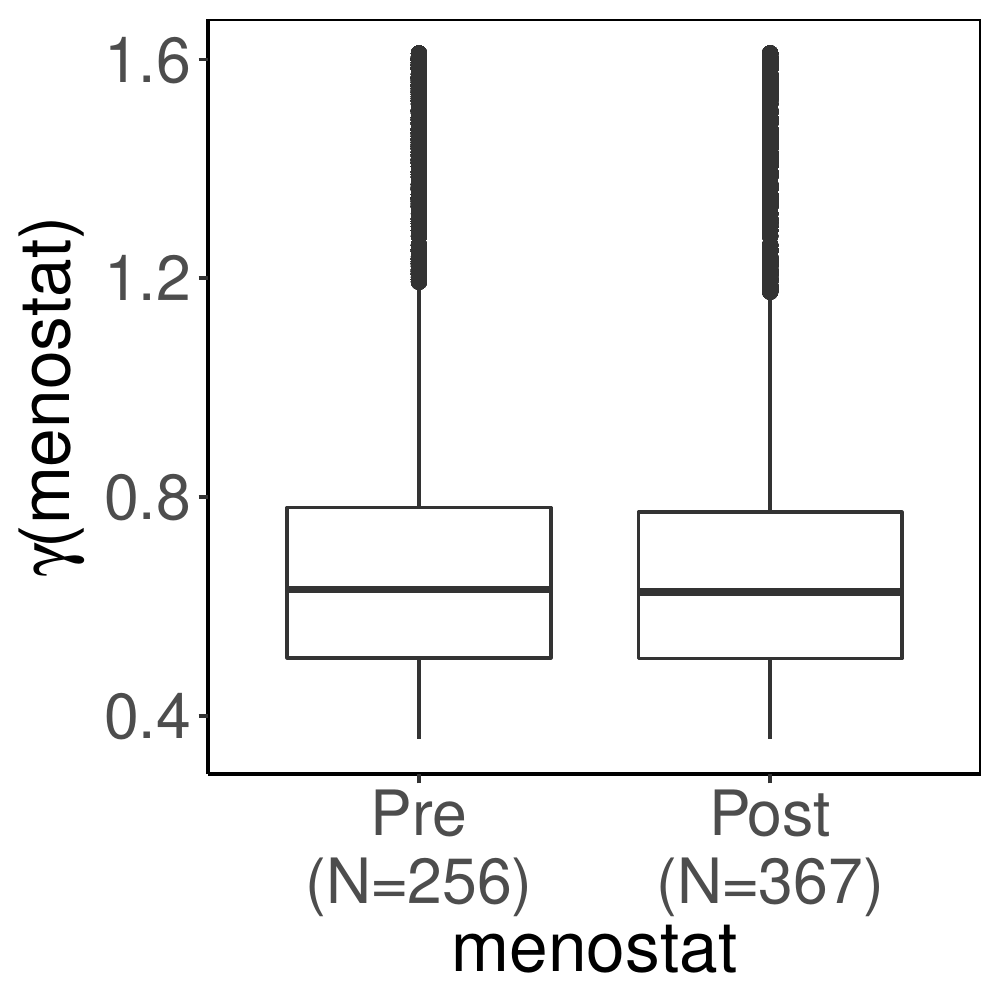}
\end{subfigure}\hfill
\begin{subfigure}[t]{.5\textwidth}
\centering
\includegraphics[height=5cm]{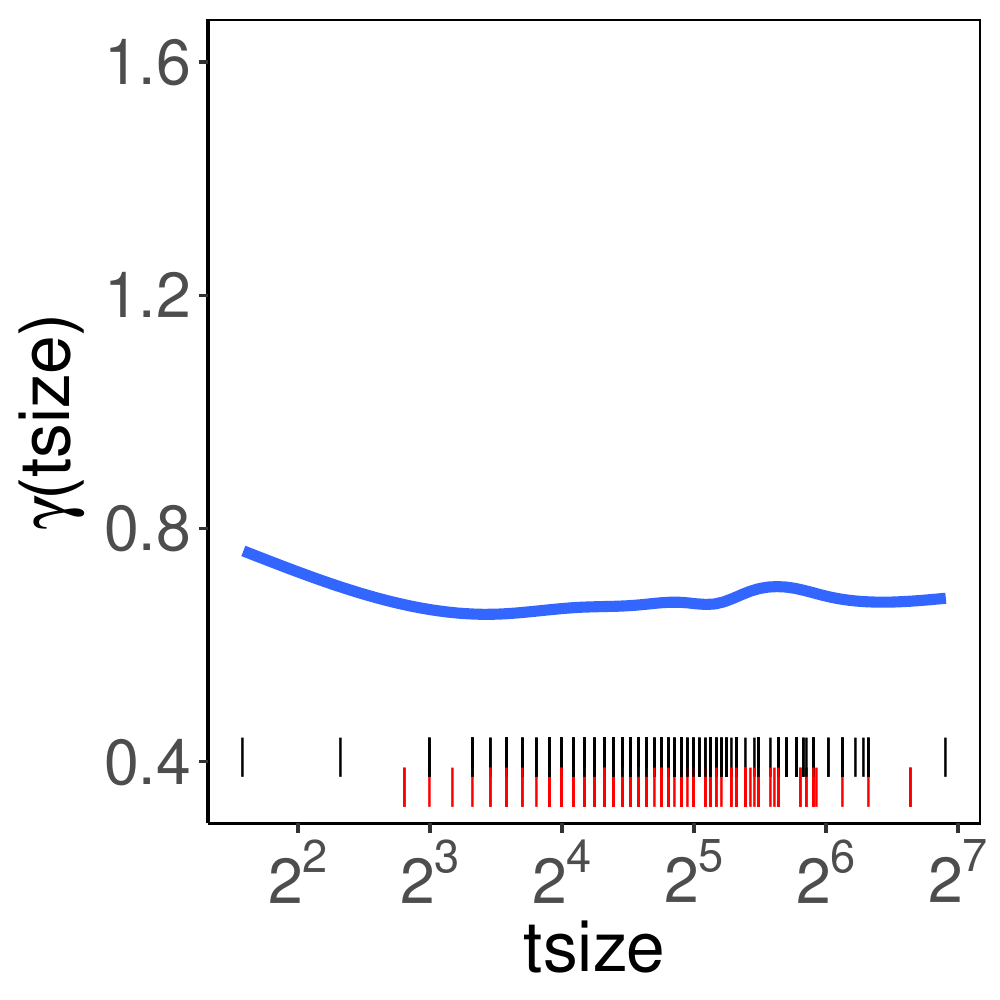}
\end{subfigure}
\caption{Partial dependence plots for predMOB in combination with doubly robust adjustment. The rug plots on the bottom of the plots for the continuous variables show the distribution of the values in the two treatment arms - patients having been treated with tamoxifen are depicted in red, patients in the treatment arm without tamoxifen in black.}
\label{fig:GBCS_Depplots}
\end{center}
\end{figure}

An important criterion for the treatment decision in breast cancer patients is the hormone receptor status of estrogen and progesterone. Cancer cells with a positive hormone receptor status are known to be responsive to therapies that lower the hormone levels or prevent the fostering of cancer cells by the respective hormone. \\
Between $1984$ and $1989$, $720$ node-positive breast cancer patients entered the \ac{GBSG} trial 2, a Comprehensive Cohort Study. \cite{Schmoor1996} The study medication consisted of chemotherapy with or without the additional use of the estrogen-receptor modulator tamoxifen. Upon trial entry, patients could decide whether they agree to be randomized to one of the two arms or whether they themselves or the treating physician shall be responsible for the treatment decision. The data set of $686$ study participants with complete information is freely available and was repeatedly utilized in the past to illustrate new statistical methods. For example, \citet{Royston2004} investigated the interaction of the estrogen receptor status and treatment with tamoxifen using an extension of the \ac{MFP} approach.\\
In the following, a predMOB forest is applied to this data in order to see whether the estrogen receptor status is found to be predictive by this approach and whether the adjustment methods yield diverging results. Known prognostic factors (age (age), menopausal status pre/post (menostat), tumor size in mm (tsize), tumor grade I - III (tgrade), progesterone (progrec) and estrogen receptor status (estrec) in fmol) for treatment with tamoxifen are selected as candidates. Variables with high variable importance measures in the resulting forest are considered to be potentially predictive. As around one third of the patients in this data set originate from the non-randomized part of the trial, the results from the unadjusted predMOB are compared against the combination of predMOB with different methods for confounder adjustment. These methods are supposed to establish a balance in the distributions of the covariates (known prognostic variables as mentioned above plus number of positive nodes (pnodes)) between patients treated with and without tamoxifen and thus reduce the risk of bias due to confounding. Figure \ref{fig:GBCS_balance} displays the absolute mean differences in the covariates between the two treatment arms. The dots in red reveal that the covariates with the largest imbalance are age, estrogen receptor and menopausal status. After using \ac{IPTW}, the mean differences in the weighted population (marked in blue) are close to zero. \\
Since the predMOB has so far only been developed for normal and binary endpoints, \ac{RFS} times are converted to a binary endpoint using a cut-off of two years. Patients experiencing an event related to \ac{RFS} within the first two years after study start shall be separated from patients who do not experience such an event within this time frame or at all. This leads to $179$ patients having an \ac{RFS} events and $514$ patients without event. For the $63$ patients censored within the first two years, the RFS status at $2$ years cannot be determined and so these patients cannot be considered for the analysis. \\
The results reported in the following are based on $1000$ single trees, each grown on a subsample of $394$ patients and $3$ variables sampled as candidates for each split. The variable importance is averaged over all trees and displayed in a barplot with the most important variable at the top (see Figure \ref{fig:GBCS_varimp}). \\
The rankings of the variables vary for the different adjustment strategies. In case of no confounder adjustment, age and menopausal status have the highest variable importance. This is probably best explained by their potential confounding character. Analogous to the simulation results, no appropriate correction can be achieved using \ac{IPTW}. The permutation importance again proposes age as a variable with a potential predictive effect. However, the values are close to zero and the mean minimal depth across the variables is generally very similar, such that none of the variables can be clearly identified. The same is true for optimal matching. The results obtained under either covariate adjustment alone or in a doubly robust approach detect a predictive effect for tumor grade as well as for the hormonal receptor status of estrogen or progesterone. Whereas the two variable importance measures produce consistent results for the other strategies, the results for the doubly robust adjustment diverge. While the permutation importance is highest for tumor grade and progesterone receptor status, the mean minimal depth advocates the alternative hormone receptor estrogen. Partial dependence plots can be used to clarify the relationship between variables with a potential predictive effect and the treatment effect. \cite{Hastie2009} These plot the prediction for the individual treatment effect as described in \ref{subsec:predMOB} against the actual value of the respective observations via a smoothing curve using e.g. cubic splines. The resulting plots show larger treatment effects with increasing levels of both of the hormone receptors (Figure \ref{fig:GBCS_Depplots}). However, the trend for estrogen seems to be stronger than for progesterone which corresponds more to the results of the mean minimal depth. Furthermore, the treatment effect is slightly smaller for patients with grade III  tumors.

The results for 2-year \ac{RFS} are basically in line with what has been reported by \citeauthor{Royston2004}. \cite{Royston2004} This can be considered as a  verification of the analysis strategy of using predMOB in combination with covariate adjustment (plus \ac{IPTW}). Furthermore, a more recent meta analysis confirmed estrogen receptor status as a predictive factor for tamoxifen in early breast cancer patients. \cite{EBCTCG2011} 

%%%%%%%%%%%%%%%%%%%%%%%%%%%%%%%%%%%%%%%%%%%%%%%%%%%%%%%%%%%%%%%%%%%%%%%%%%%%%%%%%%%%%%%%%%%%%%%%%%%%%%%%%%%%%%%%%%%%%%%

\section{Discussion}\label{sec:discuss}
Since data from a randomized treatment comparison is not always available, the idea of basing the search for predictive factors on data from observational studies or registry data seems appealing. While confounder adjustment for the average treatment effect is well established, research on unbiased estimation of treatment effect modification or causal interactions is sparse. In this work, we systematically investigate the potential bias that arises with the unadjusted application of a tree-based method for the identification of predictive factors, the predMOB, to non-randomized data and compare common adjustment strategies with respect to the achieved correction. Concerning the identification of the predictive factors by means of the variable importance of the variables in the forest, results obtained by covariate adjustment are most reliable. \ac{IPTW} fails to achieve an appropriate correction if the true predictive factor itself is associated with the treatment assignment. On the other hand, the use of matching procedures does not seem advisable in the presence of correlations. The doubly robust approach combining covariate adjustment and \ac{IPTW} shows no additional benefit over covariate adjustment alone in the investigated scenarios. With respect to the prediction of the individual treatment effect as well as the modifying effect of the true predictive factor, all adjustment methods produce unbiased results. In particular, there are no differences in the prediction accuracy when using covariate adjustment compared to \ac{IPTW}. This is in line with previous results for the estimation of the treatment effect itself. \cite{Schmoor2008, Schmoor2011} In summary, covariate adjustment emerges as the best adjustment strategy overall. Nevertheless, we would also recommend a doubly robust approach for practical applications as it might show benefits in more complex setting not covered in our simulation study. More details on the predMOB for randomized and non-randomized settings can be found in the corresponding thesis.\cite{KrzykallaDiss2020}\\
It is important to note that these conclusions only yield as long as all relevant variables have been used for the adjustment, that is, under the common assumption of no unmeasured confounding. In our simulations, we use all available variables in order to reflect a practical scenario in which there is no definite knowledge about the set of true confounders. It can be expected that the investigated adjustment strategies might fail in case one of the actual confounders has not been observed.
In epidemiological applications, causal \ac{DAG}s are a common method to identify the variables that have to be included in the adjustment of the treatment effect estimate. Unfortunately, there is no final solution on how to include effect modification in these graphs. \cite{VanderWeele2007, Weinberg2007} Even more so, there is the opinion that it should not be included in a causal graph because effect modification is depending on scale, e.g. effect modification in terms of risk difference may not be seen when evaluating the risk ratio. Alternatively, the question as to whether unmeasured confounding is present can be addressed by sensitivity analyses.\cite{Vanderweele2011, Lee2018, Mathur2022} Extensions to interaction analyses are also available. \cite{Vanderweele2012b} 
However, this additional complexity is beyond the scope for this work. \\
Another restriction of this work is that the number of potential confounders was kept in low to moderate range that can still be managed with the proposed adjustment strategies. More complex methods that are able to handle high-dimensional sets of confounders are discussed elsewhere. \cite{Franklin2015, Greenland2008, Jackson2017} \\
When it comes to the interpretation of the results, the application example shows that a (relatively) large variable importance alone does not provide a definitive decision of whether a factor is truly predictive or not. Variable importance only allows a ranking of the variables according to their contribution in the modelling of the tree. In order to get a more concrete impression, the use of partial dependence plots to visualize the relationship of the potentially predictive factor and the treatment effect is very helpful. Alternatively, one can use methods that allow variable selection based on variable importance  \cite{Kursa2010, Janitza2018} or use subsampling to obtain an estimate for the variance and corresponding confidence intervals for the permutation importance measure, as proposed by \citeauthor{Ishwaran2019}. \cite{Ishwaran2019} The latter approach requires a large training data set as the subsamples on which the random forests are constructed only comprise a relatively small proportion of the original data set (the authors recommend $\sqrt{n}$ for each subsample). To complicate matters, it has been developed for conventional random forest prediction models which can be grown on smaller sample sizes than model-based trees like the predMOB. Therefore, it is not feasible to apply it to the data example used. \\
Further research is also needed to make predMOB applicable to time-to-event data. Since the approach requires a fully parameterized model, one possibility would be to use a Weibull model as the base model. Alternatively, pseudo-values could be used to circumvent the difficulties arising from censored data.

%%%%%%%%%%%%%%%%%%%%%%%%%%%%%%

\section{Computational Details}
All computations and analyses have been implemented in R, \mbox{version~3.4.4}, \citep{R3.4.4} together with the packages partykit \mbox{(version~1.2-2) \cite{Rpartykit}} and model4you \mbox{(version~0.9-2). \cite{Rmodel4you}} For the simulation study, data has been generated using the simstudy package \mbox{(version~0.1.9). \cite{Rsimstudy}} The data of the \ac{GBSG} Trial $2$ that has been used for the application example are openly available in the TH.data package. \cite{RTHdata}

\section*{Acknowledgements}
We thank Dr. Heidi Seibold for fruitful discussions and support regarding the implementation. 

% \section*{conflict of interest}
% You may be asked to provide a conflict of interest statement during the submission process. Please check the journal's author guidelines for details on what to include in this section. Please ensure you liaise with all co-authors to confirm agreement with the final statement.

% \printendnotes

\bibliography{BibliographySiM2}

\clearpage

\end{document}

\pagestyle{plain}
  \newgeometry{
    textheight=9in,
    textwidth=177.8mm,
    top=1in,
    headheight=14pt,
    headsep=25pt,
    footskip=30pt
  }

\Huge{Supplementary Material}
\setcounter{section}{0}
\setcounter{figure}{0} 
\setcounter{table}{0} 
\renewcommand{\thesection}{S \arabic{section}}
\renewcommand{\thefigure}{S.\arabic{figure}}
\renewcommand{\thetable}{S.\arabic{table}}

% \clearpage
\section{Simulation results for the identification of predictive factors using permutation importance} \label{supp:simresults_permimp}
\begin{figure}[h]
\begin{center}
\begin{subfigure}{.49\textwidth}
\subcaption{\textbf{Scenario 0}: Null scenario \newline}
\centering
\includegraphics[height=3.8cm]{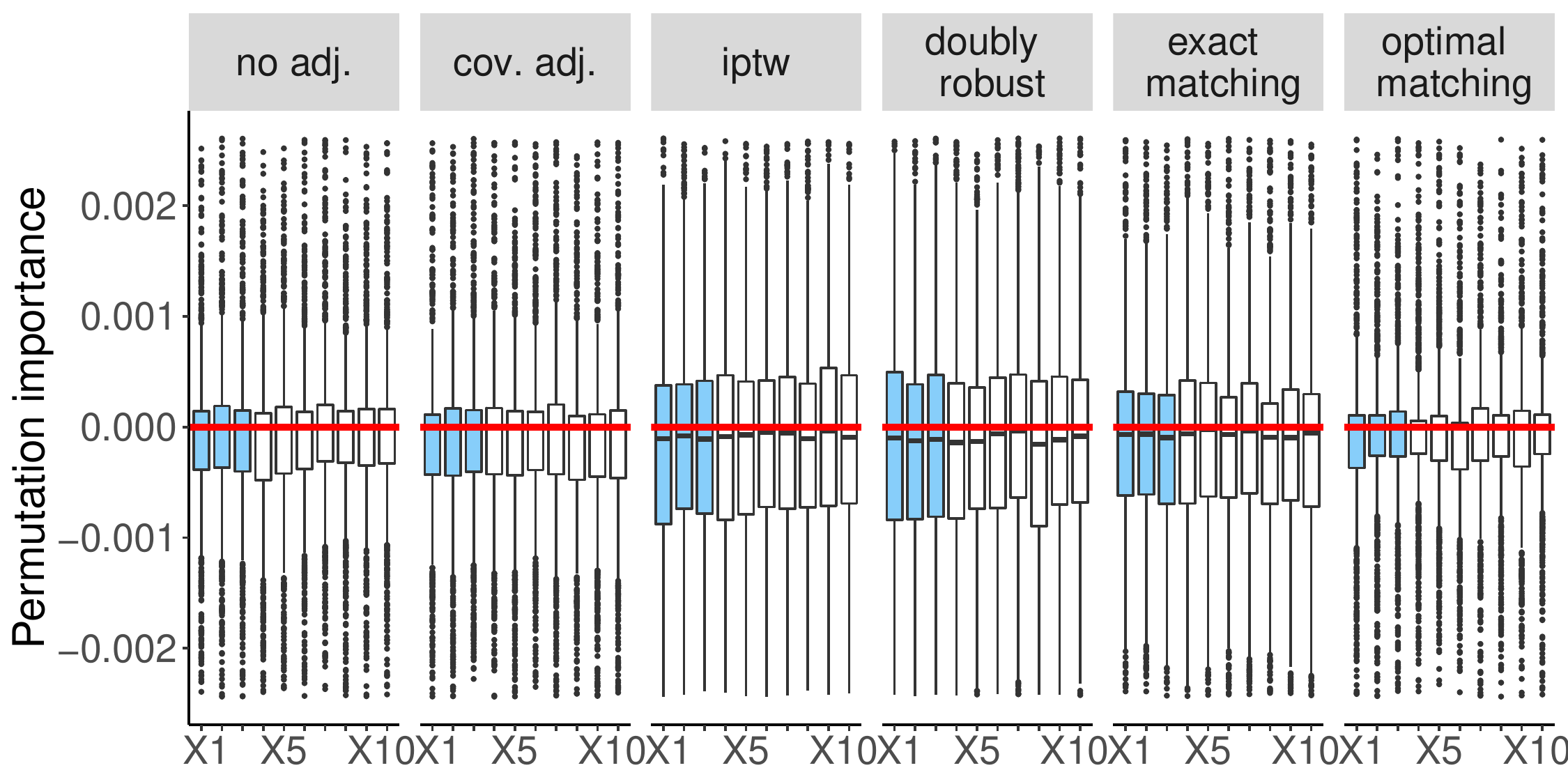}
\end{subfigure}\hfill
\begin{subfigure}{.49\textwidth}
\subcaption{\textbf{Scenario B.1}: $X_{10}$ has both a prognostic and a qualitative predictive effect}
\centering
\includegraphics[height=3.8cm]{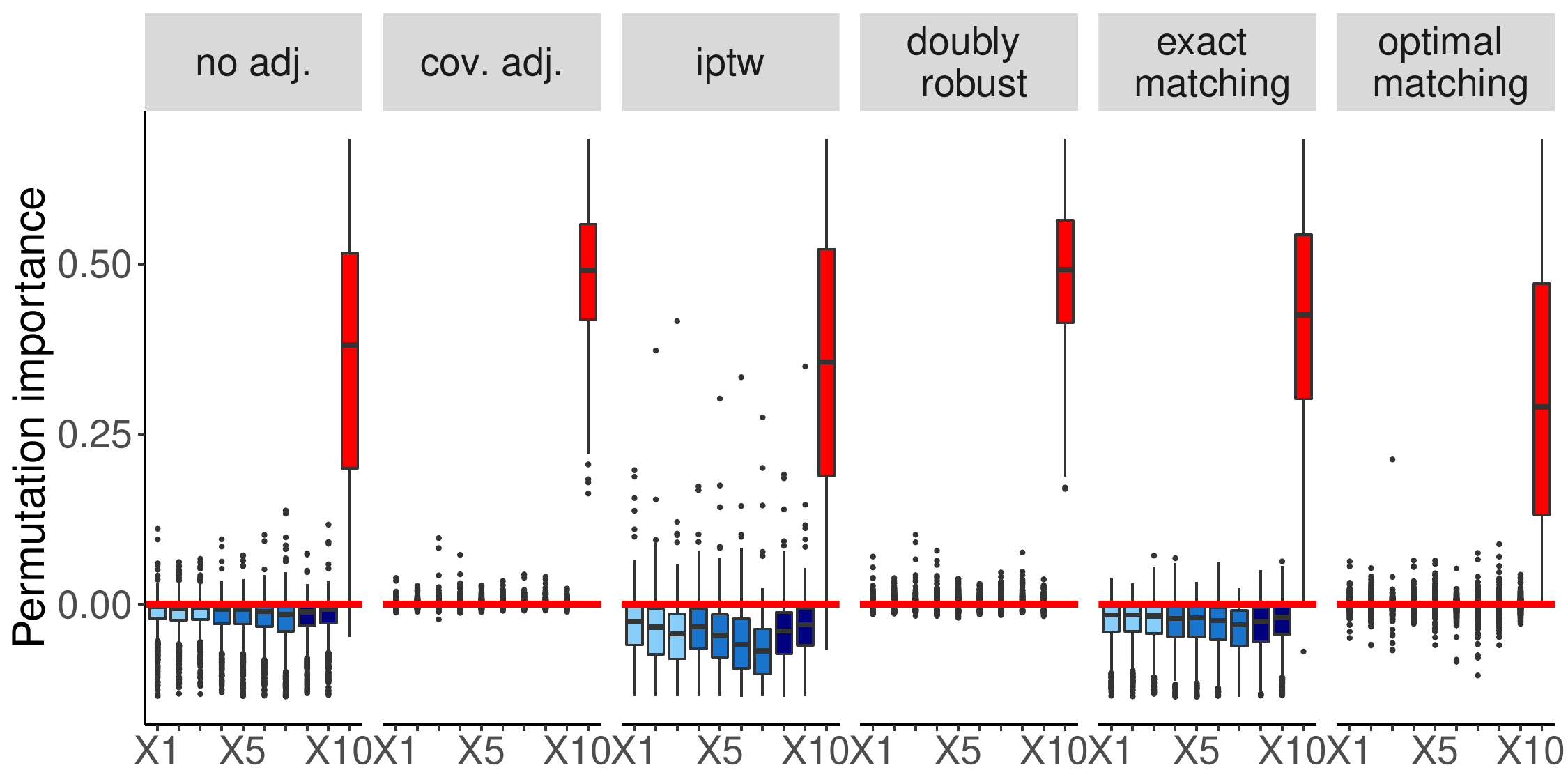}
\end{subfigure}\\[2ex]
\begin{subfigure}{.49\textwidth}
\subcaption{\textbf{Scenario B.2}: $X_{10}$ has both a prognostic and a quantitative predictive effect}
\centering
\includegraphics[height=3.8cm]{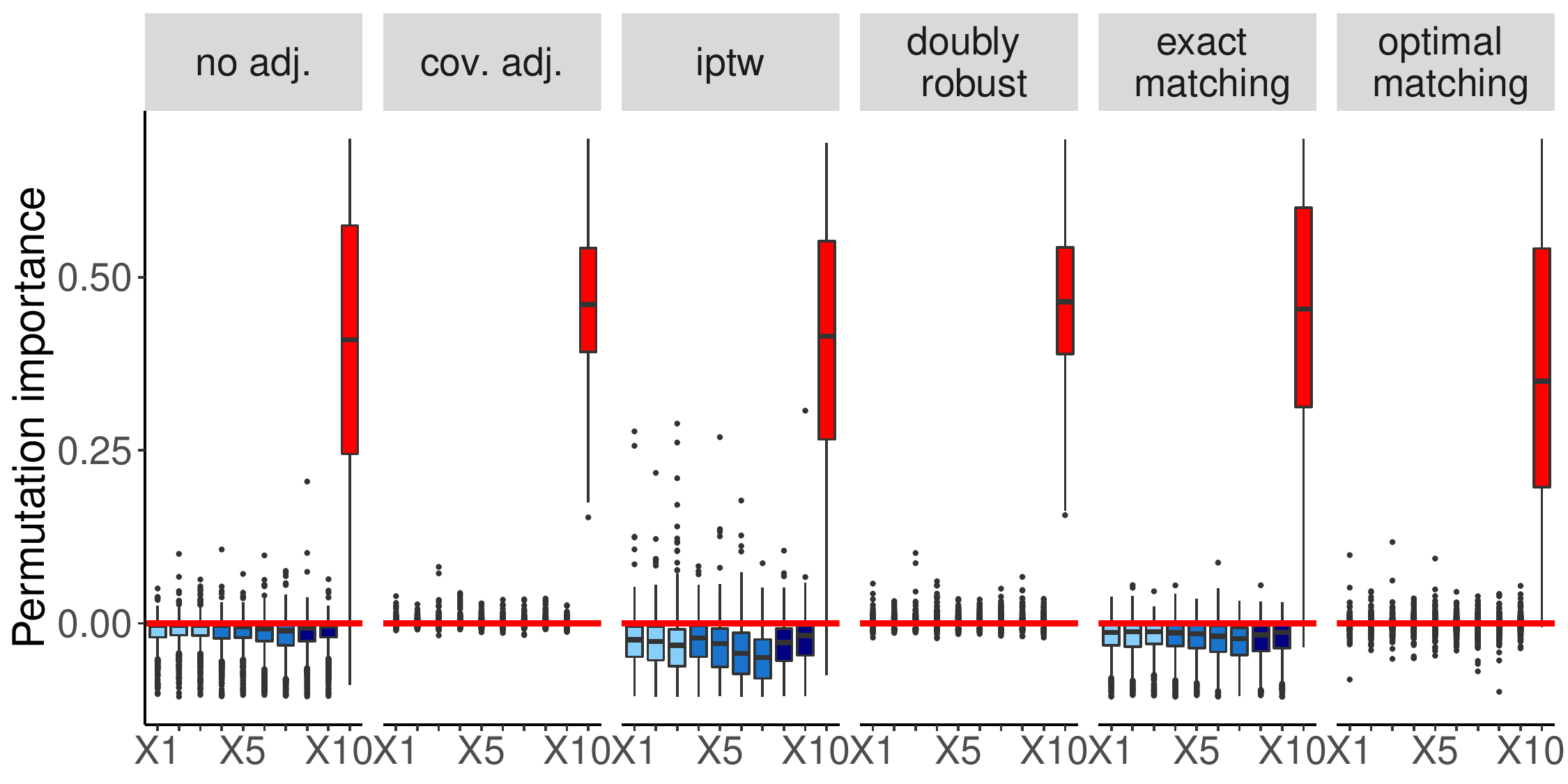}
\end{subfigure}\hfill
\begin{subfigure}{.49\textwidth}
\subcaption{\textbf{Scenario C.2}: $X_{10}$ has a quantitative predictive effect only \newline}
\centering
\includegraphics[height=3.8cm]{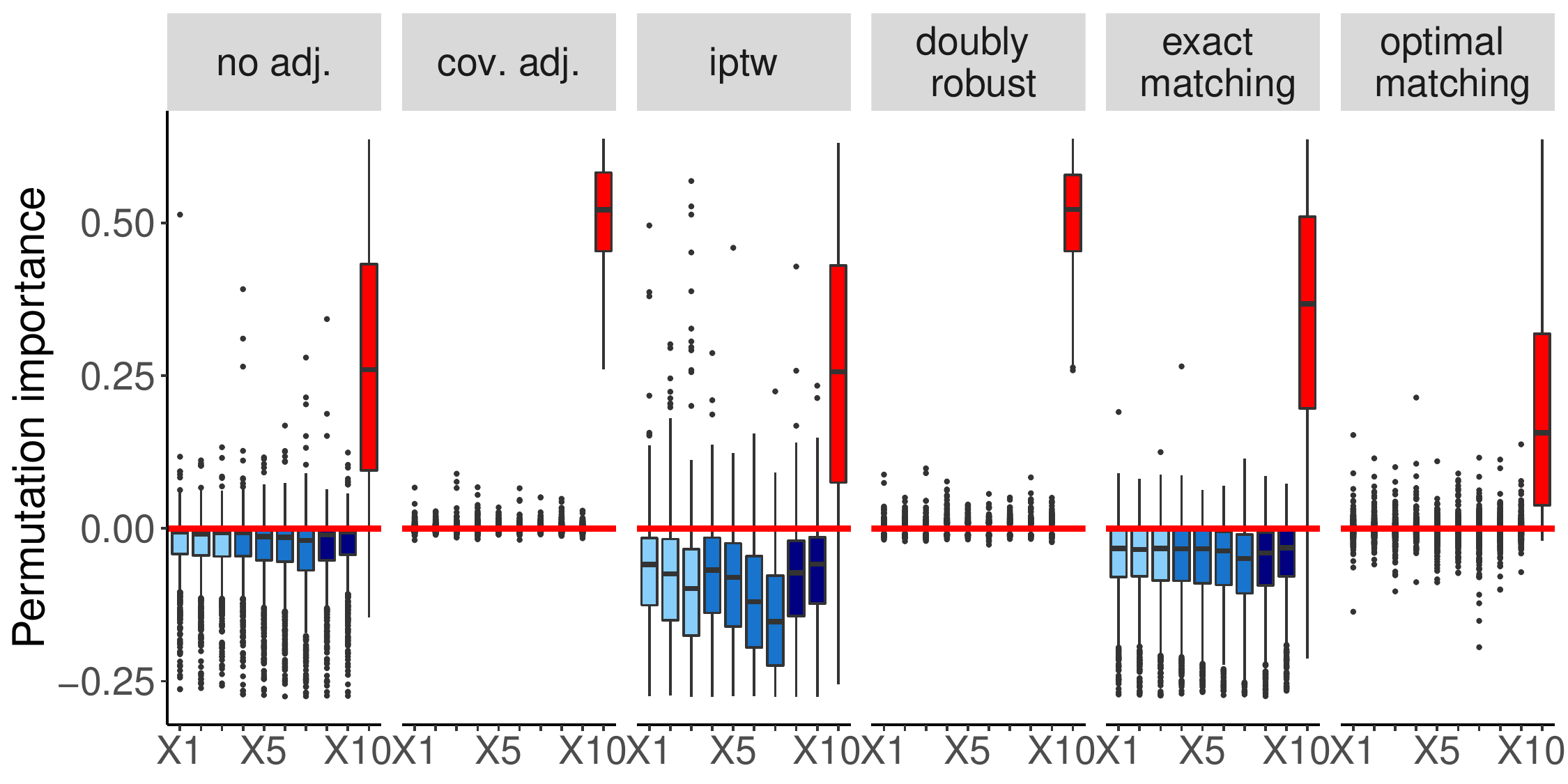}
\end{subfigure}\\[2ex]
\begin{subfigure}{.49\textwidth}
\subcaption{\textbf{Scenario D.1}: $X_{3}$ has a qualitative predictive effect \newline}
\centering
\includegraphics[height=3.8cm]{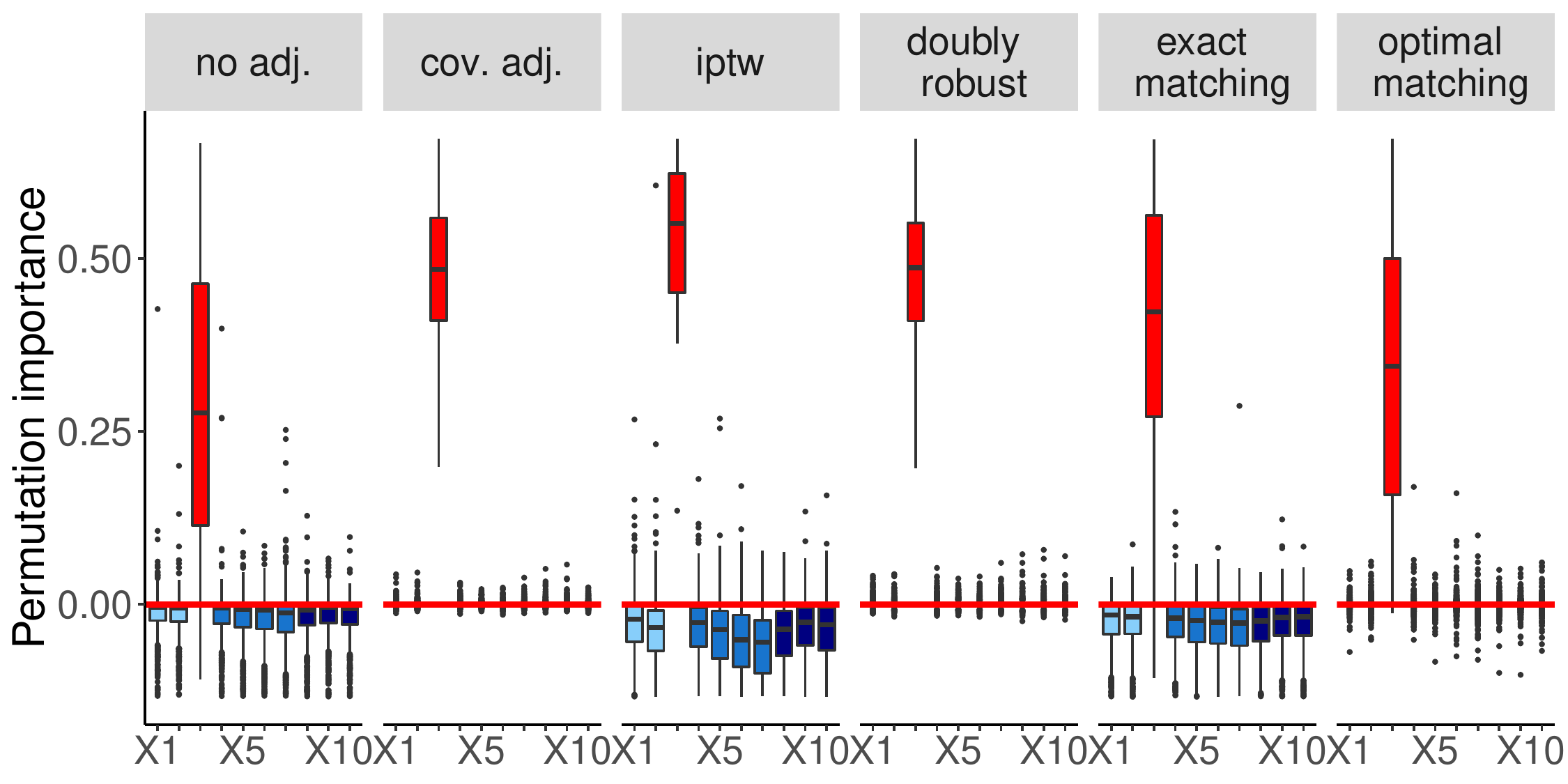}
\end{subfigure}\hfill
\begin{subfigure}{.49\textwidth}
\subcaption{\textbf{Scenario E.1}: $X_{7}$ has both a prognostic and a qualitative predictive effect}
\centering
\includegraphics[height=3.8cm]{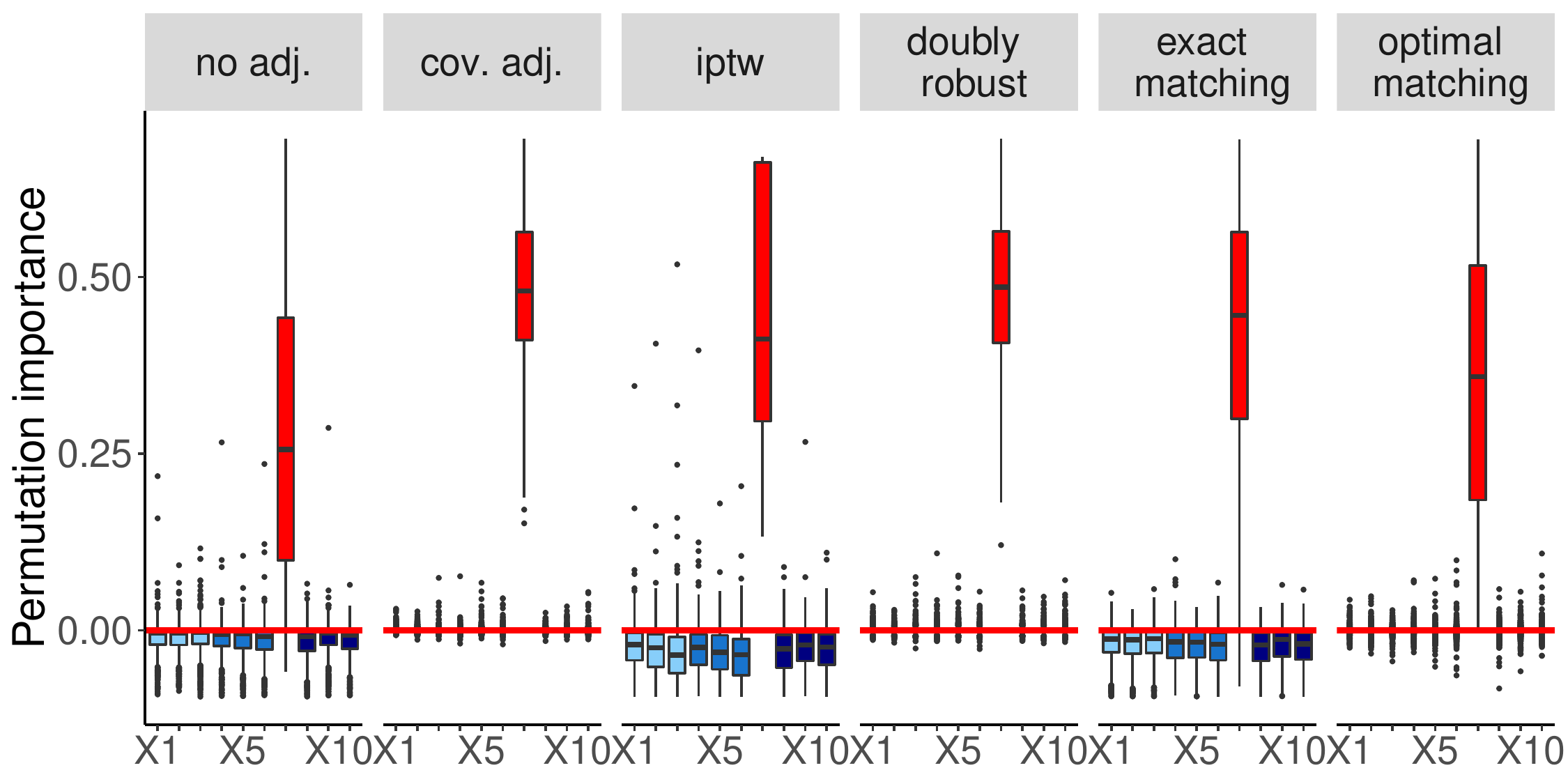}
\end{subfigure}
\end{center}
\end{figure}

\begin{figure}\ContinuedFloat
\begin{center}
\begin{subfigure}[b]{.49\textwidth}
\subcaption{\textbf{Scenario F.1}: $\beta_T=0.25, \ \beta_{10}=0.3, \ \beta_{\text{int}}=-0.5$, $n=2500$}
\centering
\includegraphics[height=3.8cm]{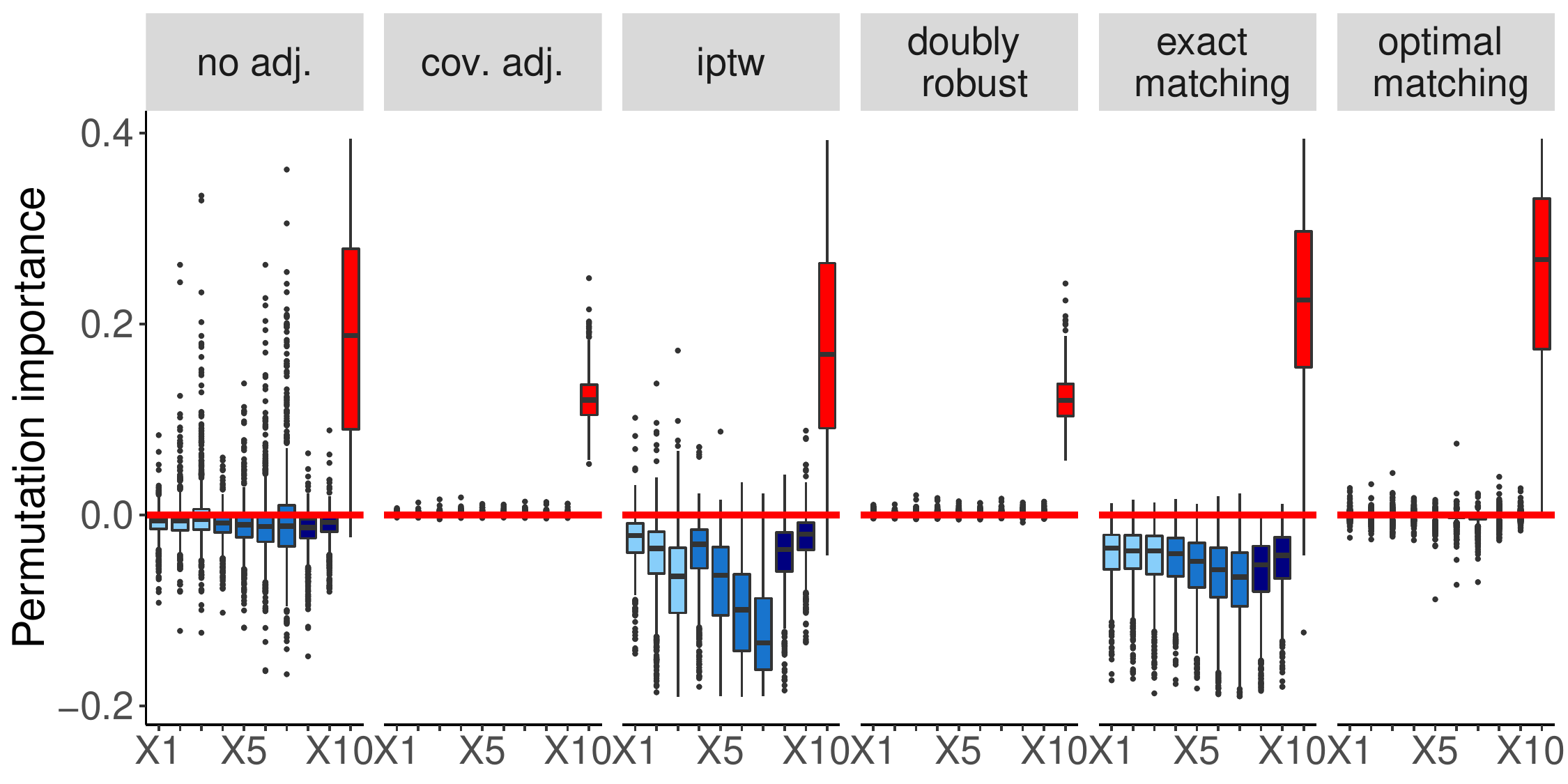}
\end{subfigure}\hfill
\begin{subfigure}[b]{.49\textwidth}
\subcaption{\textbf{Scenario F.2}: $\beta_T=0.25, \ \beta_{10}=0.3, \ \beta_{\text{int}}=-0.25$ \newline }
\centering
\includegraphics[height=3.8cm]{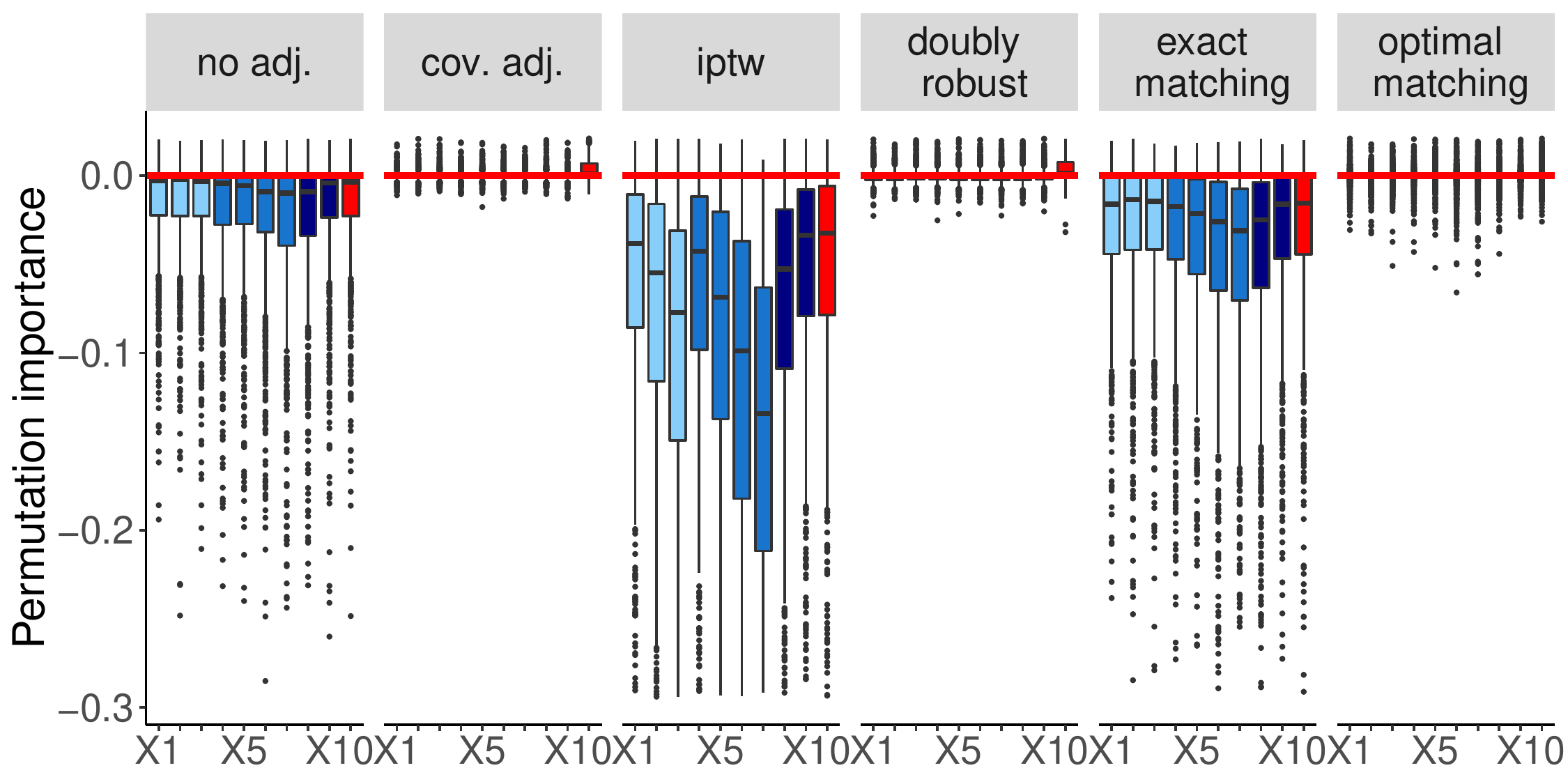}
\end{subfigure}\\[2ex]
\begin{subfigure}[b]{.49\textwidth}
\subcaption{\textbf{Scenario F.2}: $\beta_T=0.25, \ \beta_{10}=0.3, \ \beta_{\text{int}}=-0.25$, $n=2500$ }
\centering
\includegraphics[height=3.8cm]{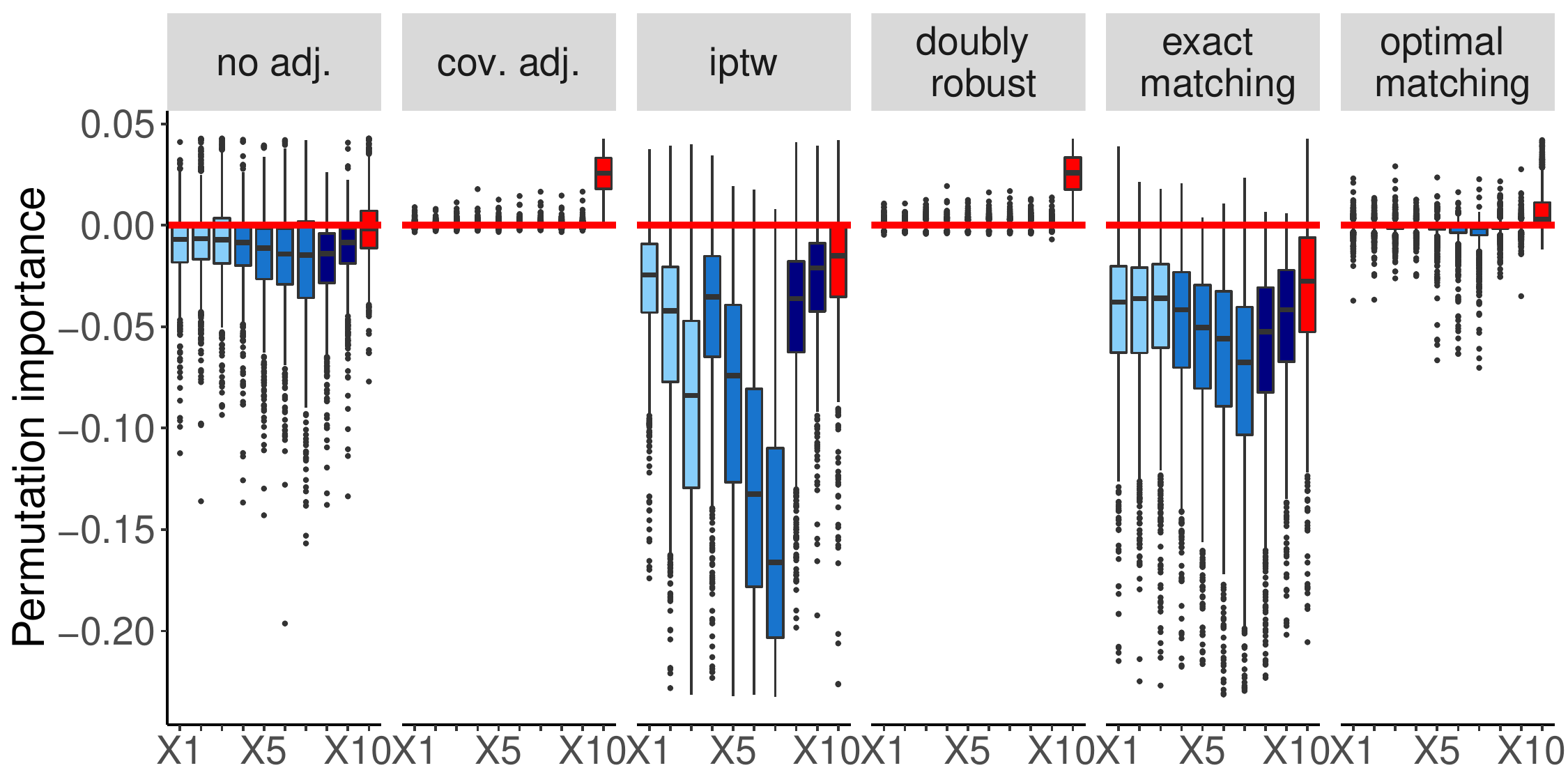}
\end{subfigure}\hfill
\begin{subfigure}[b]{.49\textwidth}
\subcaption{\textbf{Scenario G.1}: Confounding variable $X_7$ and predictive factor $X_{10}$ are positively correlated ($\Corr(X_7,X_{10})=0.5$)}
\centering
\includegraphics[height=3.8cm]{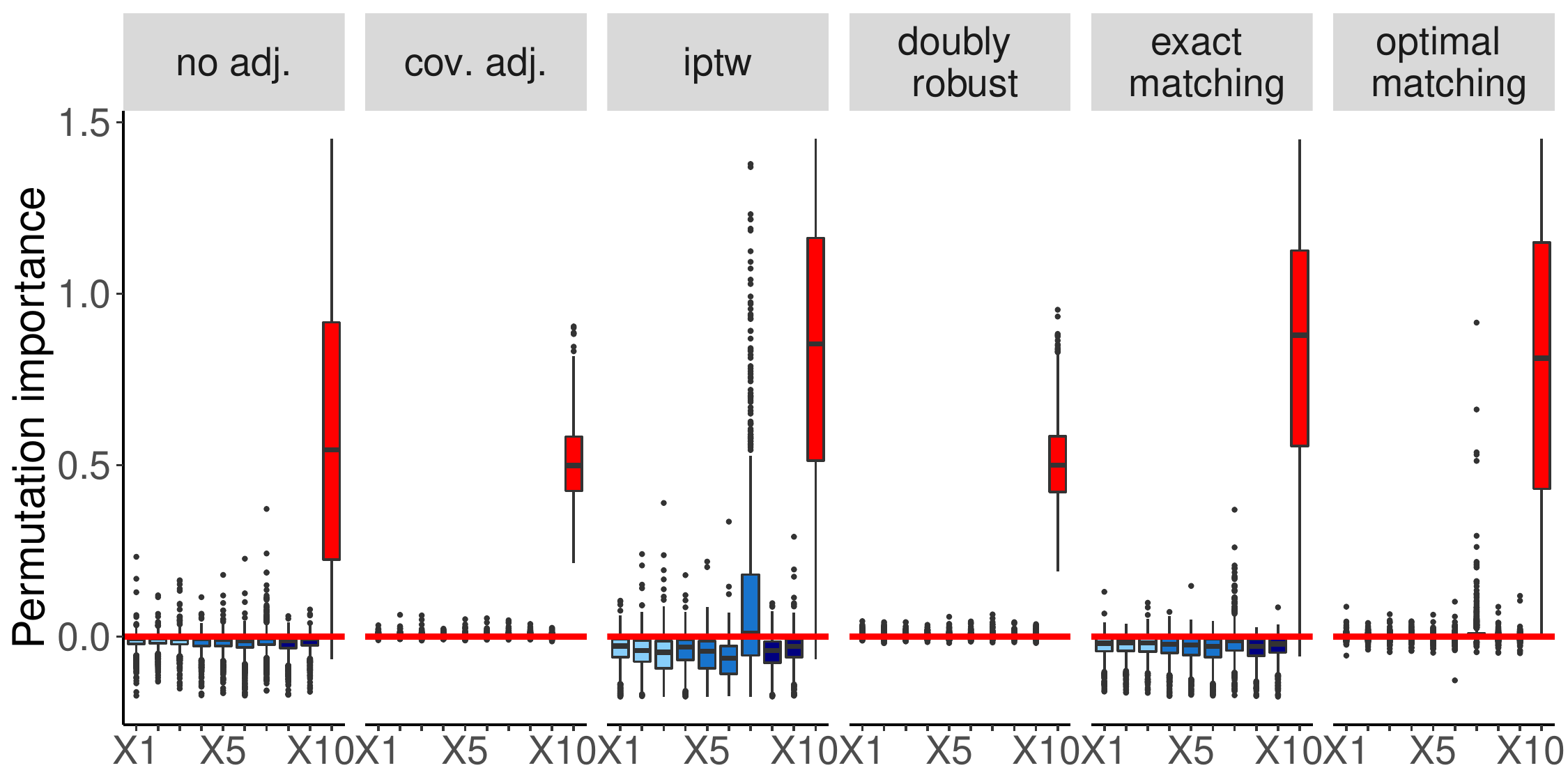}
\end{subfigure}\\[2ex]
\begin{subfigure}[b]{.49\textwidth}
\subcaption{\textbf{Scenario H.1}: $X_9$ and $X_{10}$ both predictive only with predictive effect of $X_9$ greater than that of $X_{10}$ ($\mu=0.5 T + 0.2X_4 + 0.3X_5 + 0.4X_6 + 0.5X_7 + 0.4X_8 + 0 X_9 + 0 X_{10} + 1.2 X_{10}\cdot T -1 X_{10}\cdot T$)}
\centering
\includegraphics[height=3.8cm]{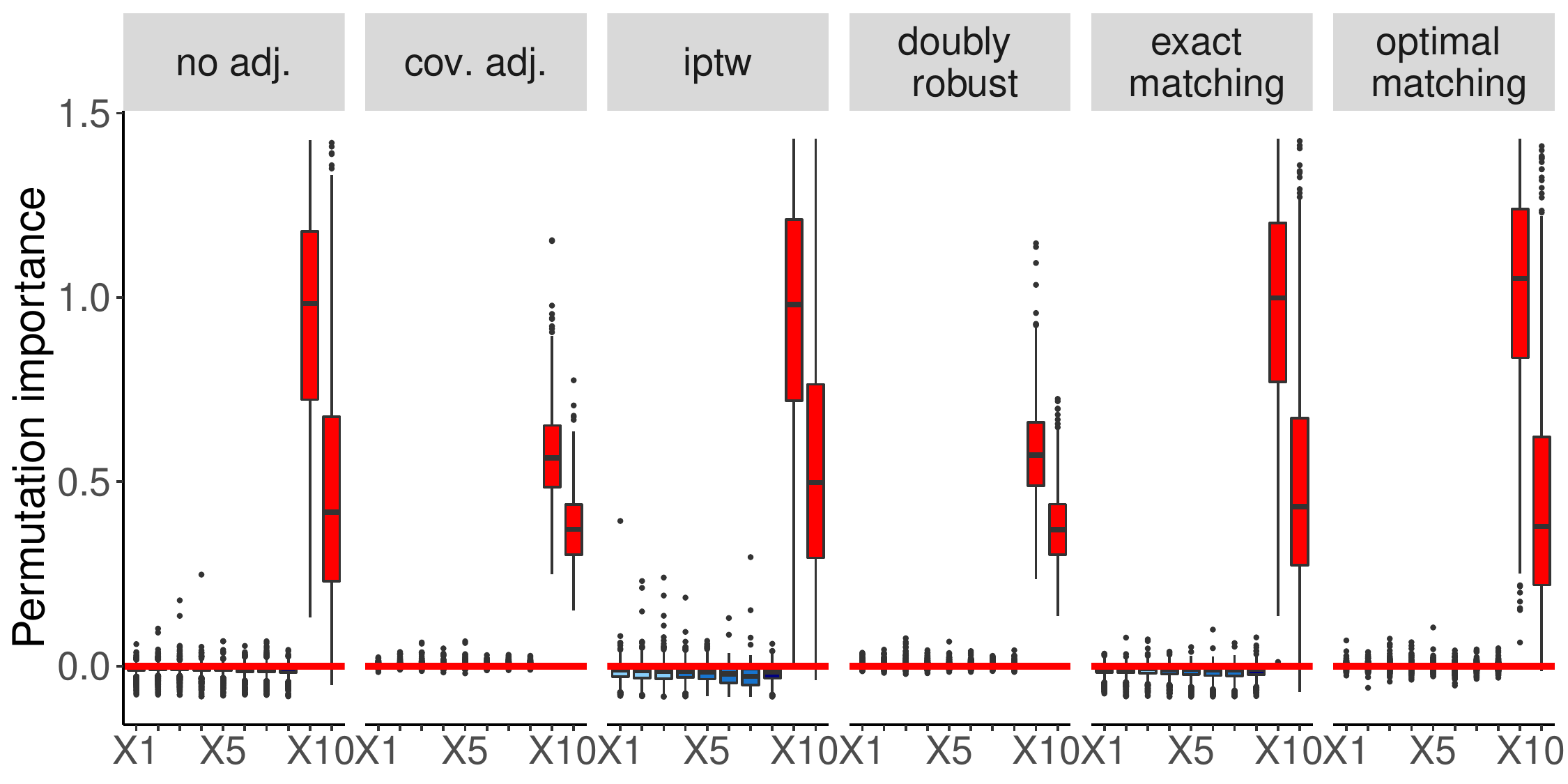}
\end{subfigure}\hfill
\begin{subfigure}[b]{.49\textwidth}
\subcaption{\textbf{Scenario H.2}: $X_9$ and $X_{10}$ both prognostic and predictive with the same predictive effect but the prognostic effect of $X_9$ is greater than that of $X_{10}$ ($\mu=0.5 T + 0.2X_4 + 0.3X_5 + 0.4X_6 + 0.5X_7 + 0.4X_8 + 0.6 X_9 + 0.3 X_{10} -1 X_{10}\cdot T -1 X_{10}\cdot T$)}
\centering
\includegraphics[height=3.8cm]{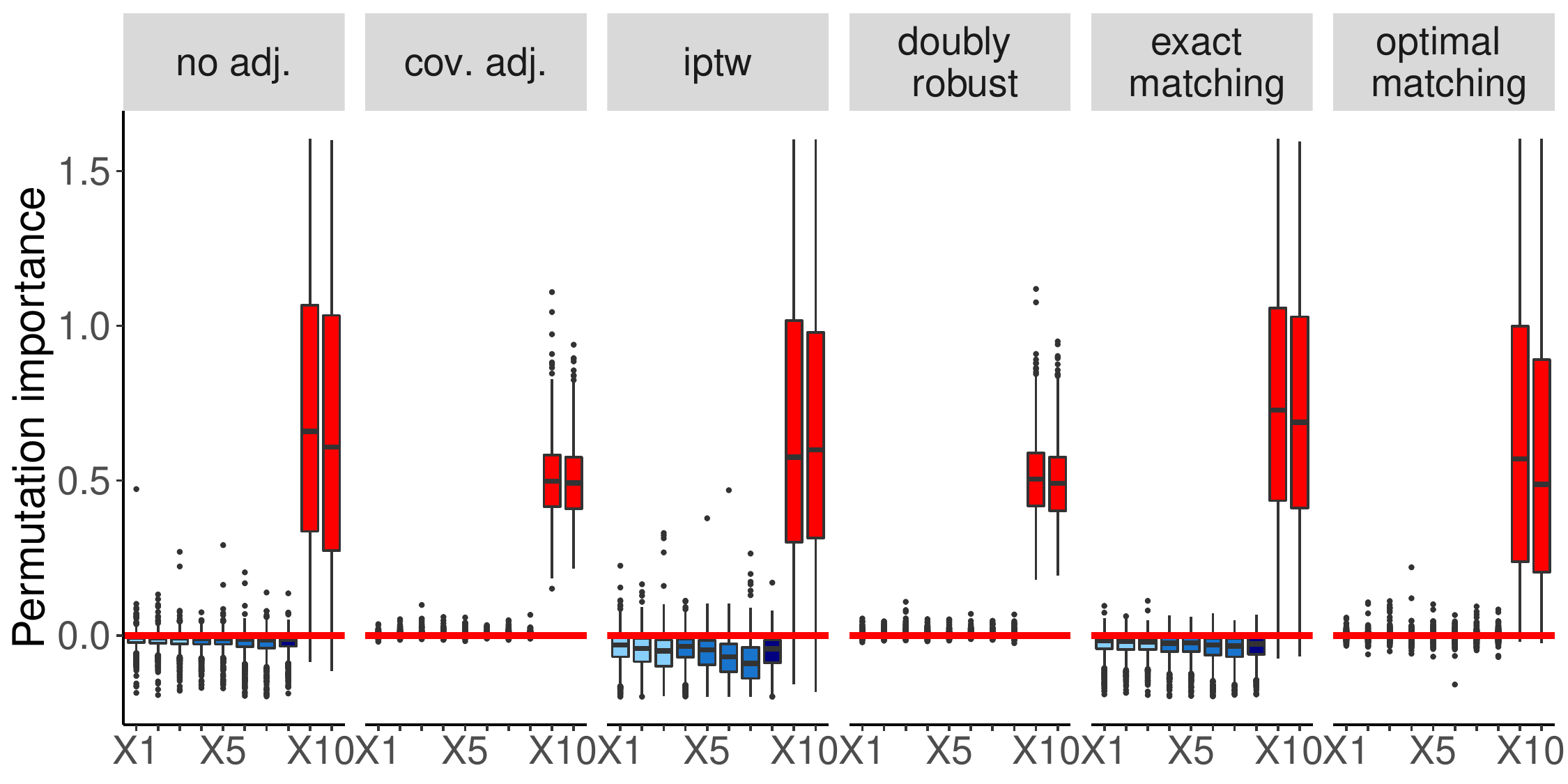}
\end{subfigure}\\[2ex]
\begin{subfigure}[b]{\textwidth}
\subcaption{\textbf{Scenario I}: Higher order predictive pattern with three-way interaction of $X_9, \ X_{10}$ and treatment\\
$Y \sim N(\mu, 0.25)$ with $\mu=0.5 T + 0.2X_4 + 0.3X_5 + 0.4X_6 + 0.5X_7 + 0.4X_8 + 0.2 X_9 -1 X_9 X_{10} T$}
\centering
\includegraphics[height=3.8cm]{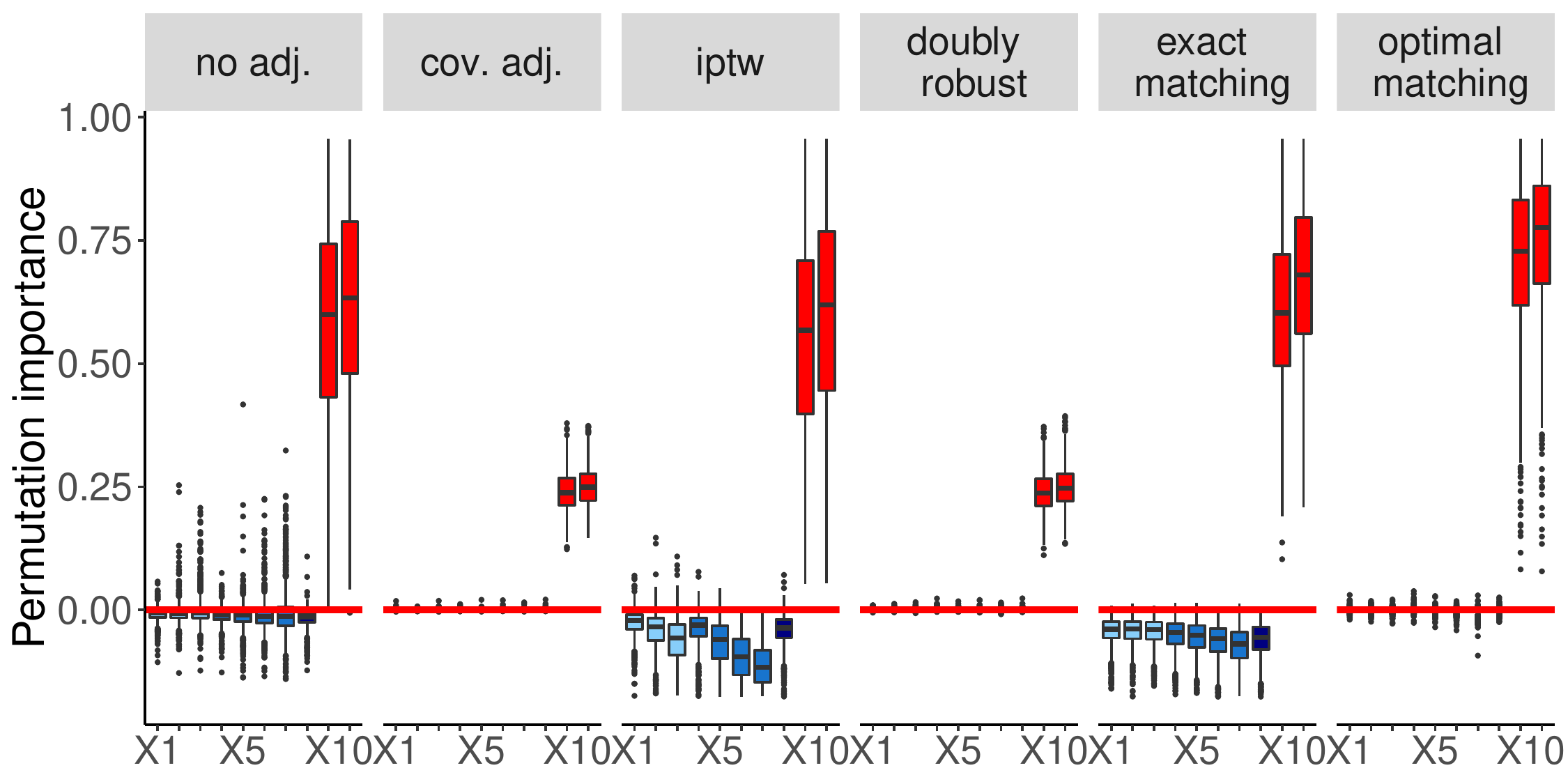}
\end{subfigure}
\caption{Permutation importance for the predMOB in combination with various adjustment methods. Instrumental variables are shown in light blue, true confounders in medium blue and factors only associated with outcome in dark blue. The boxplot for a true predictive factor is highlighted in red.}
\label{Supp:fig:Simresults_permimp}
\end{center}
\end{figure}

\clearpage
\section{Simulation results for the identification of predictive factors using mean minimal depth}\label{supp:simresults_mindepth}
\begin{figure}[htb]
\captionsetup[subfigure]{singlelinecheck=off, justification=raggedright}
\begin{center}
\begin{subfigure}{.49\textwidth}
\subcaption{\textbf{Scenario 0}: Null scenario}
\centering
\includegraphics[height=4cm]{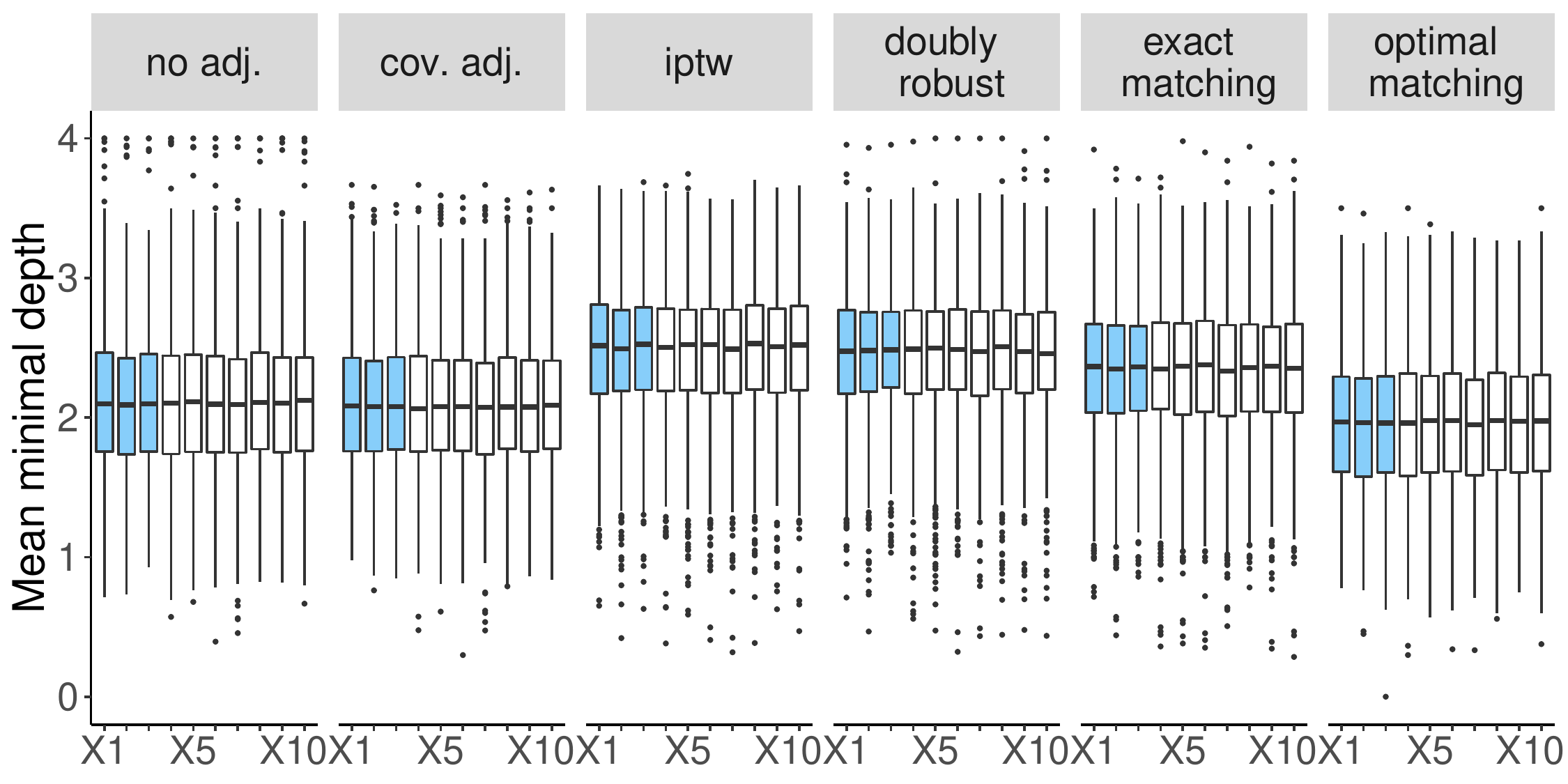}
\end{subfigure}\hfill
\begin{subfigure}{.49\textwidth}
\subcaption{\textbf{Scenario~A}: Prognostic effects only}
\centering
\includegraphics[height=4cm]{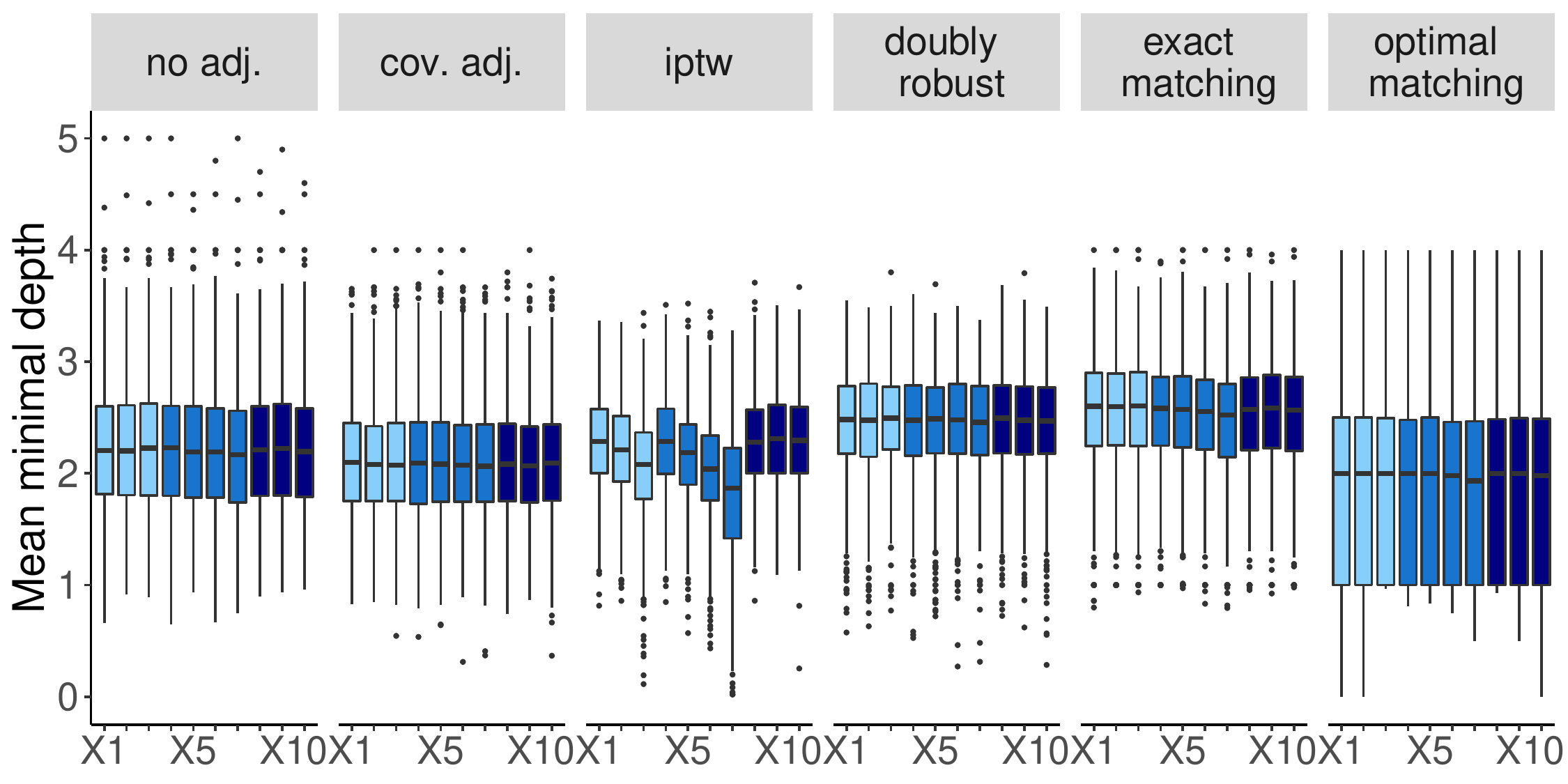}
\end{subfigure}\\[3ex]
\begin{subfigure}{.49\textwidth}
\subcaption{\textbf{Scenario B.1}: $X_{10}$ has both a prognostic and a qualitative predictive effect}
\centering
\includegraphics[height=4cm]{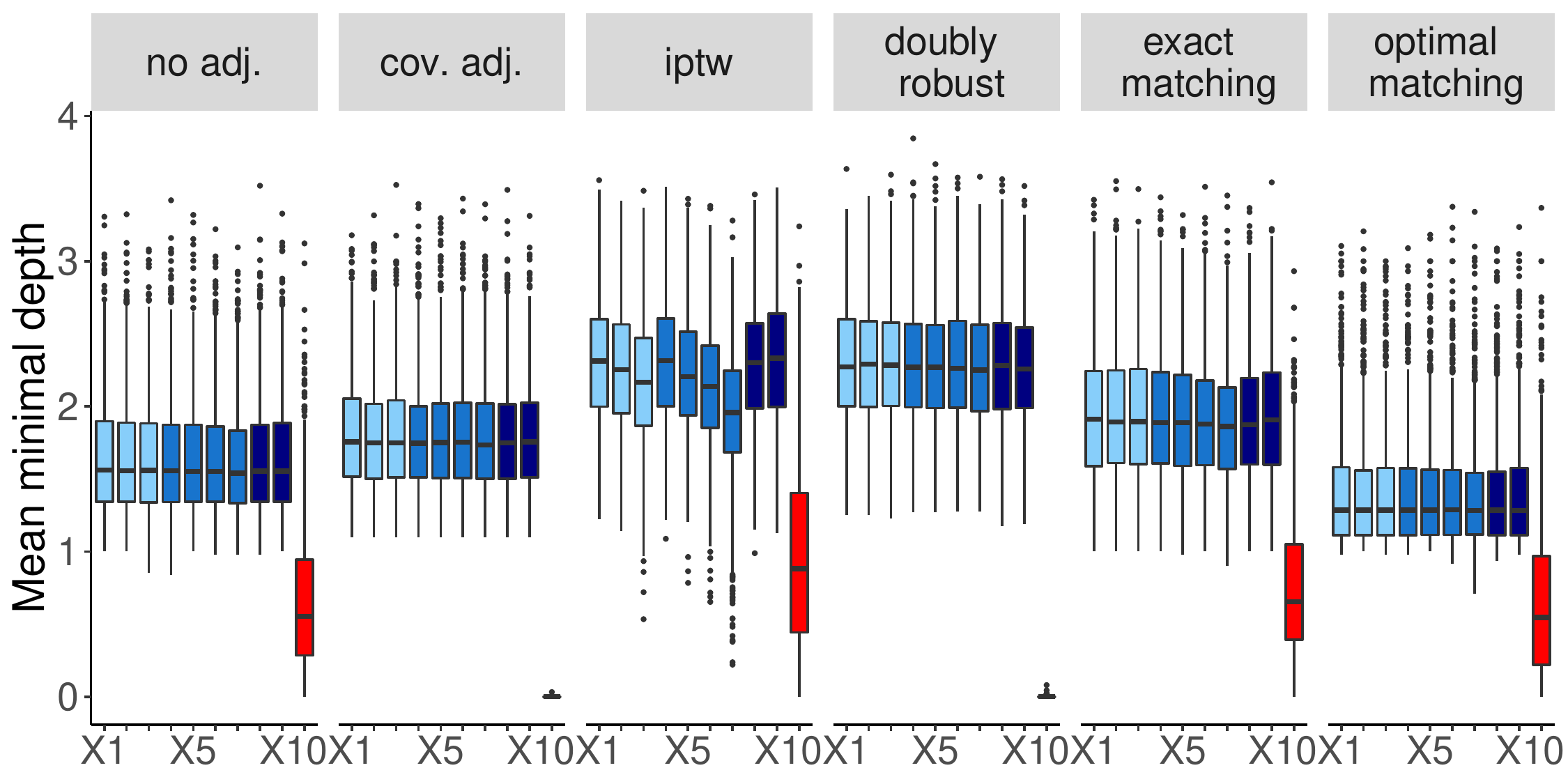}
\end{subfigure}\hfill
\begin{subfigure}{.49\textwidth}
\subcaption{\textbf{Scenario B.2}: $X_{10}$ has both a prognostic and a quantitative predictive effect}
\centering
\includegraphics[height=4cm]{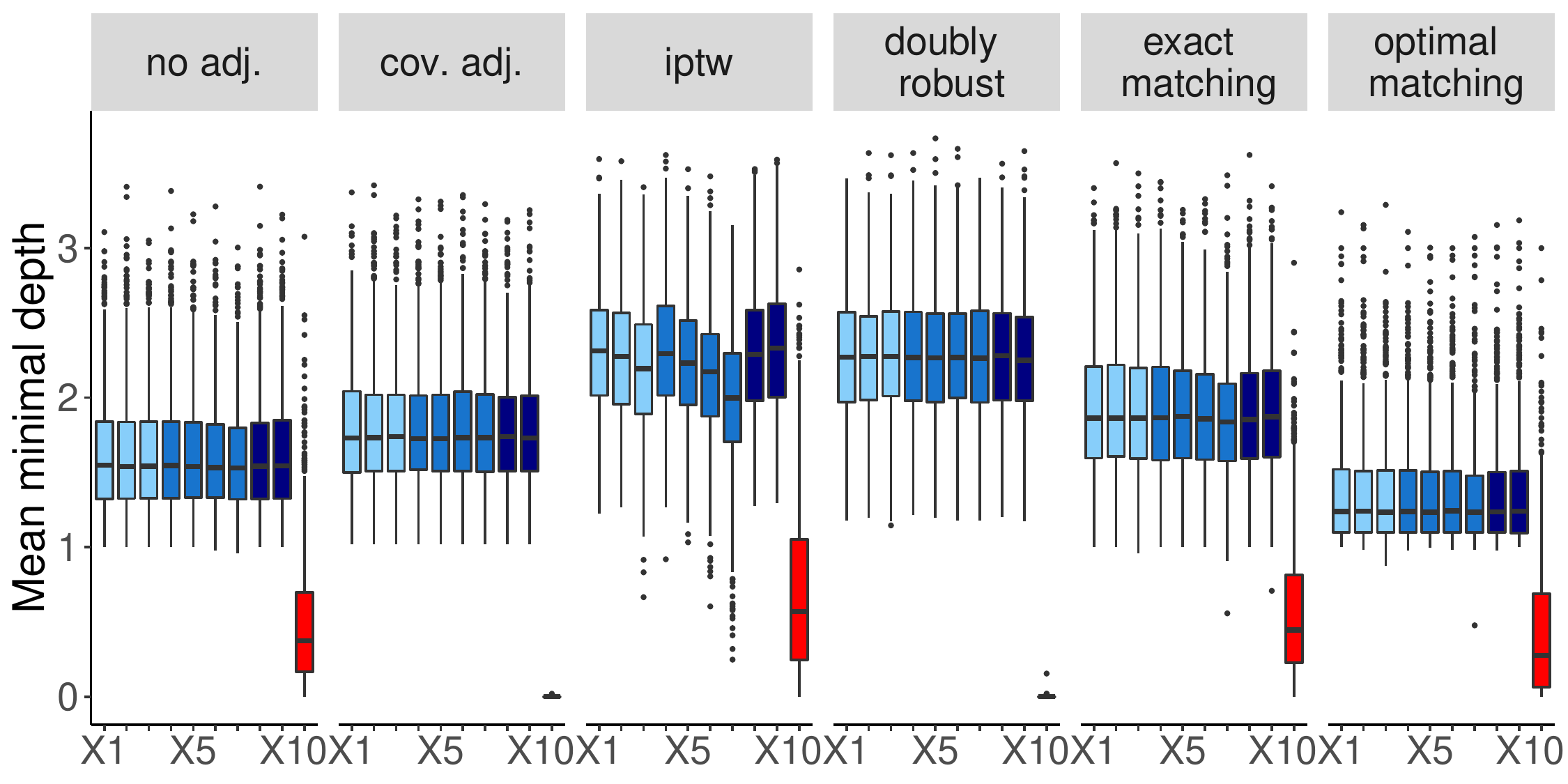}
\end{subfigure}\\[3ex]
\begin{subfigure}{.49\textwidth}
\subcaption{\textbf{Scenario~C.1}: $X_{10}$ has a qualitative predictive effect only}
\centering
\includegraphics[height=4cm]{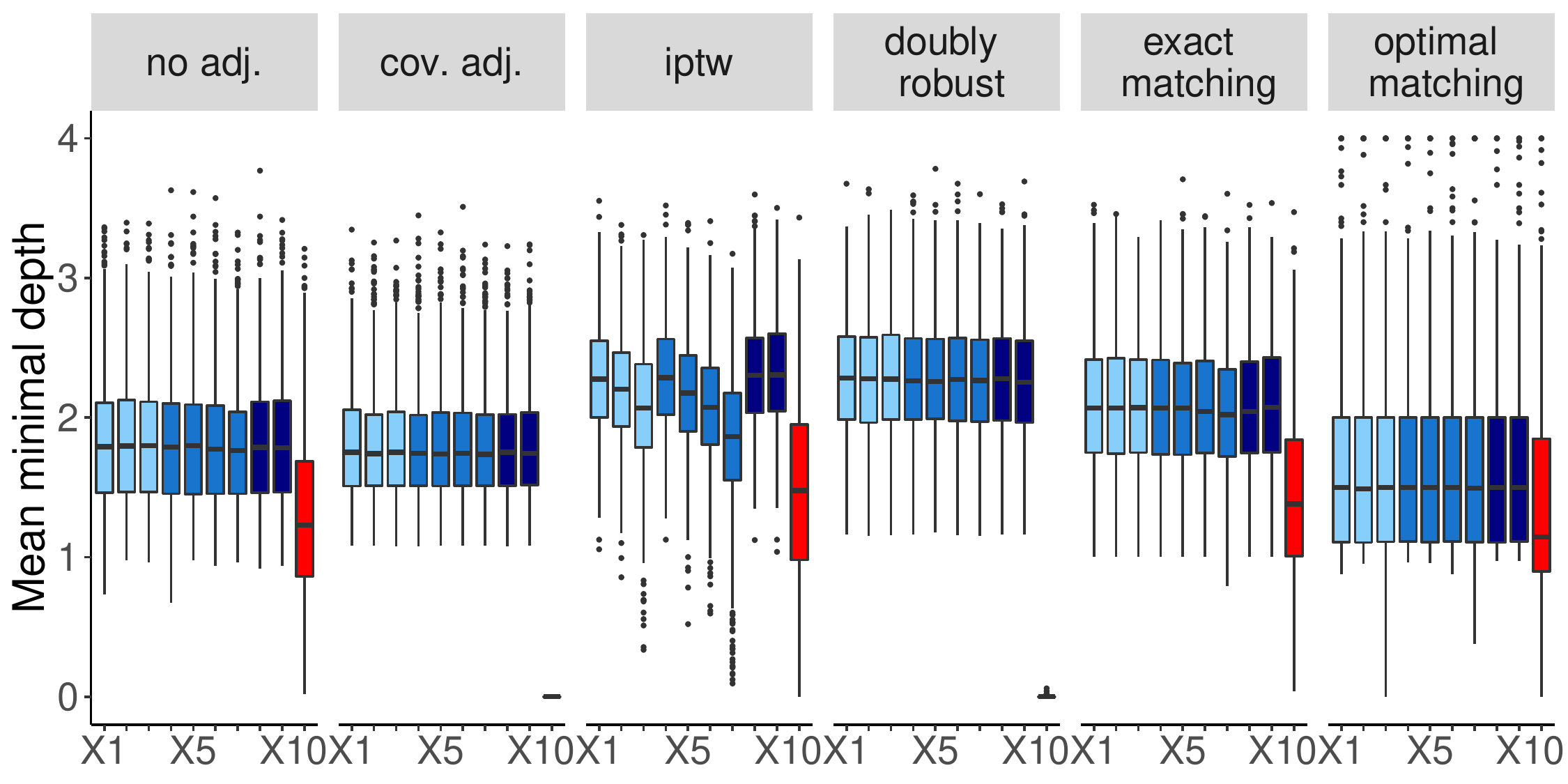}
\end{subfigure}\hfill
\begin{subfigure}{.49\textwidth}
\subcaption{\textbf{Scenario C.2}: $X_{10}$ has a quantitative predictive effect only}
\centering
\includegraphics[height=4cm]{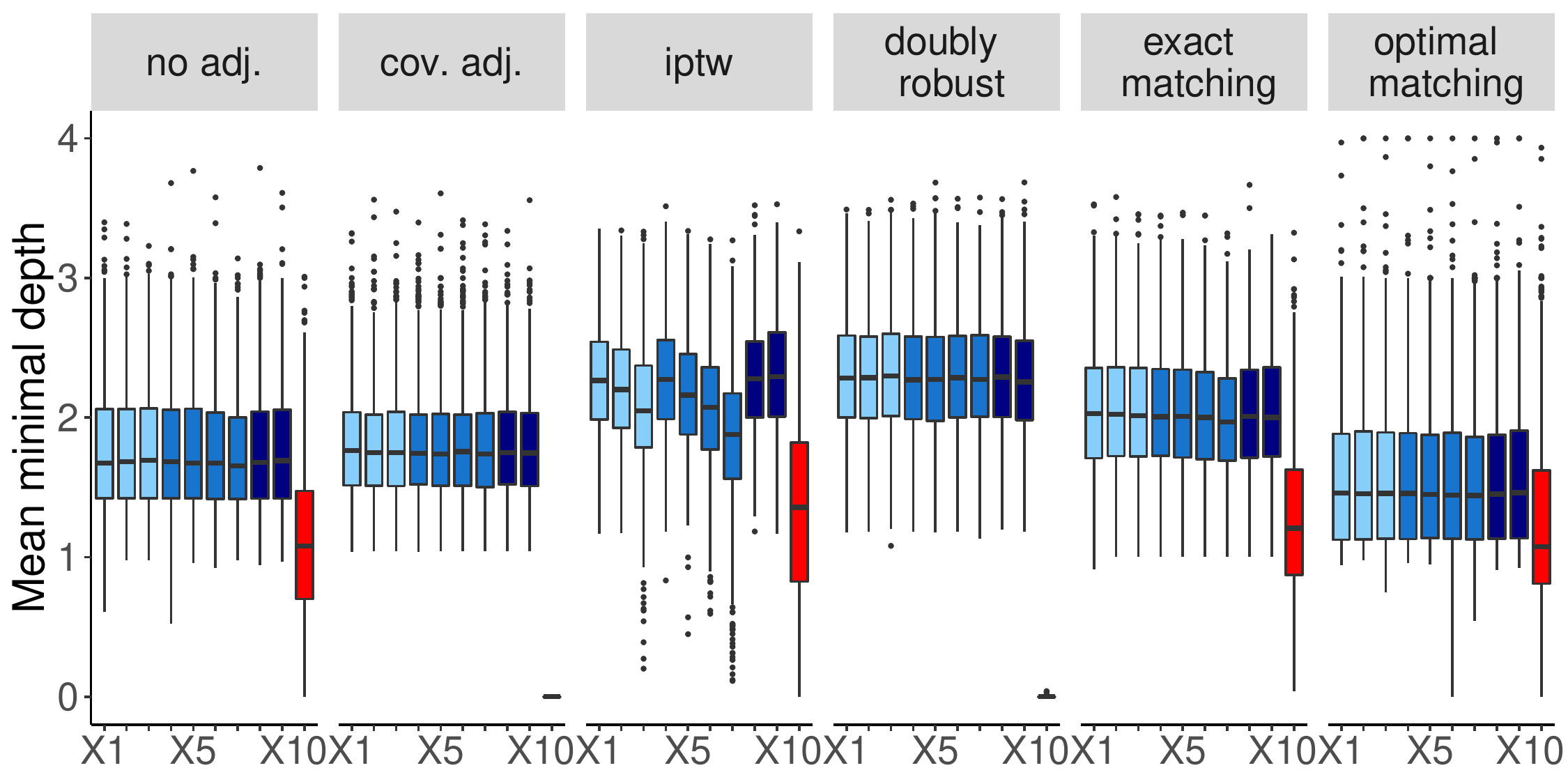}
\end{subfigure}
\end{center}
\end{figure}

\begin{figure}\ContinuedFloat
\captionsetup[subfigure]{singlelinecheck=off, justification=raggedright}
\begin{center}
\begin{subfigure}{.49\textwidth}
\subcaption{\textbf{Scenario D.1}: $X_{3}$ has a qualitative predictive effect}
\centering
\includegraphics[height=4cm]{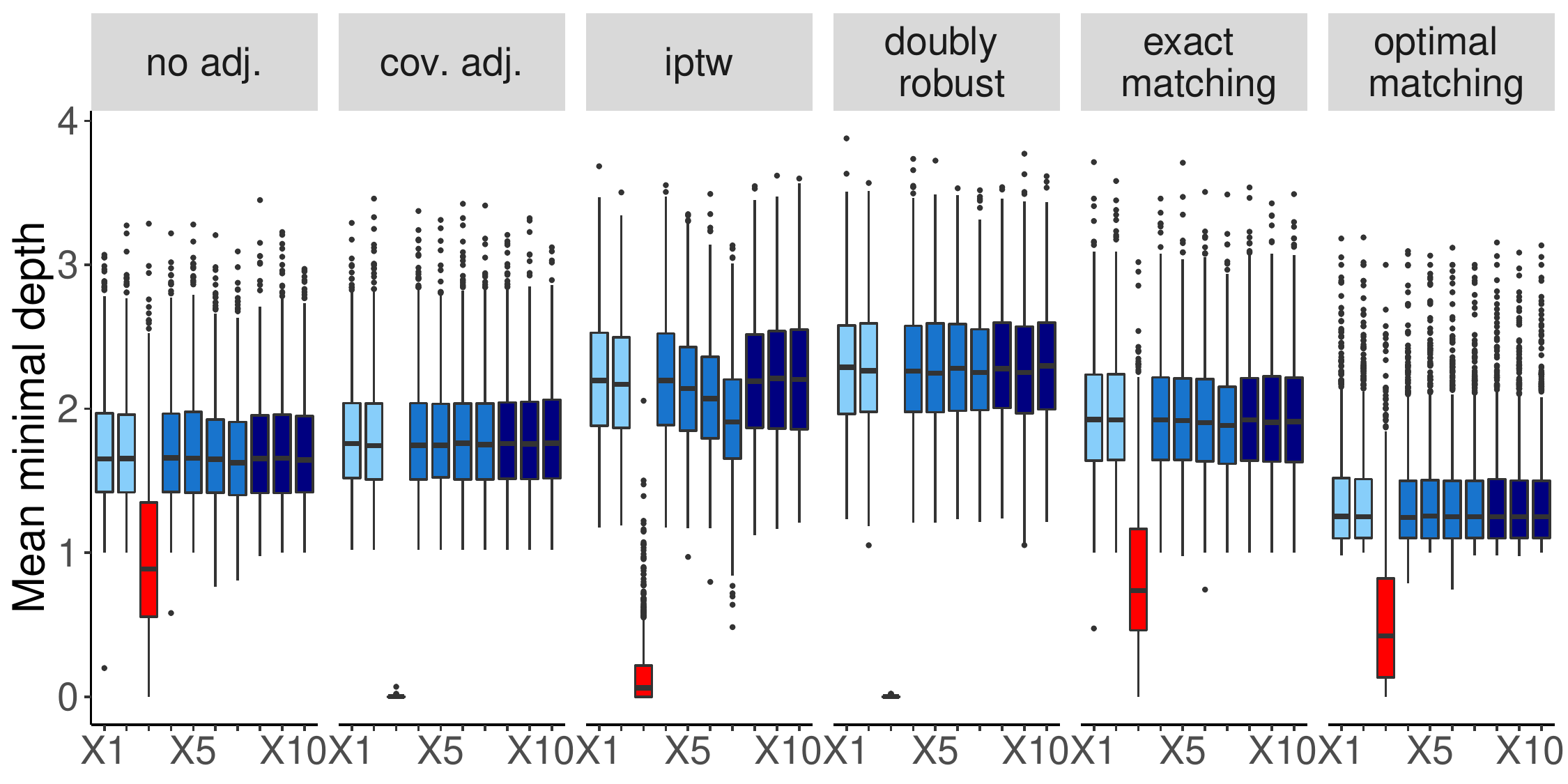}
\end{subfigure}\hfill
\begin{subfigure}{.49\textwidth}
\subcaption{\textbf{Scenario~D.2}: $X_{3}$ has a quantitative predictive effect}
\centering
\includegraphics[height=4cm]{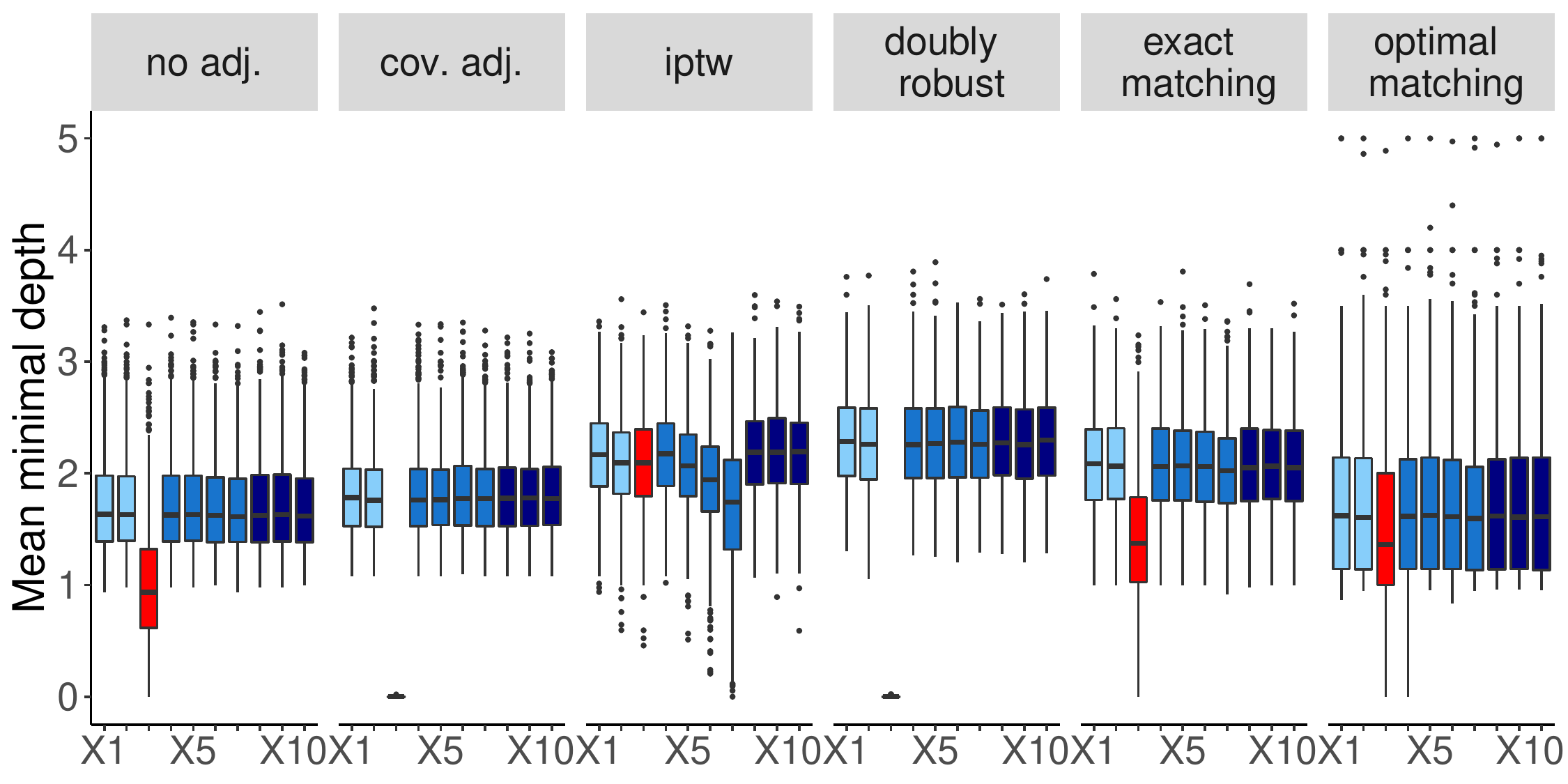}
\end{subfigure}\\[3ex]
\begin{subfigure}{.49\textwidth}
\subcaption{\textbf{Scenario E.1}: $X_{7}$ has both a prognostic and a qualitative predictive effect}
\centering
\includegraphics[height=4cm]{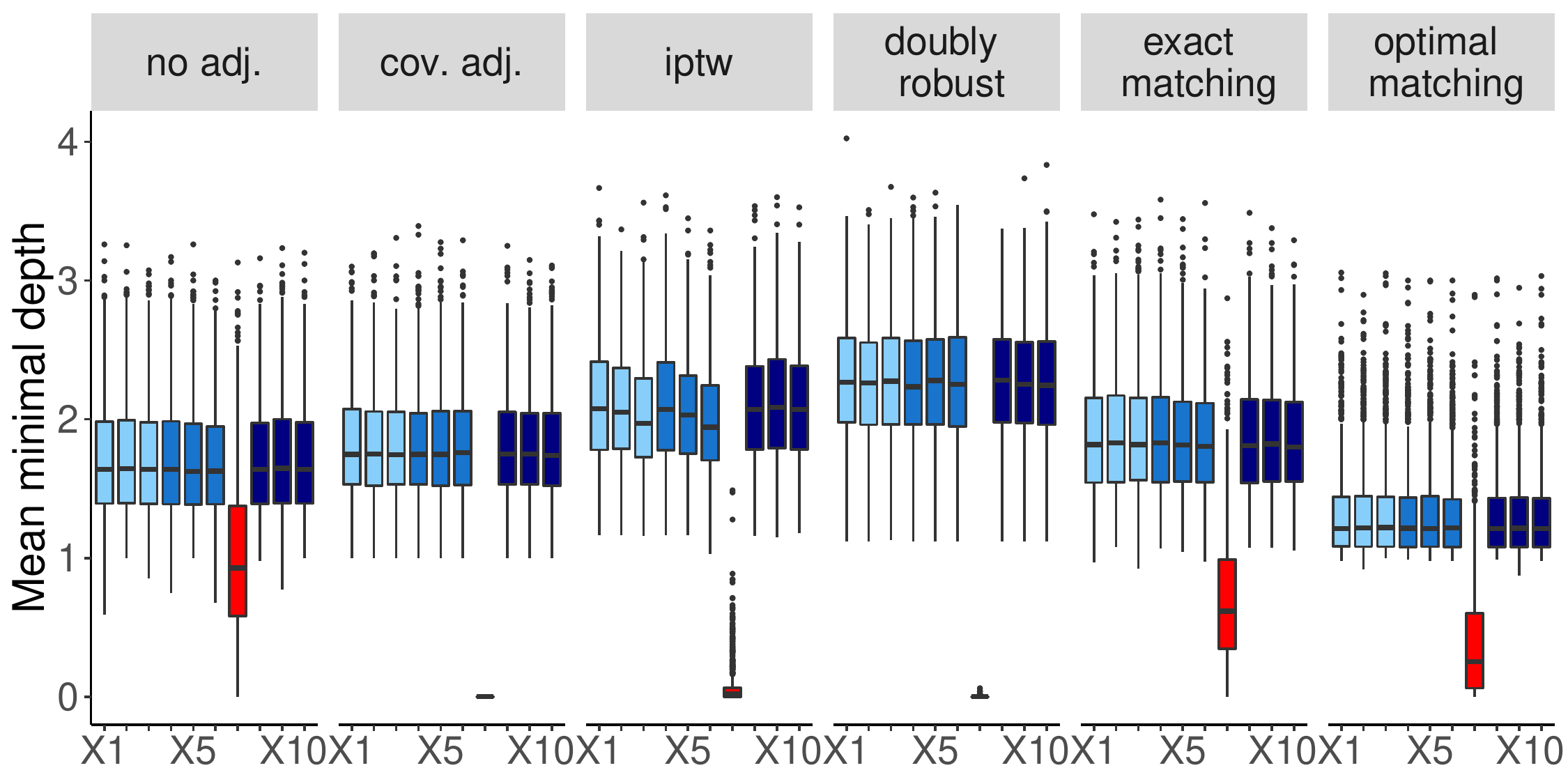}
\end{subfigure}\hfill
\begin{subfigure}{.49\textwidth}
\subcaption{\textbf{Scenario~E.2}: $X_{7}$ has both a prognostic and a quantitative predictive effect}
\centering
\includegraphics[height=4cm]{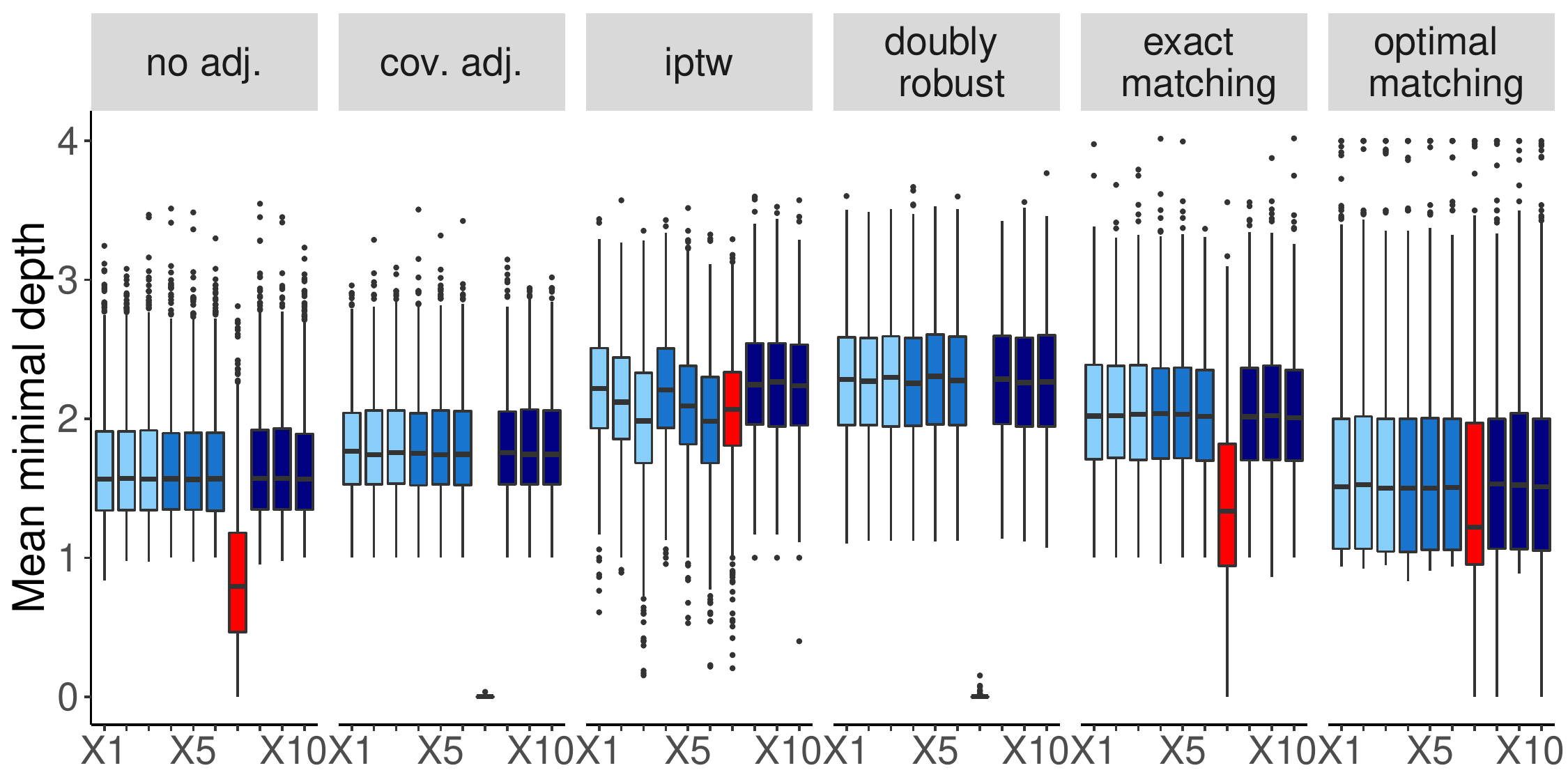}
\end{subfigure}\\[3ex]
\begin{subfigure}{.49\textwidth}
\subcaption{\textbf{Scenario~F.1}: $\beta_T=0.25, \ \beta_{10}=0.3, \ \beta_{\text{int}}=-0.5$}
\centering
\includegraphics[height=4cm]{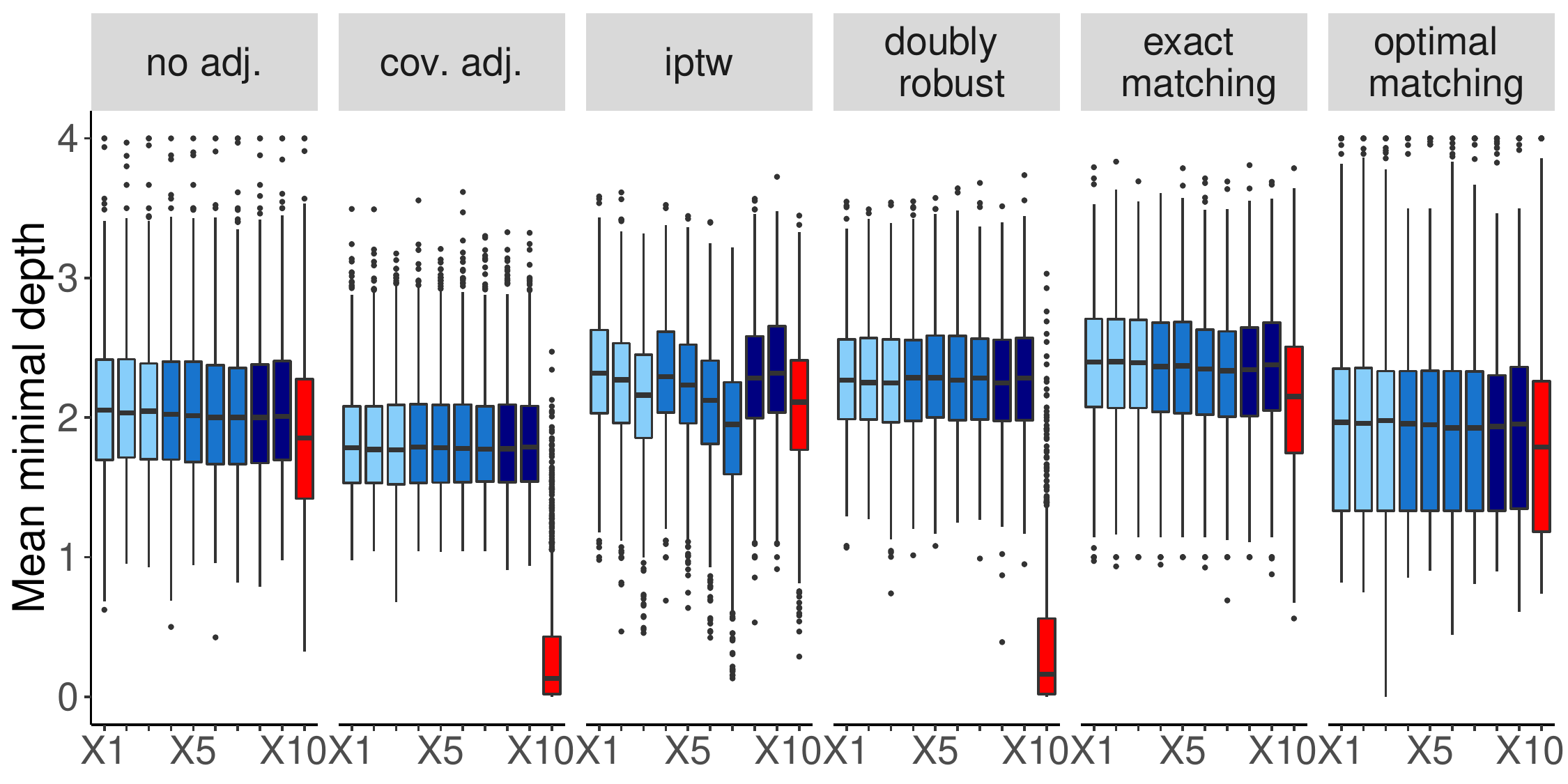}
\end{subfigure}\hfill
\begin{subfigure}{.49\textwidth}
\subcaption{\textbf{Scenario~F.2}: $\beta_T=0.25, \ \beta_{10}=0.3, \ \beta_{\text{int}}=-0.25$}
\centering
\includegraphics[height=4cm]{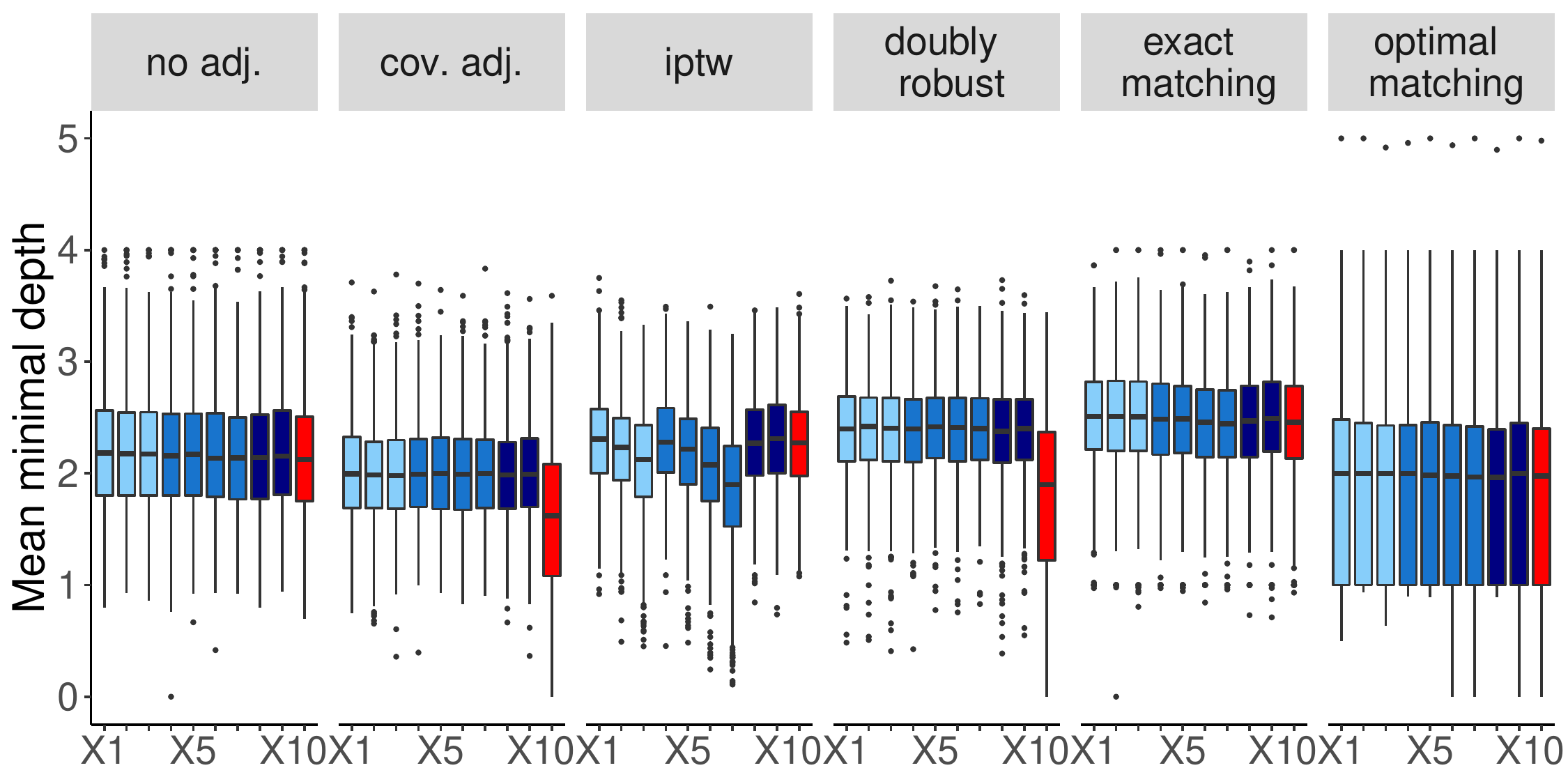}
\end{subfigure}\\[3ex]
\begin{subfigure}{.49\textwidth}
\subcaption{\textbf{Scenario~F.1} with $n=2500$}
\centering
\includegraphics[height=4cm]{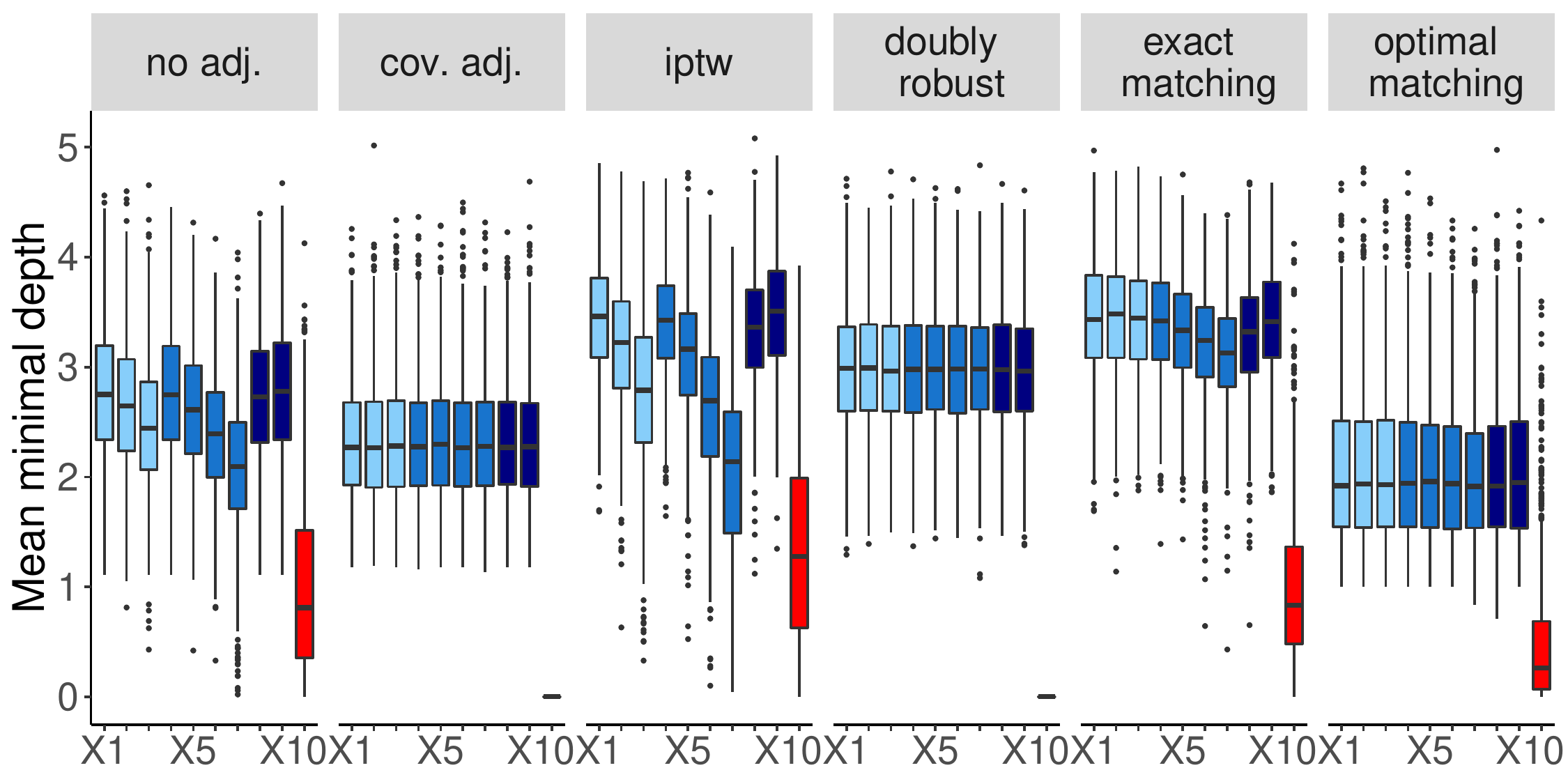}
\end{subfigure}\hfill
\begin{subfigure}{.49\textwidth}
\subcaption{\textbf{Scenario~F.2} with $n=2500$}
\centering
\includegraphics[height=4cm]{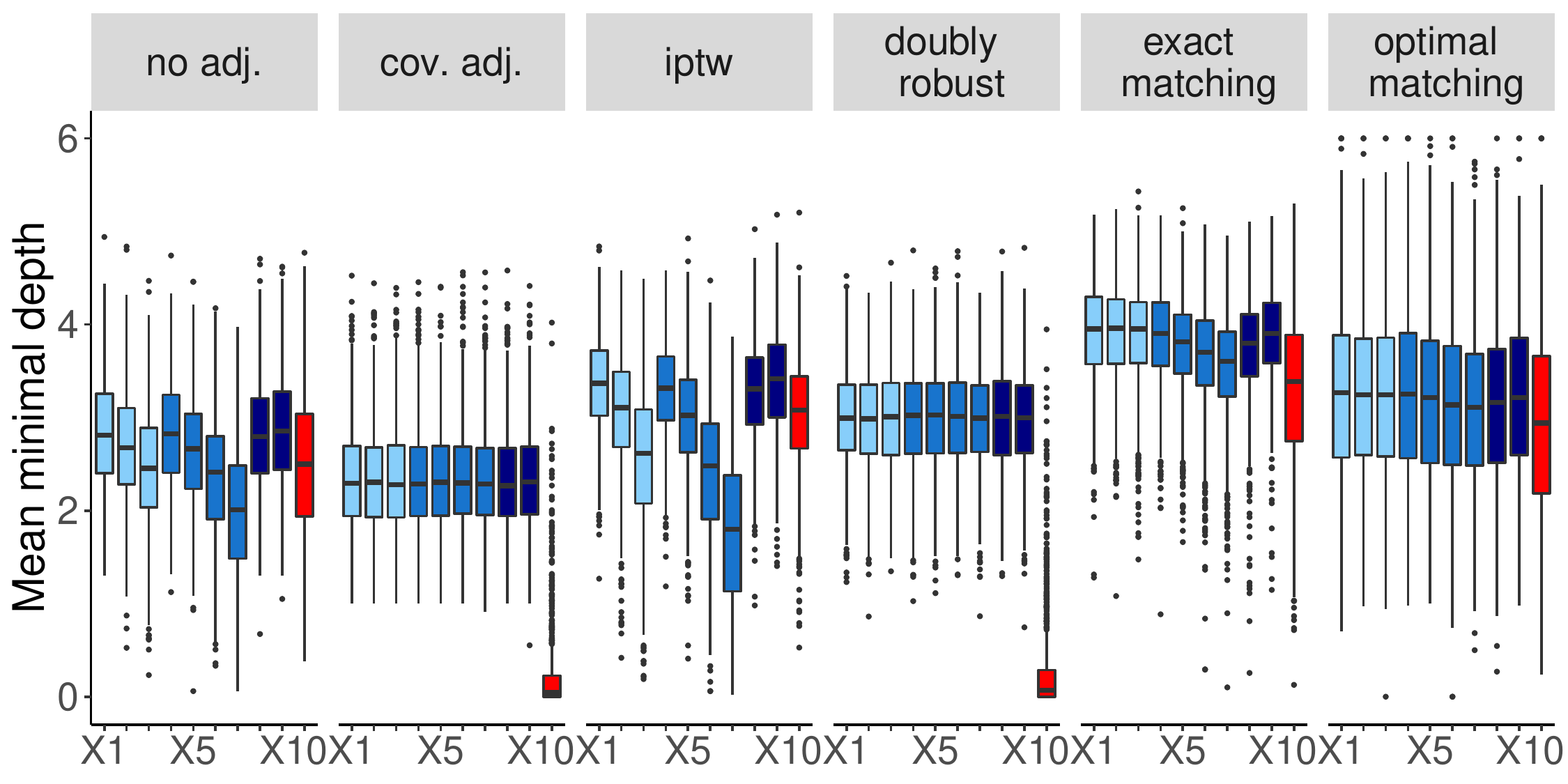}
\end{subfigure}
\end{center}
\end{figure}

\begin{figure}\ContinuedFloat
\begin{center}
\begin{subfigure}{.49\textwidth}
\subcaption{\textbf{Scenario~G.1}: Confounding variable $X_7$ and predictive factor $X_{10}$ are positively correlated ($\Corr(X_7,X_{10})=0.5$)}
\centering
\includegraphics[height=4cm]{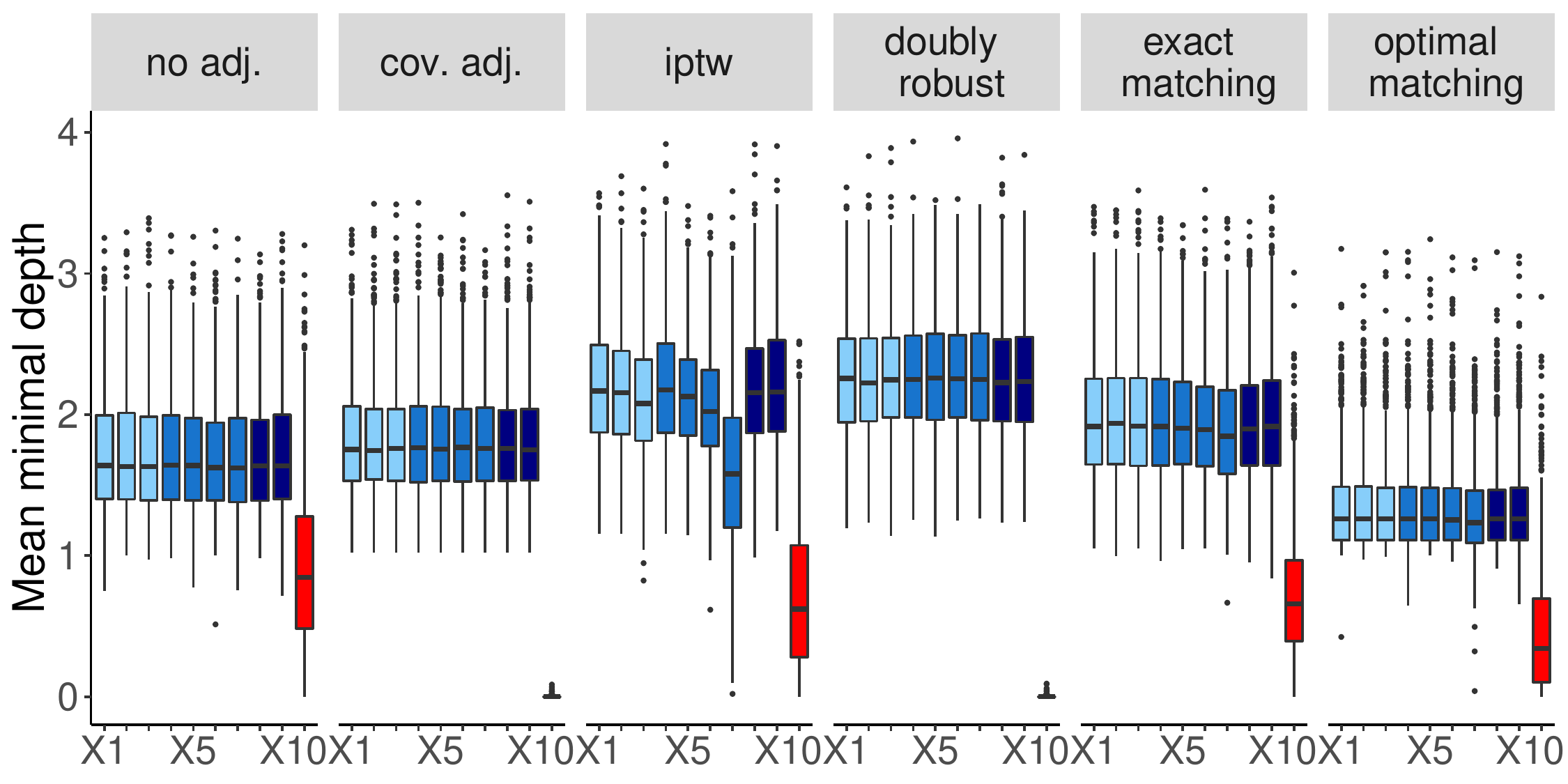}
\end{subfigure}\hfill
\begin{subfigure}{.49\textwidth}
\subcaption{\textbf{Scenario~G.2}: Confounding variable $X_7$ and predictive factor $X_{10}$ are negatively correlated ($\Corr(X_7,X_{10})=-0.7$)}
\centering
\includegraphics[height=4cm]{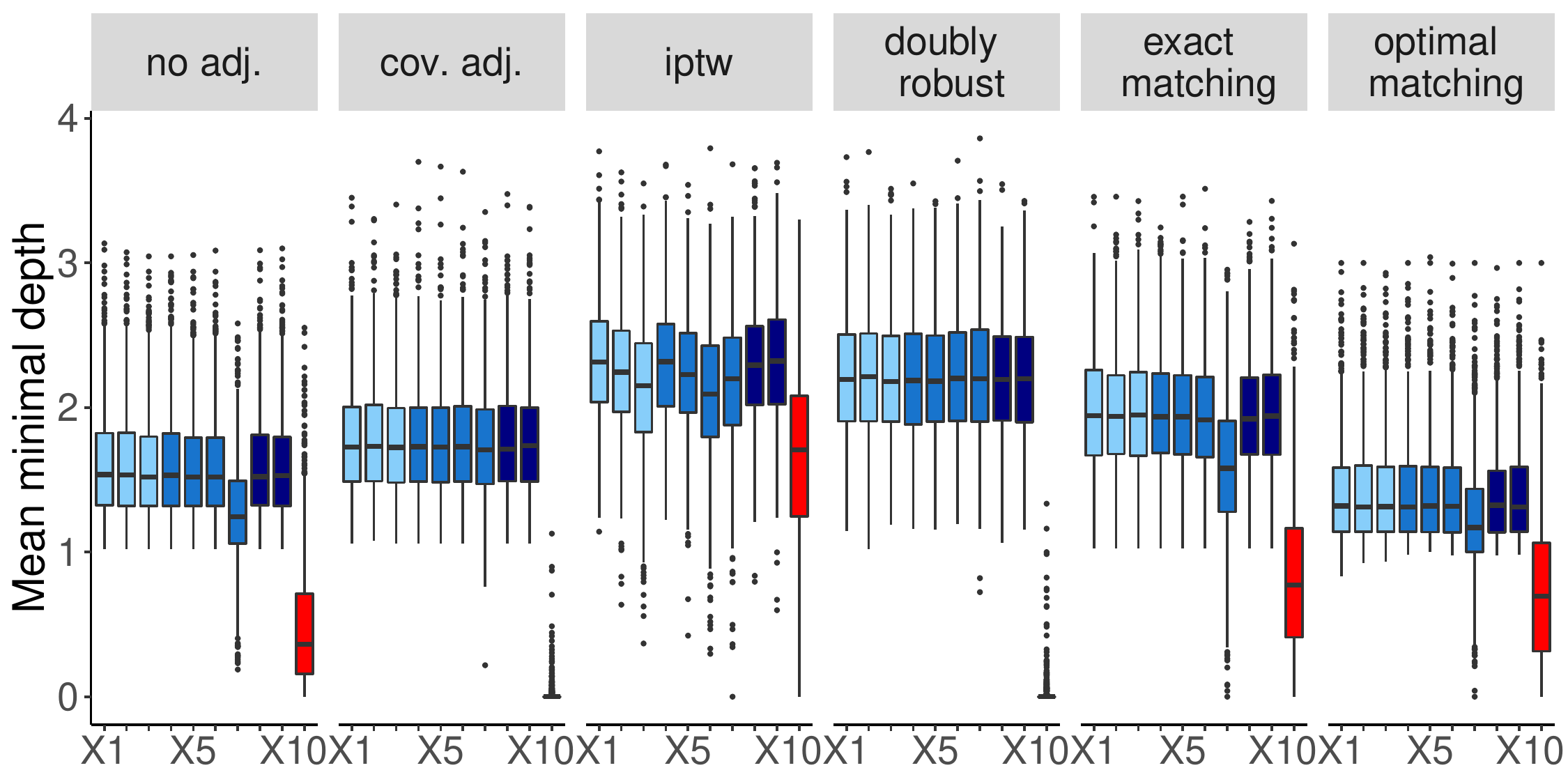}
\end{subfigure}\\[3ex]
\begin{subfigure}{.49\textwidth}
\subcaption{\textbf{Scenario~H.1}: $X_9$ and $X_{10}$ both predictive only with predictive effect of $X_9$ greater than that of $X_{10}$}
\centering
\includegraphics[height=4cm]{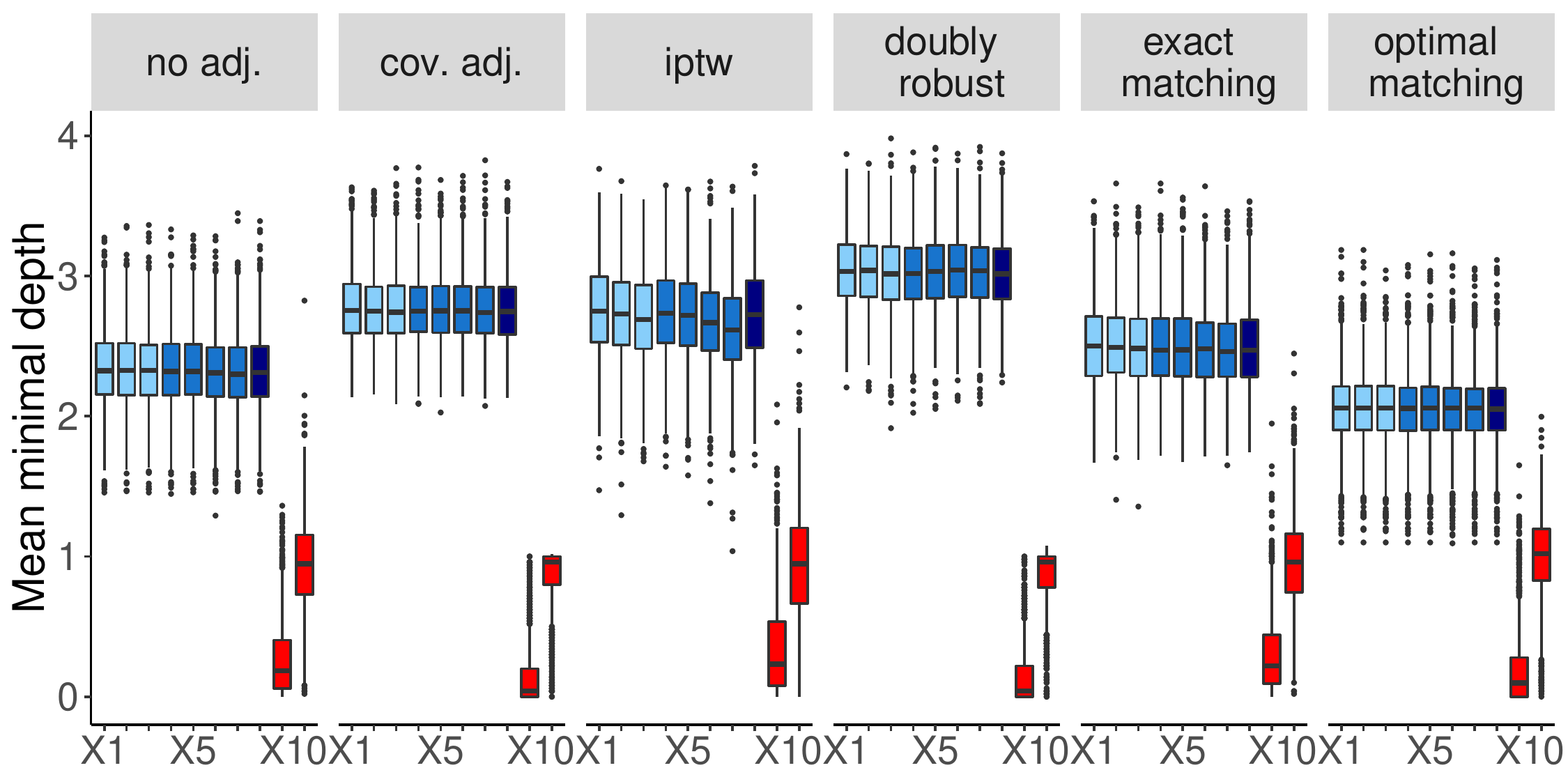}
\end{subfigure}\hfill
\begin{subfigure}{.49\textwidth}
\subcaption{\textbf{Scenario~H.2}:  $X_9$ and $X_{10}$ have the same predictive effect but the prognostic effect of $X_9$ is greater than that of $X_{10}$}
\centering
\includegraphics[height=4cm]{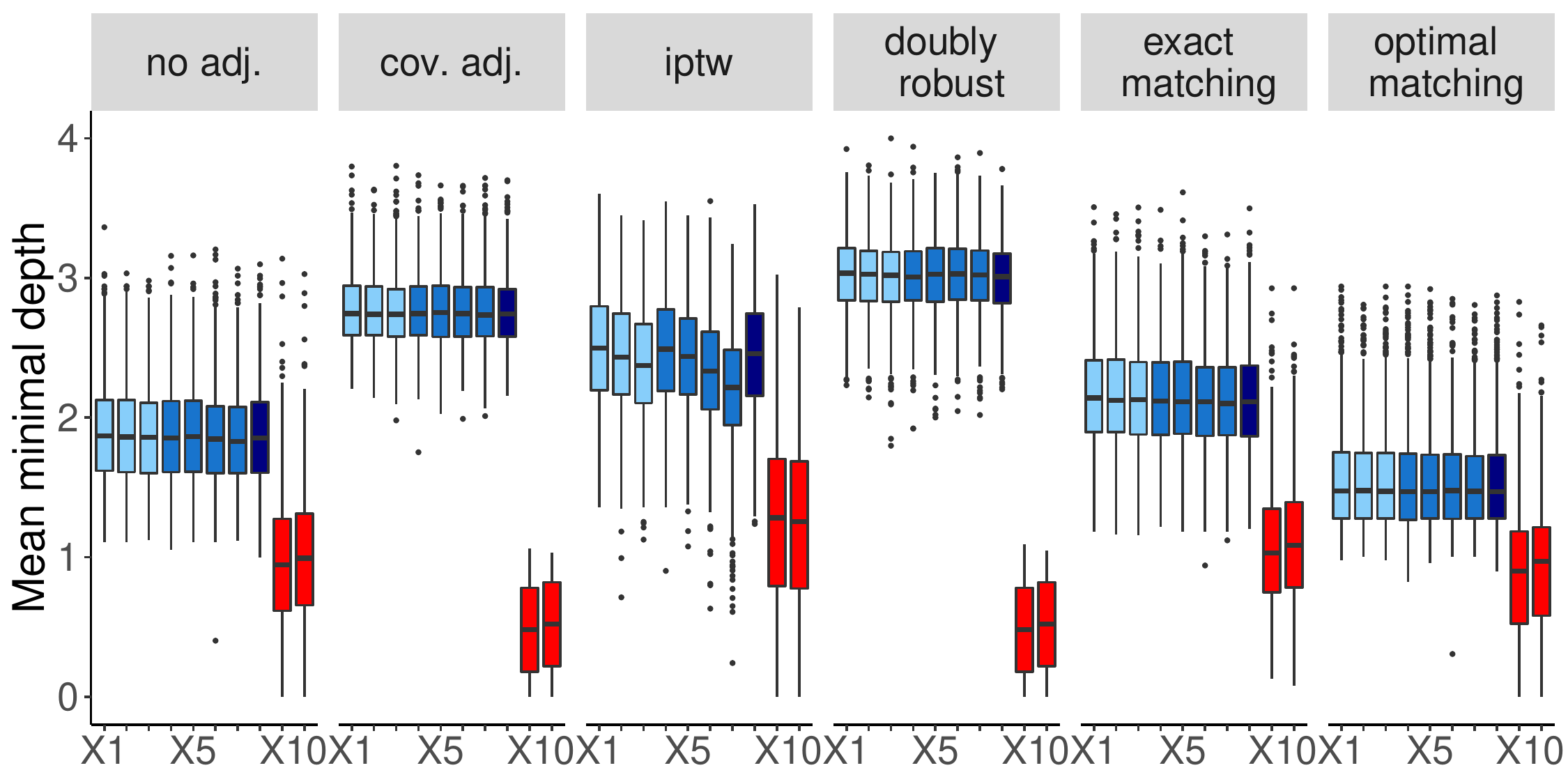}
\end{subfigure}\\[3ex]
\begin{subfigure}{\textwidth}
\subcaption{\textbf{Scenario~I}: Higher-order predictive pattern with three-way interaction of $X_9, \ X_{10}$ and treatment}
\centering
\includegraphics[height=4cm]{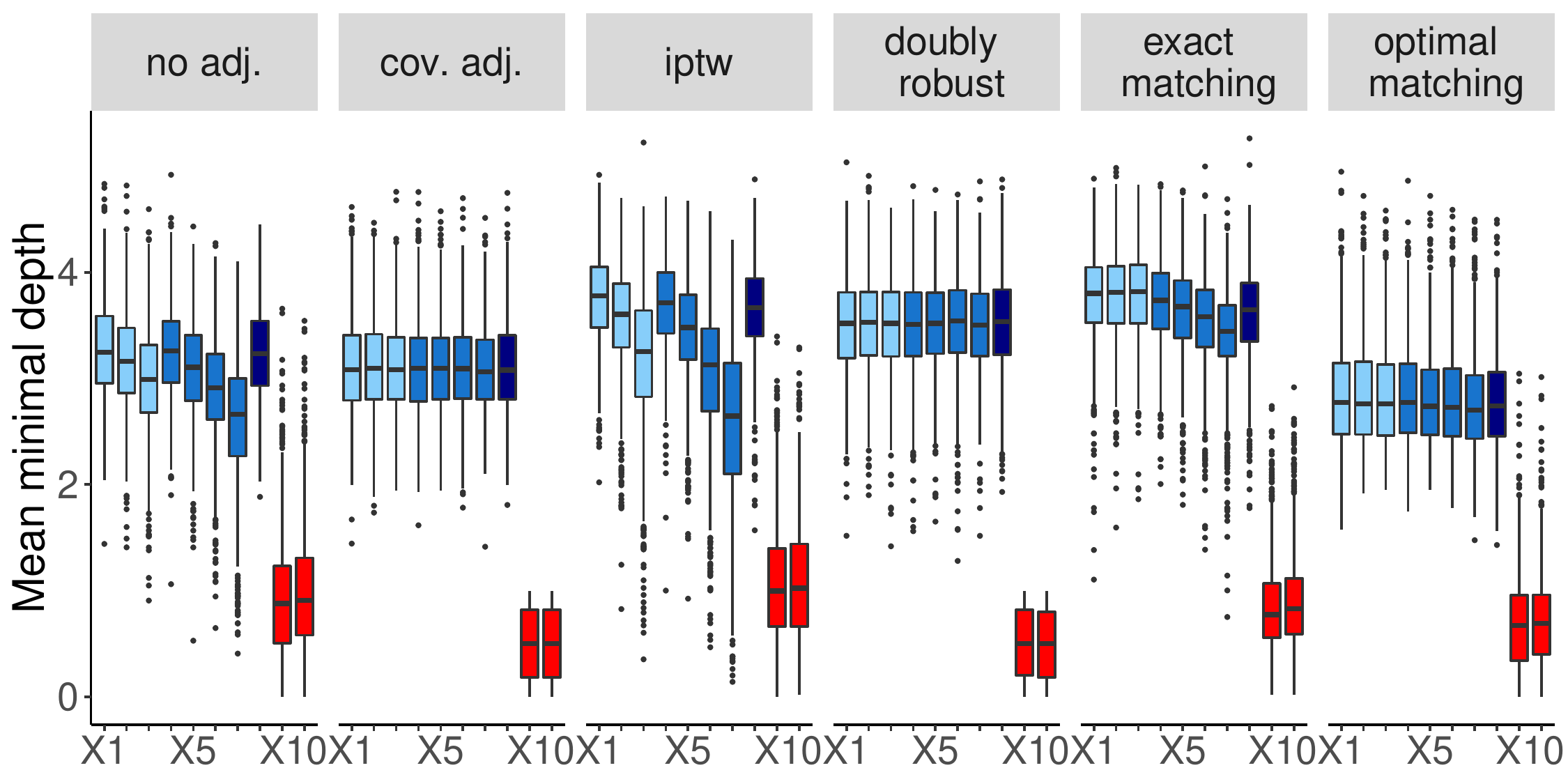}
\end{subfigure}
\end{center}
\end{figure}

\begin{sidewaysfigure}\ContinuedFloat
\centering
\begin{subfigure}{\textwidth}
\centering
\includegraphics[width=.8\textwidth]{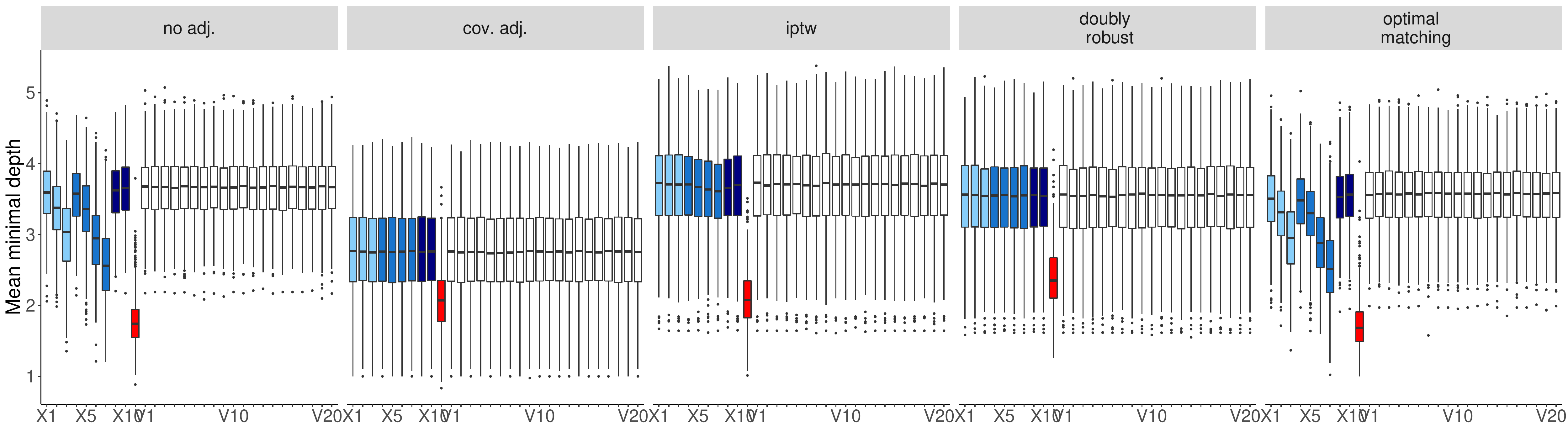}
\caption{\textbf{Scenario~J}: Additional nuisance biomarkers $V_1, \ldots, V_{20}$\\
$Y \sim N(\mu, 0.25)$ with $\mu=0.5 T + 0.2X_4 + 0.3X_5 + 0.4X_6 + 0.5X_7 + 0.4X_8 + 0.2X_9 -1 X_{10}\cdot T$}
\end{subfigure}
\caption[Mean minimal depth for the predMOB in combination with various adjustment methods in non-randomized settings]{Mean minimal depth for the predMOB in combination with various adjustment methods. Instrumental variables are shown in light blue, true confounders in medium blue and factors only associated with outcome in dark blue. The boxplot for a true predictive factor is highlighted in red.}
\label{Supp:fig:Simresults_mindepth}
\end{sidewaysfigure}